\def\cmod{\,\mathop{\rm cmod}\,}
\def\sigmaM{\sigma_{\rm M}}
\def\sigmaT{\sigma_{\rm T}}
\documentclass[aps,12pt,tightenlines,longbibliography,showpacs]{revtex4-1}

\usepackage[hidelinks]
{hyperref}
\usepackage{qcircuit}
\usepackage{amsfonts}
\usepackage{amsthm}
\usepackage{amsmath}
\usepackage{amssymb}
\usepackage{dsfont}
\usepackage{latexsym}
\usepackage{tikz}

\newtheorem*{thm}{Theorem}
\usepackage{braket}
\usepackage{graphicx}
\usepackage{xcolor}
\usepackage{empheq}
\makeatletter
\let\oldtheequation\theequation
\renewcommand\tagform@[1]{\maketag@@@{\ignorespaces#1\unskip\@@italiccorr}}
\renewcommand\theequation{(\oldtheequation)}
\makeatother

\newcommand{\id}{\mathds{1}}
\newcommand{\pp}{{\prime\prime}}

\newcommand{\appropto}{\mathrel{\vcenter{
  \offinterlineskip\halign{\hfil$##$\cr
    \propto\cr\noalign{\kern2pt}\sim\cr\noalign{\kern-2pt}}}}}

\DeclareMathOperator*{\argmax}{arg\,max}
\DeclareMathOperator*{\argmin}{arg\,min}

\newcommand{\highlight}[2]{\colorbox{#1!50}{$\displaystyle#2$}}

\newcommand{\e}{{\rm e}}
\newcommand{\T}{{\rm T}}
\newcommand{\bs}[1]{{\boldsymbol{#1}}}
\newcommand{\filledcircle}{\begin{tikzpicture}\draw (0,0) node[circle,
    fill=white,draw]{};\end{tikzpicture}}

\begin{document}

\title{Quantum Error Correction with the Toric-GKP Code}
\author{Christophe Vuillot$^{1}$\footnote{spokesperson: \href{mailto:c.vuillot@tudelft.nl}{c.vuillot@tudelft.nl}}, Hamed Asasi$^2$, Yang Wang$^1$, Leonid
  P. Pryadko$^2$, Barbara M. Terhal$^{1,3}$}
\affiliation{$^{1}$ QuTech, Delft University of Technology, P.O. Box 5046, 2600 GA Delft, The Netherlands \\
  $^{2}$ Department of Physics, University of California, Riverside,
  Riverside, California 92521, USA\\
  $^{3}$ JARA Institute for Quantum Information, Forschungszentrum
  Juelich, D-52425 Juelich, Germany} \date{\today}

\begin{abstract}
  We examine the performance of the single-mode Gottesman-Kitaev-Preskill (GKP) code and its concatenation with the toric code for a noise model of Gaussian shifts, or displacement errors.
  We show how one can optimize the tracking of errors in repeated noisy error correction for the GKP code.
  We do this by examining the maximum-likelihood problem for this setting and its mapping onto a 1D Euclidean path-integral modeling a particle in a random cosine potential.
  We demonstrate the efficiency of a minimum-energy decoding strategy as a proxy for the path integral evaluation.
  In the second part of this paper, we analyze and numerically assess the concatenation of the GKP code with the toric code. When toric code measurements and GKP error correction measurements are perfect, we find that by using GKP error information the toric code threshold improves from $10\%$ to $14\%$.
  When only the GKP error correction measurements are perfect we observe a threshold at $6\%$.

  In the more realistic setting when all error information is noisy, we show how to represent the maximum likelihood decoding problem for the toric-GKP code as a 3D compact QED model in the presence of a quenched random gauge field, an extension of the random-plaquette gauge model for the toric code.
  We present a new decoder for this problem which shows the existence of a noise threshold at shift-error standard deviation $\sigma_0 \approx 0.243$ for toric code measurements, data errors and GKP ancilla errors.
  If the errors only come from having imperfect GKP states, this corresponds to states with just 4 photons or more.

  Our last result is a no-go result for linear oscillator codes, encoding oscillators into oscillators.
  For the Gaussian displacement error model, we prove that encoding corresponds to squeezing the shift errors.
  This shows that linear oscillator codes are useless for quantum information protection against Gaussian shift errors.
 \end{abstract}

\maketitle

\tableofcontents

\section{Introduction}

Within the framework of oscillator or continuous-variable (CV) error correcting codes, one can distinguish two classes of codes.
One class generalizes qudit stabilizer codes to encode continuous degrees of freedom into a (larger) CV system \cite{braunstein:CV, LS:CV}.
We refer to these codes as \emph{linear oscillator codes}.
The other class, first introduced by Gottesman, Preskill and Kitaev (GKP) in Ref.~\cite{GKP}, and recently expanded to include many more codes \cite{michael+:bosonic, albert+:bos-codes}, encodes a discrete (finite-dimensional) system into a CV system.
Encoding and decoding for the first class of codes falls within the framework of Gaussian quantum information \cite{weedbrook+:RMP}, while the second class of codes requires using non-Gaussian states.

In this paper we propose and analyze a scalable use of the GKP code \cite{GKP} which encodes a single qubit into an oscillator.
An example of such an oscillator is a mode in a high-$Q$ microwave superconducting cavity coupled to superconducting qubits in a circuit-QED set-up.
Proposals for preparing a GKP code state in such systems exist \cite{TW:GKP}.
The CNOT gate between two GKP qubits requires about 4\,dB of squeezing in both modes and a beam-splitter (see, e.g., Ref.~\cite{TW:GKP}).
Such a beam-splitter has been recently implemented between high-$Q$ microwave cavity modes in Ref.~\cite{gao+:BS}.
Other possible physical implementations for the GKP code are the motional mode of a trapped-ion qubit \cite{fluehmann+:modular} or atomic ensembles \cite{motes+:GKP} for measurement-based CV cluster computation \cite{menicucci:ft}.

A bosonic code such as the GKP code or the recently-implemented cat code \cite{ofek+:cat} might be used to get a high-quality qubit, but the code does not provide a means to drive error rates down arbitrarily.
A {\em scalable} fault-tolerant architecture can possibly be obtained by concatenating the GKP code with a qubit stabilizer code such as the toric or surface code.
A theoretic goal is then to understand how to decode such a toric-GKP code and what is the error threshold of the architecture.
Some results on using ``analog" error information in concatenating the GKP code with a stabilizer code were obtained in Refs.~\cite{FTO:analog, thesis:yang, march:pryadko}.
A concatenation of the GKP code with the surface code was analyzed in Ref.~\cite{KTOF:concat} in the channel setting for 2D and 3D surface codes by message-passing (perfect) GKP error information to the surface code decoder.
However, this study did not look at error correction when the GKP syndrome is measured inaccurately.
Previous work has also studied the performance of the GKP code in comparison with other bosonic codes in a photon loss channel setting, not taking into account the imperfections or processing of repeated rounds of error correction \cite{albert+:bos-codes}.
Other work focused on the effect of photon loss and other sources of error on the preparation of code states \cite{DTW:sensor}. Besides its good performance compared to other bosonic codes, the GKP code is appealing since Clifford gates on the code states use only linear optical elements (including squeezing) \cite{GKP}.

In this work we first analyze repeated fault-tolerant quantum error correction for a single GKP qubit, see Section~\ref{sec:singleGKP}.
Our noise model in this analysis includes errors both on the GKP qubit as well as on the GKP ancilla qubit used in the error correction.
We show how decoding this continuous error information in discrete time steps maps onto the evaluation of a stochastic discrete-time Euclidean path integral.
We present an efficient minimum-energy decoder which chooses the path which approximately corresponds to a classical trajectory in a disordered potential. 

Second, we consider the toric-GKP code in Section \ref{sec:concat}.
Assuming that both GKP error correction and toric code correction are noiseless, we show how the use of continuous GKP
error information improves the error correction for the toric code (Section \ref{sec:noiseless}).
These results are in correspondence with the previous results in Ref.~\cite{KTOF:concat} although our likelihood function is not identical to the one in Ref.~\cite{KTOF:concat}.

In Section \ref{sec:3D} we formulate the decoding problem of repeated quantum error correction with the toric-GKP code where both the GKP syndrome and the toric code syndrome contain errors.
Since errors on the GKP qubits are intrinsic (getting perfect code states with infinite numbers of photons is unphysical), this is the physically relevant setting.
The maximum-likelihood formulation is in terms of a 3D gauge field model with quenched randomness determined by the errors (Section \ref{sec:gaugemodel}).
We discuss this model and its possible phase transitions in Section \ref{sec:phase-diag}.
Then in Section \ref{sec:decoder}, we show how to re-express this model as a random plaquette gauge model (RPGM) with a $\mathbb{Z}_2$-field coupled to an auxiliary $U(1)$-gauge field.
We then use this model to design a computationally-efficient decoder and present numerical results.

Finally, in Section \ref{sec:nogo}, we present our general no-go result for the first class of codes, namely the linear oscillator codes.
This no-go result is presented as the calculation of the probability distribution of logical errors on the encoded information after perfect maximum-likelihood decoding.
The result is in accordance with, but does not directly follow from previous no-go results on Gaussian quantum information in Ref.~\cite{NFC:nogo}.
The theorem explicitly shows that there are no linear oscillator code families of interest: there is no threshold in $\sigma_0$ below which protection of the encoded oscillators against shift errors gets better with increasing code size and the logical noise model is still Gaussian with the same $\sigma_0$, and possibly some squeezing of the logical quadratures.

The no-go result also shows that the existence of a threshold for the toric-GKP code is non-trivial.
A sufficiently large departure from Gaussian quantum information is necessary to stabilize quantum information.
In circuit-QED this departure comes exclusively from the use of the non-linear Josephson junction element. 

\section{General considerations}
\label{sec:backgroun}

\subsection{Definitions and notations}
\label{sec:notations}

We consider $n$-mode oscillator codes, which are subspaces in the $n$-mode Hilbert space $L_2(\mathbb{R}^n)$.
Such a Hilbert space can be constructed as a tensor product of $n$ single-particle Hilbert spaces $L_2(\mathbb{R})$ of complex square-integrable functions.
It supports $n$ pairs of canonically conjugated coordinate and momentum operators, $\hat p_k$ and $\hat q_k$, such that $[\hat{q}_k,\hat{p}_l]=i\delta_{kl}$.
These operators are used to define the multi-mode exponential shift operators,
\begin{equation}
  \label{eq:exponential-operator}
  \hat{U}(\bs e)\equiv \prod_{k=1}^n e^{i u_k\hat{p}_k+i v_k\hat{q}_k}, \quad \bs e\equiv (\bs u,\bs v),
\end{equation}
where $\bs u,\bs v\in\mathbb{R}^n$ are $n$-component real vectors.
It is easy to check that the product of two such operators satisfies 
\begin{equation}
  \label{eq:exponential-product}
  \hat{U}(\bs e)\hat{U}(\bs{e}') = \hat{U}(\bs e+\bs e')e^{i\omega(\bs e,\bs e')},
\end{equation}
with the phase given by the symplectic product $\omega(\bs e,\bs e')=\bs{u}\cdot\bs{v}'-\bs{v}\cdot \bs{u}'$.
The set $\mathcal{H}_n$ of all such operators with arbitrary phases is closed under multiplication, it forms an irreducible representation of the Heisenberg group $\mathbb{H}^n$ acting in $L_2(\mathbb{R}^n)$.
Just as for the $n$-qubit Hilbert space and the Pauli group $\mathcal{P}_n$, any operator acting in $L_2(\mathbb{R}^n)$ can be represented as a linear combination of elements of $\mathcal{H}_n$.
Furthermore, the product \ref{eq:exponential-product} of two exponential operators, up to a phase, can be represented in terms of the sum of the corresponding vectors, $\bs e''=\bs e+\bs e'$.
This map to $\mathbb{R}^{2n}$ is an analogue of the symplectic representation of $\mathcal{P}_n$ used in the theory of quantum codes.

An $n$-mode GKP code, ${\cal Q}$, is a CV stabilizer code defined in terms of an Abelian stabilizer group ${\cal S}\subset \mathcal{H}_n$ with elements in the form \ref{eq:exponential-operator}, such that $\hat{U}({\bs 0})\equiv\id$ is the only element in $\mathcal{S}$ proportional to the identity.
Namely, the code $\mathcal{Q}\subseteq L_2(\mathbb{R}^n)$ is the common $+1$-eigenspace of all elements of $\mathcal{S}$, 
\begin{equation}
  \label{eq:stabilizer-code}
  \mathcal{Q}=\{\ket\psi\in L_2(\mathbb{R}^n)\;|\;\hat{S}\ket\psi=\ket\psi, \;\forall \hat{S}\in\mathcal{S}\}. 
\end{equation}
The structure of such Abelian subgroups and the implications for $\mathcal{Q}$ are described in Appendix~\ref{sec:twoclasses}.
In the following, we will assume the representation of such a group in terms of some number $r$ of its members chosen as generators, $\mathcal{S}=\langle \hat{S}_1,\ldots,\hat{S}_r\rangle$, $\hat{S}_j\in\mathcal{H}_n$.

The formalism of qubit stabilizer codes \cite{gottesman-thesis,Hostens-Dehaene-DeMoor-2005} carries over entirely to such CV stabilizer codes and errors from $\mathcal{H}_n$ play the special role played by Pauli errors in the qubit case.
Given an error $\hat{E}\in\mathcal{H}_n$, one can compute its syndrome, $\bs{q}\equiv \bs{q}(\hat{E})$, whose components are given by the extra phases in the commutation relations with the stabilizer generators, $\hat{E}\hat{S}_j=\hat{S}_j\hat{E}e^{iq_j}$.
The set of errors which commute with all elements of the stabilizer group is called the centralizer ${\rm C}(\mathcal{S})$; these errors have a trivial syndrome, $\bs{q}=\bs{0}$.
Of these, any error that is a member of the stabilizer group acts trivially on code states, while the remaining errors $\hat{L}\in{\rm C}(\mathcal{S})\setminus\mathcal{S}$ act non-trivially within the code, they are called \emph{logical operators}.
An error $\hat{E}\in\mathcal{H}_n$ that does not commute with all stabilizer generators has a non-trivial syndrome and it takes $\mathcal{Q}$ into an orthogonal subspace $\hat{E}\mathcal{Q}\equiv\{\hat{E}\ket\psi\;|\;\ket\psi\in\mathcal{Q}\}$.
Two errors that differ by an element of the stabilizer group, $\hat{E}^\prime=\hat{E}\hat{S}$, $\hat{S}\in\mathcal{S}$, have the same syndrome, and as such, are called mutually \emph{degenerate}.
They act identically on the code and are also called \emph{equivalent}.
Two errors that differ by a logical operator, $\hat{E}^\prime=\hat{E}\hat{L}$, $\hat{L}\in{\rm C}(\mathcal{S})\setminus\mathcal{S}$, also have the same syndrome but they act differently on the code.
The set of inequivalent logical operators ${\rm L}({\cal S})$ is formed by the cosets of $\mathcal{S}$ in the centralizer ${\rm C}(\mathcal{S})$.
If we ignore the phases, the set of cosets ${\rm L}(\mathcal{S})$ actually forms a group, the group of logical operators.

By a slight abuse of notation, and when the global phase is irrelevant, we will often refer to an operator $\hat{U}(\bs{e})\in\mathcal{H}_n$ directly by its symplectic vector component, $\bs{e}\in\mathbb{R}^{2n}$.
For example, we can refer to a logical operator $\bs{c}\in{\rm L}(\mathcal{S})$ when we ought to write $\hat{U}(\bs{c})\in{\rm L}(\mathcal{S})$.
In accordance with classical codes terminology, we also refer to $\bs{c}\in{\rm L}(\mathcal{S})$ as a \emph{codeword}.
Furthermore, for two equivalent errors, ${\bs e}$ and $\bs{e}^\prime = \bs{e}+\bs{s}$, where ${\bs s}\in\mathcal{S}$,
we use $\bs e'\simeq \bs e$ to denote their equivalence, and $[\bs e]$ to denote the entire equivalence class,
\begin{equation}
[\bs e] = \left \{\bs{e}+\bs{s}:\forall\bs{s}\in\mathcal{S}\right \}.\label{eq:eqclass}
\end{equation}

Throughout this work we consider the independent Gaussian displacement channel ${\cal N}(\rho)$ with standard deviation $\sigma_0$:
\begin{equation}
{\cal N}(\rho)=\int_{-\infty}^{\infty}{\rm d}u \int_{-\infty}^{\infty}{\rm d}v\;\mathbb{P}_{\sigma_0}(u)\mathbb{P}_{\sigma_0}(v)e^{iu\hat{p}+iv\hat{q}} \rho \; e^{-iu\hat{p}-iv\hat{q}}, 
\label{eq:GDC}
\end{equation}
where $\rho$ is a single-mode density matrix and $\mathbb{P}_{\sigma_0}(x)$ the Gaussian probability density function with mean zero and variance $\sigma_0^2$, i.e. $\mathbb{P}_{\sigma_0}(x)=(2\pi \sigma_0^2)^{-1/2} e^{-x^2/2\sigma_0^2}$.
We will refer to $\sigma_0$ as the \emph{bare} standard deviation, this is because we will often consider \emph{scaled} observable, e.g. $\hat{P} = 2\alpha\hat{p}$ for which the corresponding effective rescaled standard deviation is $\sigma = 2\alpha\sigma_0$.
Even though this channel may not necessarily be the one which is physically most relevant, it is, like the Pauli error model, a convenient model which allows us to numerically and analytically model approximate GKP code states with finite levels of squeezing, see further motivation in Section~\ref{sec:setupsingle}.

As a convention we use bold italic symbols, such as $\bs{u}$, to denote row vectors.
We use hatted symbols, such as $\hat{P}, \hat{Q}$, for quantum operators and un-hatted symbols for the corresponding eigenvalues, such as $P$ and $Q$.
We will consider modulo values for real numbers quite often where, for convention, we chose the remainder to be in a symmetrical interval around 0.
For example, given $\phi\in\mathbb{R}$, writing $q=\phi \cmod 2\pi$ means that $q\in[-\pi, \pi)$ and $\phi = q + 2\pi k$ for some $k\in\mathbb{N}$.
In conventional notation $q := (\phi + \pi)\,\bmod\,2\pi - \pi$ and $k := \lfloor(\phi + \pi)/2\pi\rfloor$.
We also denote a range of integers as $[n]\equiv\{1,\dots,n\}$. We will refer to single-mode $q$-type errors as displacements of the form $\exp(i \eta \hat{q})$ for some $\eta$. Such errors induce shifts in $p$ and are alternatively called shift-in-$p$ errors. Similarly, for $p$-type errors which induce shifts in $q$. 

\subsection{Maximum-likelihood vs.\ minimum-energy decoding}
\label{sec:ml-decoding}

A (classical) binary linear code \cite{MS-book} of length $n$ encoding $k$ bits is a linear space of dimension $k$ formed by binary strings of length $n$, ${\cal C}\subseteq{\mathbb{F}}_2^n$.
For such a code, maximum likelihood (ML) syndrome-based decoding amounts to finding the most likely error which results in the given syndrome.
Generally, there are $2^{n-k}$ distinct syndromes and $|{\cal C}|=2^k$ codewords.
It is not hard to find a vector $\bs{e}$ which produces the correct syndrome; ML decoding can then be done by comparing the probabilities of errors $\mathbb{P}(\bs{e}+\bs{c})$, where $\bs{c}\in{\cal C}$ goes over all the codewords.
In the simplest case of the binary symmetric channel, the probabilities scale exponentially with error weight which can be thought of as the ``energy'' associated with the error.
Thus, for linear binary codes under the binary symmetric channel, ML decoding is the same as the minimum-weight, or \emph{minimum-energy} (ME) decoding.

Syndrome-based ML decoding for a qubit stabilizer code can be done similarly.
The main difference here is the degeneracy: errors that differ by an element of the stabilizer group are equivalent, they can not and need not be distinguished.
As a result, the probability $\mathbb{P}(E)$ of an $n$-qubit Pauli error $E\in\mathcal{P}_n$ needs to be replaced by the total probability to have any error equivalent to $E$.
In the case of Pauli errors which are independent on different qubits, quite generally, this probability can be interpreted as a partition function of certain random-bond Ising model \cite{dennis+:top,KP:spinglass}.
Exactly which statistical model one gets, depends on the code.
For a qubit square-lattice toric code with perfect stabilizer measurements the partition functions are those of 2D Random-Bond Ising model (RBIM).
Similarly, for the toric code with repeated noisy measurements, the partition function is that of a random-plaquette gauge model (RPGM) in three dimensions \cite{dennis+:top}, where the ``time'' dimension enumerates syndrome measurement cycles.  More general models are discussed, e.g. in Refs.~\cite{KP:spinglass,Dumer-Kovalev-Pryadko-bnd-2015}.

Instead of computing the partition functions proportional to the total probabilities of errors in different sectors, one could try finding a single most-likely error compatible with the syndrome.  It is the latter method that is usually called the ME decoding for a quantum code.
Indeed, in terms of the statistical-mechanical analogy, for ML decoding one needs to minimize the free energy, minus the logarithm of the partition function.
In comparison, for ME decoding, one only looks at a minimum-energy configuration (not necessarily unique); this ignores any entropy associated with degenerate configurations.
While the ME technique is strictly less accurate than ML decoding, in practice the difference may be small.

The two approaches are readily extended to GKP codes, both in the \emph{channel model} where perfect stabilizer measurement is assumed, and in the more general \emph{fault-tolerant} (FT) case where repeated measurements are used to offset the stabilizer measurement errors.
The latter case can be interpreted in terms of a larger \emph{space-time} code dealing with both the usual quantum errors and the measurement errors \cite{dennis+:top,Landahl-2011,Dumer-Kovalev-Pryadko-bnd-2015}.
One important aspect is that the quantum errors accumulate over time, while measurement errors in different measurement rounds are independent from each other.
This leads to an extended equivalence between combined data-syndrome errors which is similar to degeneracy.
The corresponding generators can be constructed, e.g., by starting with a single-oscillator error, followed by measurement errors on all adjacent stabilizer checks that result in zero syndrome, followed by the error which exactly cancels the original error.
Because of this cancellation, such an invisible error has no effect and should be counted as a part of the degeneracy group of the larger space-time code.

The following discussion applies to a GKP code in either the channel model or the fault-tolerant model.
In both cases we denote as $\mathcal{S}\subseteq \mathcal{H}_n$ the degeneracy group of the code.
In the channel model, $\mathcal{S}$ is exactly the stabilizer group, acting on the data oscillators.
In the fault-tolerant case, $\mathcal{S}$ is the degeneracy group of the space-time code, acting on both data oscillators as well as ancillary oscillators used to measure the syndrome.
Consider a multi-oscillator error, ${\bs e}\in\mathbb{R}^{2n}$, see Eq.~\ref{eq:exponential-operator}, and the corresponding probability density $\mathbb{P}(\bs e)$.
The probability is assumed to have a sharp [exponential or Gaussian, cf.\ Eq.~\ref{eq:GDC}] dependence on the components of $\bs e$; for this reason we can also write
\begin{equation}
\mathbb{P}(\bs e)=\exp\big(-H(\bs e)\big),\label{eq:prob-energy}
\end{equation}
where $ H(\bs e)$ is the dimensionless \emph{energy} associated with the error operator $\hat{U}(\bs e)$.
Syndrome-based ML decoding can be formulated as follows.
The error $\bs e$ yields the syndrome $\bs{q}(\bs e)$.
Given a logical operator $ {\bs c}\in {\rm L}(\mathcal{S})$, denote $\mathbb{P}([\bs e+\bs c]|\bs q)$ as the probability for any error in the class $[\bs e+\bs c]$, see \eqref{eq:eqclass}, conditioned on the syndrome $\bs{q}$.
This probability can be written as
\begin{eqnarray} 
\nonumber \mathbb{P}([\bs e+\bs c]| \bs{q})&=&\int_{\bs s\in\mathcal{S}} {\rm d}{\bs s}\;\mathbb{P}(\bs e+\bs c+\bs s|\bs{q})\\
  &=&\frac{1}{\mathbb{P}(\bs{q})}\int_{\bs s\in\mathcal{S}} {\rm d}\bs s\;\mathbb{P}(\bs e+\bs s+\bs c) \nonumber\\
  &=&\frac{1}{\mathbb{P}(\bs{q})}\int_{\bs s\in\mathcal{S}} {\rm d}\bs s\;\e^{-H(\bs{e}+\bs s+\bs c)}\equiv\frac{1}{\mathbb{P}(\bs{q})} Z_{\bs c}(\bs e),\label{eq:part}
\end{eqnarray}
where an appropriate integration measure should be used, and $\mathbb{P}(\bs q)$ is the net probability density to obtain the syndrome $\bs{q}=\bs{q}(\bs e)$.
Among all inequivalent codewords $\bs c\in {\rm L}(\mathcal{S})$, we select the most likely, i.e., with the largest $Z_{\bs c}(\bs e)$.
The probability of leaving a logical error $\bs{c}$ after ML decoding is the net probability of all the errors $\bs e$ for which the sector $[\bs e - \bs c]$ is the most likely, so
\begin{equation}
\mathbb{P}^{(\rm ML)}(\bs c)=\int_{\bs e:\;\forall \bs b\not\simeq(-\bs c),\, Z_{-\bs c}(\bs e) > Z_{\bs b}(\bs e)}{\rm d}\bs e \,\mathbb{P}(\bs e),\label{eq:succ-prob-init}
\end{equation}
[here we disregarded the contribution from sectors $\bs b\not\simeq (- \bs c)$ equiprobable with $(- \bs c)$, $Z_{- \bs c}(\bs e)= Z_{\bs b}(\bs e)$].
The probability of success of ML decoding can then be expressed as $\mathbb{P}^{(\rm ML)}_{\rm succ} = \mathbb{P}^{(\rm ML)}(\bs{0})$.
It is easy to see that any other decoding algorithm gives success probability that is not higher than that of ML decoding.
Indeed, a different algorithm would swap some errors $\bs e$ for $\bs e+\bs c$, which may reduce the measure in the corresponding analog of Eq.~\ref{eq:succ-prob-init}.

Furthermore, given an error $\bs{e}$, the probability $\mathbb{P}(\bs q) $ to obtain the syndrome $\bs{q}=\bs{q}(\bs e)$ can be written as $\mathbb{P}(\bs q)=\int_{\bs b\in{\rm L}(\mathcal{S})}{\rm d}\bs{b}\,Z_{\bs b}(\bs e)$, using the appropriate integration measure for the logical operators $\bs{b}$.
For this error $\bs{e}$ we denote $\bs{c}_{\rm max}(\bs{e})$ as its corresponding most likely sector,
\begin{equation}
	\bs{c}_{\rm max}(\bs{e}) =\argmax_{\bs{c}\in{\rm L}(\mathcal{S})}Z_{\bs{c}}(\bs{e}).
\end{equation}
The probability of a logical error $\bs{c}$ after ML decoding \ref{eq:succ-prob-init} can then be rewritten as an expectation by multiplying and dividing by $\mathbb{P}(\bs{q})$, changing variables, and re-summing, resulting in 
\begin{equation}
\label{eq:succ-decoding}
\mathbb{P}^{(\rm ML)}(\bs c) = \left\langle\mathbb{P}\big([\bs e+ \bs c_{\rm max} + \bs c]|\bs{q}(\bs e)\big)\right\rangle = \int \mathrm{d}\bs e\,\mathbb{P}(\bs e) \,{Z_{\bs c_{\rm max} + \bs c}(\bs e)\over\mathbb{P}(\bs{q})} = \int \mathrm{d}\bs e\,\mathbb{P}(\bs e) \,{Z_{\bs c_{\rm max} + \bs c}(\bs e)\over\int{\rm d}{\bs b}\,Z_{\bs b}(\bs e)}.
\end{equation}
ML decoding is successful if the most likely error is actually the one that happened, which corresponds to the trivial sector $\bs c\simeq \bs 0$ being dominant over all other sectors $\bs 0\not\simeq \bs c\in{\rm L}(\mathcal{S})$.
Given the error probability distribution $\mathbb{P}(\bs e)$, we say that a sequence of discrete GKP codes of increasing length $n$ is in the \emph{decodable phase} if $\mathbb{P}_{\rm succ}^{\rm(ML)}\equiv \mathbb{P}^{\rm (ML)}(\bs{0})\to1$ with $n\to\infty$.

With the definitions \ref{eq:prob-energy} and \ref{eq:part}, $Z_{\bs{c}_{\rm max}}(\bs e)$, can be interpreted as a partition function of a classical model in the presence of
quenched randomness determined by the actual error $\bs e$.
The partition function $Z_{\bs{c}_{\rm max} + \bs c}(\bs e)$ differs by an addition of a defect, e.g. a homologically non-trivial domain wall at the locations specified by non-zero components of the codeword $\bs c$.
Having already $\bs{c}_{\rm max}(\bs{e})\not\simeq\bs{0}$ means that the disorder, $\bs{e}$, energetically favors the domain wall $\bs{c}_{\rm max}$.
In the following, we will also consider the free energy, $F_{\bs{c}}(\bs e)$,
\begin{equation}
F_{\bs{c}}(\bs e)\equiv-\ln Z_{\bs{c}_{\rm max}+\bs c}(\bs e),\label{eq:free-energy}
\end{equation}
as well as the corresponding average $\langle F_{\bs c}\rangle\equiv \int {\rm d}\bs e\,\mathbb{P}(\bs e)\,F_{\bs c}(\bs e)$.
It follows from the Gibbs inequality that below the error-correction threshold for the noise parameters in $\mathbb{P}(\bs{e})$, the free energy increment $\Delta F_{\bs c}\equiv F_{\bs c}(\bs e)-F_{\bs{0}}(\bs e)$ associated with a logically-distinct ``incorrect'' class ($\bs c\not\simeq\bs 0$) necessarily diverges with $n$ for any error, $\bs e$, likely to happen \cite{KP:spinglass}.
More precisely, if ML decoding is asymptotically successful with probability one, $\mathbb{P}^{(\rm ML)}_\mathrm{succ}\to1$, the average free energy increment, $\langle \Delta F_{\bs c}\rangle $ associated with any non-trivial codeword $\bs c\not\simeq{\bs 0}$ must diverge for $n\to\infty$.
Such a divergence can be seen as a signature of a phase transition in the corresponding model. 

As is the case of the surface codes \cite{dennis+:top}, the partition functions $Z_{\bs c}(\bs e)$ are evaluated at a temperature that is not a free parameter but depends on the distribution $\mathbb{P}(\bs e)$.
For the sake of understanding the physics of the corresponding models, we could relax this, e.g. by additionally rescaling the energy $H(\bs e)\to \beta H(\bs e)$, cf.~Eq.~\ref{eq:prob-energy}, in the definition \ref{eq:part} of the partition function $Z_{\bs c}(\bs e)$, while keeping the original error probability distribution in the average \ref{eq:succ-decoding}.
This amounts to using ML decoder with incorrect input information, thus the corresponding success probability is not expected to increase,
\begin{equation}
  \mathbb{P}_\mathrm{succ}^{\rm(ML)}(\beta)\le \mathbb{P}_\mathrm{succ}^{\rm(ML)}(\beta=1)\equiv\mathbb{P}_\mathrm{succ}^{\rm(ML)},\label{eq:decoding-cmp}
\end{equation}
similar to decoding away from the Nishimori line in the case of qubit stabilizer code \cite{dennis+:top,KP:spinglass}.
In particular, the limit $\beta\to\infty$ corresponds to ME decoding, where we are choosing the codeword $\bs c$ to minimize the function
\begin{equation}
H_{\bs c}(\bs e)\equiv \min_{\bs s\in\mathcal{S}} H(\bs e+\bs c+\bs s).\label{eq:min-energy}
\end{equation}

\section{Protecting a Single GKP Qubit}
\label{sec:singleGKP}

\subsection{Set-up}
\label{sec:setupsingle}

The single-mode GKP code \cite{GKP} is a prescription to encode a qubit ---a two-dimensional Hilbert space--- into the Hilbert space of an oscillator using a discrete subgroup of displacement operators ${\cal H}_1$ as the stabilizer group. One chooses the two commuting displacement operators, $S_p=e^{2i \alpha \hat{p}}$ and $S_q=e^{i 2 \pi \hat{q}/\alpha}$, where $\alpha\neq0$ is any real number.
For this encoded qubit the (logical) Pauli operators are $Z=e^{i \pi \hat{q}/\alpha}$ (with $Z^2=S_q$) and $X=e^{i \alpha \hat{p}}$ (with $X^2=S_p$).
One can verify that $XZ =-ZX$.
The oscillator observables, $\hat{P}=2\alpha \hat{p}$ and $\hat{Q}=2\pi \hat{q}/\alpha$, can both take any value $2\pi k$ for $k \in \mathbb{Z}$ on ideal codewords.
The codeword $\ket{\overline{0}}$ (respectively $\ket{\overline{1}}$) is distinguished by $k$ being even (respectively odd).
The action of phase space translations $S_p$ (respectively $S_q$) on the eigenvalues of $\hat{Q}$ (respectively $\hat{P}$) is $Q \rightarrow Q+4 \pi$ (respectively $P \rightarrow P+4 \pi$).
The action of $X$ (respectively $Z$) is $Q \rightarrow Q+2 \pi$ (respectively $P \rightarrow P+2 \pi$).

A visual representation can be obtained by imagining the variables $Q$ and $P$ as a torus in phase space with both handles of circumference $2\pi$.
In this representation $S_p$ lets $Q$ wind around the handle exactly twice, while $X$ lets $Q$ go around the handle exactly once.
A correctable error constitutes a shift in $Q$ by less than half the circumference.
In this convenient representation, a logical error thus occurs when the winding number is odd, and no error occurs when the winding number is even.
The shifts in $P$ corresponds to windings around the other handle of the torus.

We will assume that the oscillator undergoes noise modelled as a Gaussian displacement channel with \emph{bare} standard deviation $\sigma_0$, see Eq.~\eqref{eq:GDC}.
The effect on the scaled observables $\hat{P}$ and $\hat{Q}$ is to map $P \rightarrow P+\epsilon_p$ and $Q \rightarrow Q+\epsilon_q$ where $\epsilon_p$ and $\epsilon_q$ are drawn from Gaussian distributions with \emph{rescaled} variances
\begin{equation}
\sigma_P^2=4\alpha^2 \sigma_0^2\;\text{\ and\ }\;\sigma_Q^2=\frac{4 \pi^2 \sigma_0^2}{\alpha^2}.
\label{eq:scaled-variances}
\end{equation}
For symmetry reasons, $\alpha$ is chosen to be $\sqrt{\pi}$ and we write $\sigma = \sigma_P = \sigma_Q$.
Given perfect measurements of $S_{p}$, the error $\epsilon_{q}$ can be corrected if $\left\vert\epsilon_{q}\cmod4\pi\right \vert< \pi$.

\begin{figure}[htbp]
		\centering
	\begin{minipage}{0.2\textwidth}
		\Qcircuit @C=.8em @R=0em {
			& \gate{\mathcal{N}} & \gate{\rm EC_{GKP}\left
				(\mathcal{N}_{\rm M}\right )} & \qw & & \equiv & }
	\end{minipage}
	\begin{minipage}{0.6\textwidth}
		\Qcircuit @C=.8em @R=.7em {
			& \gate{\cal{N}} & \qw &\qw &\qw &\qw & \ctrl{1} & \qw                          & \qw & \qw & \qw                                &  \qw & \targ      & \qw & \qw & \qw & \qw & \\
			&&&&& \lstick{\ket{\overline{+}}} & \targ    & \gate{{\cal N}_{\rm M}} &  \measureD{\hat{q}} & & & \lstick{\ket{\overline{0}}} & \ctrl{-1} & \gate{{\cal N}_{\rm M}} \qw & \measureD{\hat{p}} \gategroup{1}{4}{2}{15}{1.3em}{--}\\
		}     
	\end{minipage}
	\caption{A single round of fault-tolerant GKP syndrome measurement for both $q$ and $p$ shifts.
		Here $\ket{\overline{+}}$ is the $+1$ eigenstate of $S_q$ and $X$, and $\ket{\overline{0}}$ is the $+1$ eigenstate of $S_p$ and $Z$.
		The CNOT gate is the logical CNOT for the GKP code which induces the transformation $q_{\rm target} \rightarrow q_{\rm control}+q_{\rm target}$ (while $p_{\rm control} \rightarrow p_{\rm control}-p_{\rm target}, q_{\rm control}\rightarrow q_{\rm control},p_{\rm target}\rightarrow p_{\rm target}$).
		Each measurement is a perfect homodyne measurement of $\hat{q}$ or $\hat{p}$.
		$\mathcal{N}$ are $\mathcal{N}_{\rm M}$ are Gaussian displacement channels in Eq.~\eqref{eq:GDC} which model shift errors on the encoded state in each round of error correction, respectively shift errors in the homodyne measurement.
	}
	\label{fig:EC_GKP}
\end{figure}
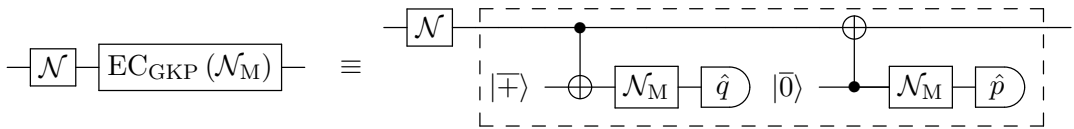

In order to measure stabilizer generators $S_p$ and $S_q$ we consider the fault-tolerant Steane measurement circuits \cite{GKP} in Fig.~\ref{fig:EC_GKP}, where encoded $\ket{\overline{+}}$ or $\ket{\overline{0}}$ ancillas, CNOTs and $\hat{q}$ or $\hat{p}$ measurements are used.

For simplicity, we consider the ancilla preparations, the CNOT and the $\hat{q}$ and $\hat{p}$ measurements to be perfect and only add Gaussian displacement channels on the data qubit and on the ancilla qubit right before its measurement.
Doing this ignores the back propagation of $q$-type errors to the data due to an imperfect ancilla \cite{GK:osc}, but if we treat $p$-type and $q$-type error correction independently, then this back-propagation does not fundamentally alter the noise model.
We will keep the freedom of choosing different standard deviations for the data and the ancilla errors and denote as $\sigmaM$ the \emph{scaled} standard deviation for the ancilla errors.

What is important is that our error model covers dominant sources of imperfections stochastically.
Any physically-realistic GKP code state has finite photon number $\overline{n}$ and one reasonable model of such finite-photon GKP state is a coherent superposition of Gaussian displacement errors on a perfect code state, see Eqs.~(40),(41) in Ref.~\cite{GKP}.
The quality of such an approximate GKP state can be given by an effective squeezing parameter $\Delta$ with $\overline{n} \sim \frac{1}{2 \Delta^2}-\frac{1}{2}$.
Assuming a coherent superposition of Gaussian displacements can be replaced by a Gaussian mixture of displacements on a perfect state we can identify $\Delta^2=2 \sigma_0^2$.
If errors are dominated by such a finite squeezing/finite photon number, we could use $\sigma_0 \sim \frac{1}{\sqrt{2(2\overline{n}+1)}}$ to interpret our numerical data.
For example, $\overline{n}=4$ gives $\sigma_0 \approx 0.236$.

Besides this, it has been shown that photon loss with rate $\gamma$ followed by an amplification or pumping step produces the Gaussian displacement channel with $\sigma_0^2=\frac{\gamma}{1-\gamma}$ \cite{albert+:bos-codes}.
For example, the rate $\gamma=0.02$ corresponds to $\sigma_0=0.14$.
Such an amplification step on the GKP data qubit could be added in each step of error correction.

\begin{figure}[htbp]
	\centering
	\begin{minipage}{0.9\textwidth}
		\Qcircuit @C=.8em @R=.7em {
			&&	\lstick{\ket{\overline{\Psi}}}& \gate{{\cal N}} & \ctrl{1} & \qw & \qw & \qw & \qw & \qw & \gate{{\cal N}} & \ctrl{1} &  \qw & \qw & & \cdots &  & &\gate{{\cal N}}& \qw & \ctrl{1} & \gate{{\cal N}} & \measureD{\hat{q}} \\
			&&	\lstick{\ket{\overline{+}}}& \qw & \targ  & \gate{{\cal N}_{\rm M}} & \measureD{\hat{q}} & & & &\lstick{\ket{\overline{+}}} & \targ & \gate{{\cal N}_{\rm M}} & \measureD{\hat{q}} & &\cdots &  &
			& \lstick{\ket{\overline{+}}} & \qw & \targ & \gate{{\cal N}_{\rm M}} & \measureD{\hat{q}} \\
	}         \end{minipage}
	\caption{Repeated rounds of error correction for the GKP code to detect and keep track of error shifts in $\hat{q}$ followed by a final destructive measurement of the data (modeled as $\mathcal{N}$ followed by a perfect measurement of $\hat{q}$).
		No explicit corrections based on the measured values are shown.}
	\label{fig:repeated-QEC}
\end{figure}
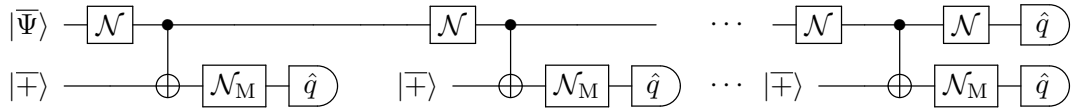

Since the measurement outcomes in Fig.~\ref{fig:EC_GKP} are inaccurate, they cannot be used to infer a correction which maps the state back to the code space.
In order to perform error correction one has to measure frequently and try to use the record of measurements to stay as close as possible to the code space without incurring logical errors to preserve the codeword. Figure~\ref{fig:repeated-QEC} shows this repeated measurement protocol for $p$-type errors (or shift-in-$q$ errors). 

We will analyse only $p$-type data errors with scaled standard deviation $\sigma$ and measurement errors with scaled standard deviation $\sigmaM$, cf.\ Eq.~\ref{eq:scaled-variances}.
The analysis for $q$-type errors would be similar.
When considering the realization of a particular shift error we will use the following notation: $\epsilon_t\in\mathbb{R}$ is the shift error occurring on the data before the $t^{\rm th}$ measurement, $\delta_t\in\mathbb{R}$ is the measurement error occuring at the $t^{\rm th}$ step.
Furthermore, $q_t\in[-\pi,\pi)$ is the $t^{\rm th}$ measurement outcome for the rescaled variable $\hat{Q}$ and $\phi_t=\epsilon_1+\cdots +\epsilon_t$ is the cumulative shift on the data.
The relations between these quantities are
\begin{equation}
q_t = \phi_t+\delta_t\cmod2\pi,\qquad\phi_t - \phi_{t-1} = \epsilon_t,\qquad \phi_0\equiv 0.\label{eq:qphie}
\end{equation}

We consider a total of $M$ rounds of GKP measurements indexed by $t\in[M]$.
Of these, the last measurement is assumed perfect, incorporating any measurement error into the corresponding shift error.
Specifically, we write $\phi_M=\phi_{M-1} + \epsilon_M $, $ q_M=\phi_M \cmod 2\pi$, so that $\delta_M=0$.
This last measurement can be thought of as a destructive measurement performed directly on the data without the use of an ancilla, as one would do to retrieve the encoded information.
As such, the last data error $\epsilon_M$ can equivalently be thought of as the last measurement error on the destructive measurement of the data.
Having this last perfect measurement permits to map back to the code space and easily define successful or failed error correction.
Specifically, we are trying to determine the parity of $k_M$ in the relation $q_M=\phi_M+2\pi k_M$; error correction is successful as long as we determined the parity correctly.
We denote the set of $M$ measurements as $\bs{q}$ and $M$ cumulative shift errors as $\bs{\phi}$.

To get some intuition, imagine that we apply a single round of error correction of Fig.~\ref{fig:repeated-QEC} and $\mathcal{N}_{\rm M}$ is the identity channel.
The ancilla qubit $\ket{\overline{+}}$ is a uniform sum of delta functions with $Q=q=0 \cmod 2 \pi$, hence we represent
the measurement outcome compactly as $q \in [-\pi,\pi)$.
An incoming logical X on the data qubit is pushed (through the CNOT) onto the ancilla qubit where it translates $q$ by a full $2\pi$-period, hence logical information is not observed.
One corrects a shift of up to $\pi$ (at most half-a-logical) by shifting $Q$ back by the least amount to make it again equal to $0 \cmod 2\pi$.  

\subsection{Decoding Strategies}

We start by describing the maximum-likelihood strategy.
Given the measurement record, one would like to compute the conditional probabilities for different classes of errors which are distinguished by their logical action.
In this case of correcting a single qubit against shift errors in $q$, one has to decide whether there was an $X$ error or there was none.
Knowing the details of the error model, namely $\sigma$ and $\sigmaM$, one can write down the probability of these two
classes.
Formally, they are given by
\begin{equation}
  \mathbb{P}(0\vert\bs{q}) = \int_{I_0}\mathbb{P}(\bs{\phi}\vert\bs{q})\mathrm{d}\bs{\phi}, \qquad\mathbb{P}(1\vert\bs{q}) = \int_{I_1}\mathbb{P}(\bs{\phi}\vert\bs{q})\mathrm{d}\bs{\phi},  
  \label{eq:probasuccess}  
\end{equation}
where the integration covers all possible realizations of the shift errors described by $\bs{\phi}$, and $I_0$ (respectively, $I_1$) limits the integral to realizations leaving no $X$ error (respectively, leaving an $X$ error).
Since the last measurement is assumed perfect, $I_0$ and $I_1$ are characterized by $\delta\left (\phi_M -q_M+ 2\pi k_M\right )$, with any even $k_M$ in $I_0$ and any odd $k_M$ for $I_1$.  

In practice, to do decoding for the given measurement history, $\bs q$, one needs to compare the probabilities \ref{eq:probasuccess}.
ML decoding algorithm suggests that a logical $X$ correction is needed if $\mathbb{P}(1\vert\bs{q})>\mathbb{P}(0\vert\bs{q})$.
Of course, this does not guarantee success in each particular trial.
If we take just one measurement round, $M=1$, which corresponds to measuring the data directly, we get
\begin{align}
  \mathbb{P}_{M=1}(0\vert q_1) &=\sum_{k_1 \text{even}} \int_{-\infty}^\infty d\epsilon_1\;\mathbb{P}(\epsilon_1\vert q_1)\delta(\epsilon_1 - q_1 + 2k_1\pi)\nonumber\\ 
  &\propto\sum_{k_1 \text{ even}}\mathbb{P}_\sigma(q_1-2\pi k_1),\label{eq:succ}
\end{align}
and $\mathbb{P}_{M=1}(1\vert q_1)$ is given by the complementary sum over odd $k_1$, which makes the normalization the full sum with $k_1$ running over all integer values.

To compute these probabilities in general, we apply Bayes' rule:
\begin{equation}
  \mathbb{P}(\bs{\phi}\vert\bs{q}) = \frac{\mathbb{P}(\bs{q}\vert\bs{\phi})\mathbb{P}(\bs{\phi})}{\mathbb{P}(\bs{q})}.
  \label{eq:bayes}  
\end{equation}
Then the probability for some outcome $\bs{q}$ given data errors $\bs{\phi}$ can be computed from the measurement error model and the probability for some data error $\bs{\phi}$ from the data error model.
The normalization, $\mathbb{P}(\bs{q})$, can be computed by integrating the numerator over every $\bs{\phi}$.
Using Eq.~\eqref{eq:qphie}, we have from Eq.~\eqref{eq:bayes} 
\begin{align}
  \mathbb{P}(\bs{\phi}\vert\bs{q})\mathbb{P}(\bs{q})&\nonumber\\ 
  \propto\sum_{\bs{k}\in\mathbb{Z}^{M}}&\left[\prod_{t=1}^{M-1}\exp\left(-\frac{\left (q_t - \phi_t + 2\pi k_t\right)^2}{2\sigmaM^2}\right)\prod_{t=1}^{M}\exp\left(-\frac{\left(\phi_t - \phi_{t-1}\right)^2}{2\sigma^2}\right)\delta(\phi_M-q_M-2\pi k_M)\right].\label{eq:fullexpr} 
\end{align}
Recalling Eq.~\ref{eq:part}, we write the corresponding complementary probabilities \ref{eq:probasuccess} in
terms of partition functions,
\begin{eqnarray}
  \label{eq:Z0-one-gkp-def} 
  \mathbb{P}(0\vert\bs q)&=&\frac{Z_0(\bs q)}{\mathbb{P}(\bs{q})},\quad\mathbb{P}(1\vert\bs q)=\frac{Z_1(\bs q)}{\mathbb{P}(\bs{q})},\\
  Z_c(\bs q)&=&N^{-1}\int {\rm d}\bs \phi\sum_{\bs k\in\mathbb{Z}^{M-1}}\exp\left(-\sum_{t=1}^{M-1}\frac{\left(q_t - \phi_t+2\pi k_t\right)^2}{2\sigmaM^2}\right)\exp\left(-\sum_{t=1}^{M}\frac{\left (\phi_t -\phi_{t-1}\right)^2}{2\sigma^2}\right) \nonumber \\ 
  & & \hskip1.5in \times \sum_{k_M\in\mathbb{Z}}\delta(\phi_M-q_M-2\pi c-4\pi k_M), \quad c=0,1,
  \label{eq:Z0-one-gkp}
\end{eqnarray}
with $N$ a normalization constant \footnote{Strictly speaking the normalization constant, $N$, diverges due to $\mathbb{P}(\bs{q})$ being an infinite sum of delta peaks. In practice we always compute the quantities $Z_c(\bs q)$ with a cutoff and can adjust $N$ such that $1 = Z_0(\bs q)+Z_1(\bs q)$.}.
In this special case picking $\bs{e} = \bs{q}$, for a candidate error is always a valid choice, that is why we can write directly $Z_c(\bs{q})$.

The evaluation of the Gaussian integrals in Eq.~\ref{eq:Z0-one-gkp} (see Appendix~\ref{sec:gauss-int}) gives
\begin{equation}
Z_c(\bs{q}) ={(2\pi)^{(M-1)/2}\over N(\det B)^{1/2}}\sum_{\bs{k}=(\bs{\tilde{k}},c+2m) \colon \bs{\tilde{k}} \in
  \mathbb{Z}^{M-1},\; m\in\mathbb{Z}}\exp\left (-\frac{1}{2}\left(\bs{q} + 2\pi\bs{k}\right)A\left(\bs{q} + 2\pi\bs{k}\right)^\T\right),\label{eq:mldexpr}
\end{equation}
where $B$ and $A$ are symmetric positive-definite matrices given explicitly by Eqs.~\ref{eq:mat-B} and \ref{eq:mat-A}.
The sums with $c\in\{0,1\}$ can be numerically computed by setting a cut-off $K$, restricting every $k_t$ to the interval $-K\leq k_t\leq K$.
The number of terms then still exponentially increases with the number of rounds.
We used Eq.~\ref{eq:mldexpr} with a cut-off $K=2$ for up to $M=7$ rounds for the results shown in Fig.~\ref{fig:expdecay} and Fig.~\ref{fig:single_gkp_plots}.
Intuitively, this cut-off corresponds to only considering events where the measurement shift errors let one wind around the torus at most twice in each round.
This is pretty reasonable since these errors follow a Gaussian distribution with small variance.

Generally, a more clever way to calculate the sum in Eq.~\eqref{eq:mldexpr} is to express it in terms of a genus-$M$ Riemann theta function \cite{nist-theta-f}, and then transform the matrix so that the summation terms can be rearranged in decreasing order, stopping at a desired precision.
However, this requires solving a shortest vector problem with the eigenvectors of $A$ and is therefore also computationally difficult \cite{Deconinck-etal-2004,Frauendiener-Jaber-Klein-2017}.

In addition to the formally exact but hard to calculate expressions \ref{eq:Z0-one-gkp}, \ref{eq:mldexpr} for the conditional probabilities, we would like to consider a class of approximate minimum-energy solutions of the corresponding optimization problem.
To this end, we define a $2\pi$-periodic potential, $V_\sigma(x)=V_\sigma(x+2\pi)$,
\begin{equation}
  \label{eq:Villain}
  \exp\big(-V_\sigma(x)\big)\equiv \sum_{k\in\mathbb{Z}}e^{-(x+2\pi k)^2/2\sigma^2}.
\end{equation}

The periodicity of the sum of the Gaussians implies that one should be able to approximate $V_\sigma(x)$ by its principal Fourier harmonic,
\begin{equation}
  \label{eq:villain-form}
  V_\sigma(x)\approx A_0-\beta_V(\sigma)\cos x,
\end{equation}
where $\beta_V$ is Villain's effective inverse temperature parameter, and the overall shift $A_0$ is irrelevant.
Such a simplified form is exactly the approximation used by Villain \cite{Villain-75}, but ``in reverse.''
Indeed, for large $\beta$, one has \cite{janke}
\begin{equation}
e^{\beta \cos x} \approx e^{\beta}\sum_{k\in\mathbb{Z}}\exp\left(-\frac{\beta(x+2\pi k)^2}{2}\right), 
\label{eq:villain}
\end{equation}
which gives $\beta_V(\sigma)=1/\sigma^2$, $\sigma\ll1$.

With the defined periodic potential, the logarithm of the non-singular part of Eq.~\ref{eq:fullexpr} acquires a form of a discrete-time Euclidean action, cf. Eq.~\ref{eq:prob-energy},
\begin{equation}
  \label{eq:GKP-path-integral}
{H}(\bs{\phi};\bs{q})\equiv -\log[\mathbb{P}(\bs q|\bs\phi)\mathbb{P}(\bs \phi)]=\sum_{t=1}^{M} {(\phi_{t}-\phi_{t-1})^2\over 2\sigma^2}+\sum_{t=1}^{M-1}V_{\sigmaM}(q_t-\phi_t)+{\rm const}. 
\end{equation}
With the given values $\phi_{0}=0$ and $\phi_M$, the corresponding extremum can be found by solving the equations
\begin{equation}
\phi_{t+1}-2\phi_t+\phi_{t-1}+\sigma^2 V_{\sigmaM}'(q_t-\phi_t)=0,\quad t\in[M-1],\label{eq:GKP-FK-eqn}
\end{equation}
where $V_{\sigmaM}'(x)$ denotes the derivative of the potential in Eq.~\ref{eq:Villain}.
These equations can be readily solved one-by-one, starting with $\phi_{0}=0$ and some $\phi_1\equiv \phi$; the boundary condition $\phi_M=q_M+2\pi k_M$ can be satisfied by scanning over different values of $\phi_1$ in a relatively small range around zero, with the global minimum subsequently found by comparing the resulting values of the sum in Eq.~\ref{eq:GKP-path-integral}.
Then, any even value of $k_M$ corresponds to no logical error, while an odd $k_M$ indicates an $X$ error to be corrected.
While such a minimization technique gives the exact ME solution, in practice it is rather slow.
Namely, with increasing non-linearity $\sigma^2/\sigmaM^2$ and increasing length $M$ of the chain, a small change in $\phi_1$ may strongly affect the configuration of the entire chain.
Respectively, it is easy to miss an extremum corresponding to the global minimum.
This numerical complexity of minimizing Eq.~\ref{eq:GKP-path-integral} is a manifestation of chaotic behavior inherent in the equations \ref{eq:GKP-FK-eqn}.

Indeed, the problem of minimizing the energy \ref{eq:GKP-path-integral} can be interpreted as a disordered version of a generalized Frenkel-Kontorova (FK) model \cite{FrenkelKontorova}, where a chain of masses coupled by springs lies in a periodic potential.
In our setup, random shifts $q_t$ can be traded for randomness in the initial (unstretched) lengths of the springs, $q_t-q_{t-1}$.
The original FK model, with $V_{\sigma}(x)$ replaced by a harmonic function, is obtained if one uses the Villain approximation ``in reverse'', see Eq.~\ref{eq:villain-form}.
Even in the absence of disorder, the FK model is an example of a minimization problem with multiple competing minima which can be extremely close in energy.
The corresponding equations \ref{eq:GKP-FK-eqn}, viewed as a two-dimensional map $(\phi_{t-1},\phi_t)\to(\phi_t,\phi_{t+1})$, are a version of the Chirikov-Taylor area-preserving map from a square of size $2\pi$ to itself \cite{Kivshar-Benner-Braun-book-2008}, one of the canonical examples of emergent chaos.

\begin{figure}[htbp]
	\center
	\includegraphics[scale=0.4]{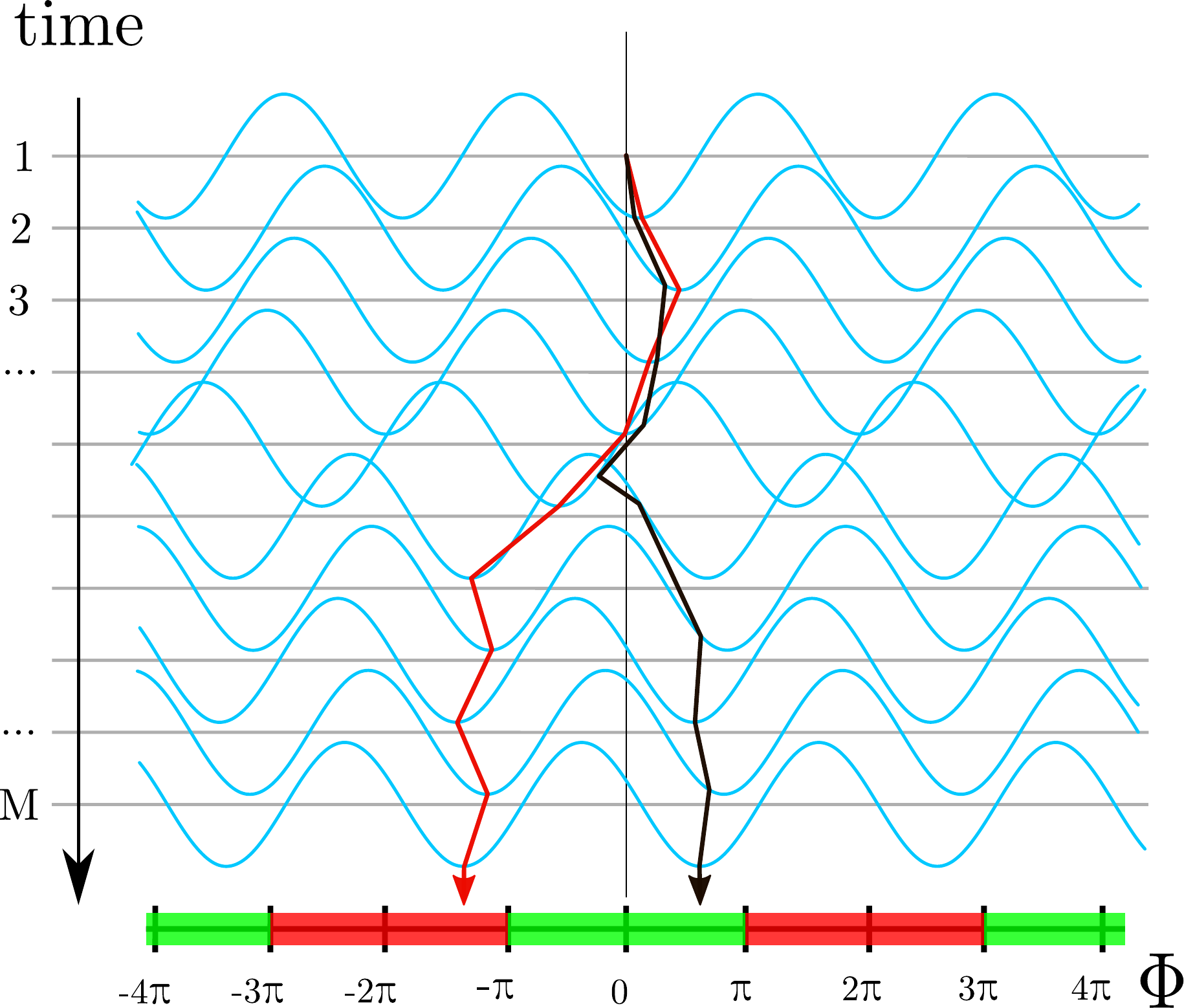}
	\caption{(Color online) Sketch of GKP decoding problem and its solution strategy.
		For each round of error correction $t\in[M]$ one has a Villain potential $V_{\sigmaM}(q_t-\phi_t)$, depicted in blue, with the minimum centered at the measured value $q_t$.
		Green and red intervals along the horizontal axis denote the sets $I_0$ and $I_1$ in Eq.~\ref{eq:probasuccess}, they correspond to realizations leaving no error or an $X$ error, respectively.
		In a memoryless decoding strategy one decides how to shift the code state after each measurement based on the value of $q_t$.
		In a maximum-likelihood decoding procedure one evaluates the path integral in Eq.~\eqref{eq:mldexpr}.
		In a minimum-energy decoder one determines an optimal path for the sequence $\phi_t$, $t\in[M]$, given the random potential and the quadratic ``kinetic energy'' term proportional to $(\phi_{t}-\phi_{t-1})^2$, see Eq.~\ref{eq:GKP-path-integral}.
		Red and black lines show two decoding trajectories which start at the same point but have different winding numbers.
		Upon a final decoding step, choosing the black trajectory leads to deciding that no logical $X$ error has taken place, since the final value $\phi_M$ lies in the green region.
		Choosing the red trajectory leads to deciding that a logical $X$ has taken place since the final value $\phi_M$ lies in the red region.
		Examples of actual trajectories are shown in Fig.~\ref{fig:dynprogcompare}.}
	\label{fig:winding-path-strategy}
\end{figure}

For this reason, and also in an attempt to come up with a numerically efficient decoding algorithm, we have designed the following approximate \emph{forward-minimization} technique.
For each $t<M$, starting from $t=1$, given the present value $\phi_{t-1}$, one determines the next value $\phi_t$ such that $\frac{\partial H}{\partial\phi_t}\bigr|_{\phi_{t+1}=\phi_t}=0$.
Given the syndrome $q_t$, this implies
\begin{equation}
  \phi_t = \argmin_{\phi}\left[\frac{\left(\phi-\phi_{t-1}\right)^2}{2} -\sigma^2 V_{\sigmaM}(q_t-\phi)\right ]\;\Rightarrow\;\phi_t={\sigma^2}V_{\sigmaM}'(q_t-\phi_t)+\phi_{t-1}. 
  \label{eq:path-min}
\end{equation}
At the end, after one obtains $\phi_{M-1}$, one chooses a $k_M$ such that $q_M + 2\pi k_M$ is the closest to $\phi_{M-1}$.
The parity of thus chosen $k_M$ then tells if a logical error happened.
This strategy is illustrated in Fig.~\ref{fig:winding-path-strategy}.

These equations are certainly different from the exact extremum equations \ref{eq:GKP-FK-eqn}, and the configuration found by this forward minimization technique necessarily has the energy higher than the exact minimum.
On the other hand, empirically, the corresponding energy difference is typically small, much smaller what one gets, if the correct minimum is missed by the formally exact technique based on Eqs.~\ref{eq:GKP-FK-eqn}.
Even though this technique is only an approximation, it is fast and is accurate enough in practice.
To ensure that the approximation doesn't hurt the performance of our decoder we have compared it to a rigorous dynamic programming approach, see Appendix~\ref{sec:dynprog}.
This comparison shows that the dynamic programming approach has very little advantage while it is substantially slower in its execution.

A very simple decoding strategy that one might also try is to trust every measurement outcome and immediately correct each round.
This doesn't require any memory so we refer to it as the \emph{memoryless} decoder.
Intuitively, this method is risky as every round transfers the measurement errors to the data, increasing the variance of the effective error model acting on the data.

\subsection{Numerical Results}
\label{sec:num-single}

\begin{figure}[htbp]
	\begin{minipage}[l]{0.45\textwidth}
		\centering
		\includegraphics[width=\textwidth]{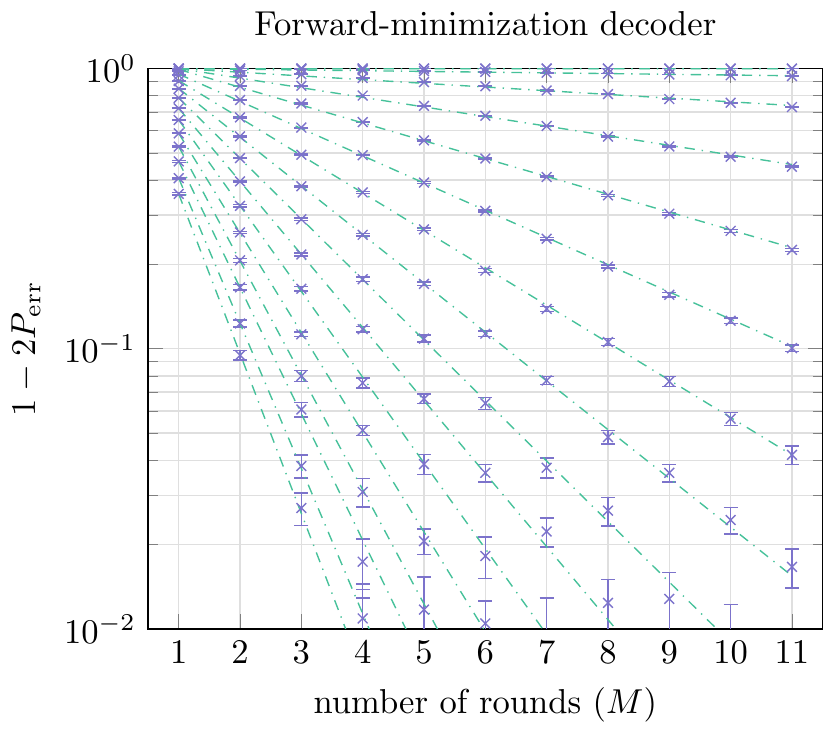}
	\end{minipage}\qquad
	\begin{minipage}[r]{0.45\textwidth}
		\centering
		\includegraphics[width=\textwidth]{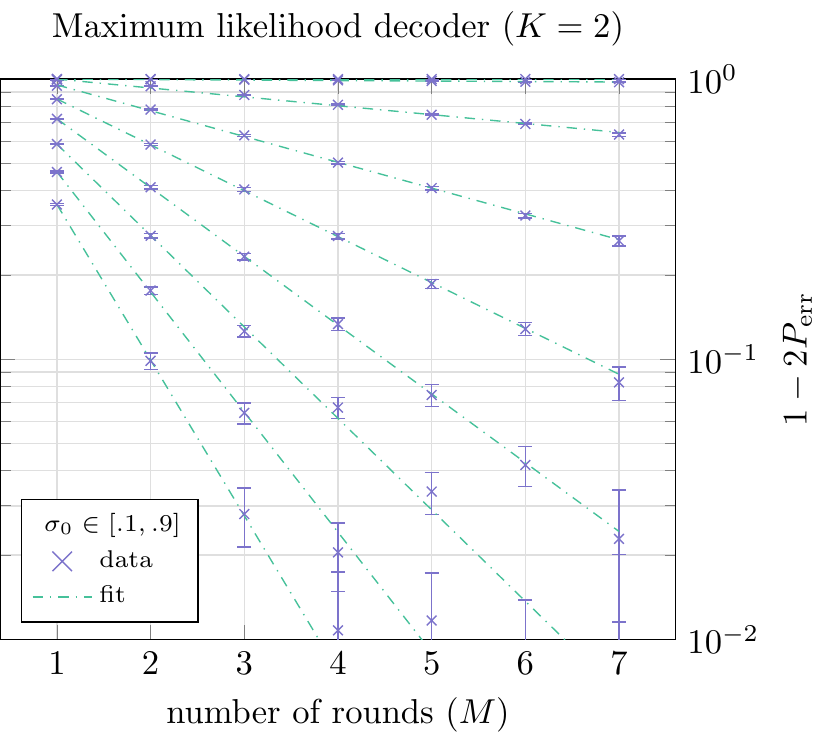}
	\end{minipage}
	\caption{(Color online) Some of the numerical results for the forward-minimization and maximum-likelihood decoders (with cut-off $K=2$).
		To observe the exponential decay towards $1/2$ we plot $1-2P_{\rm err}$ on a log scale for different number of rounds and different bare standard deviation $\sigma_0$.
		Low hundred thousands of trials are performed for each data point and the confidence intervals at $95\%$ are shown.
		For each $\sigma_0$ we fit an exponential decay, the slope gives us an effective logical error rate per round.
		On both plots the value for $\sigma_0$ varies between $.1$ and $.9$, on the left by $.05$ increments, on the right by $.1$ increments.
		All effective logical error rates are plotted in Fig.~\ref{fig:single_gkp_plots}.}
	\label{fig:expdecay}
\end{figure} 

\begin{figure}[htbp]
		\centering
		\includegraphics[width=0.95\textwidth]{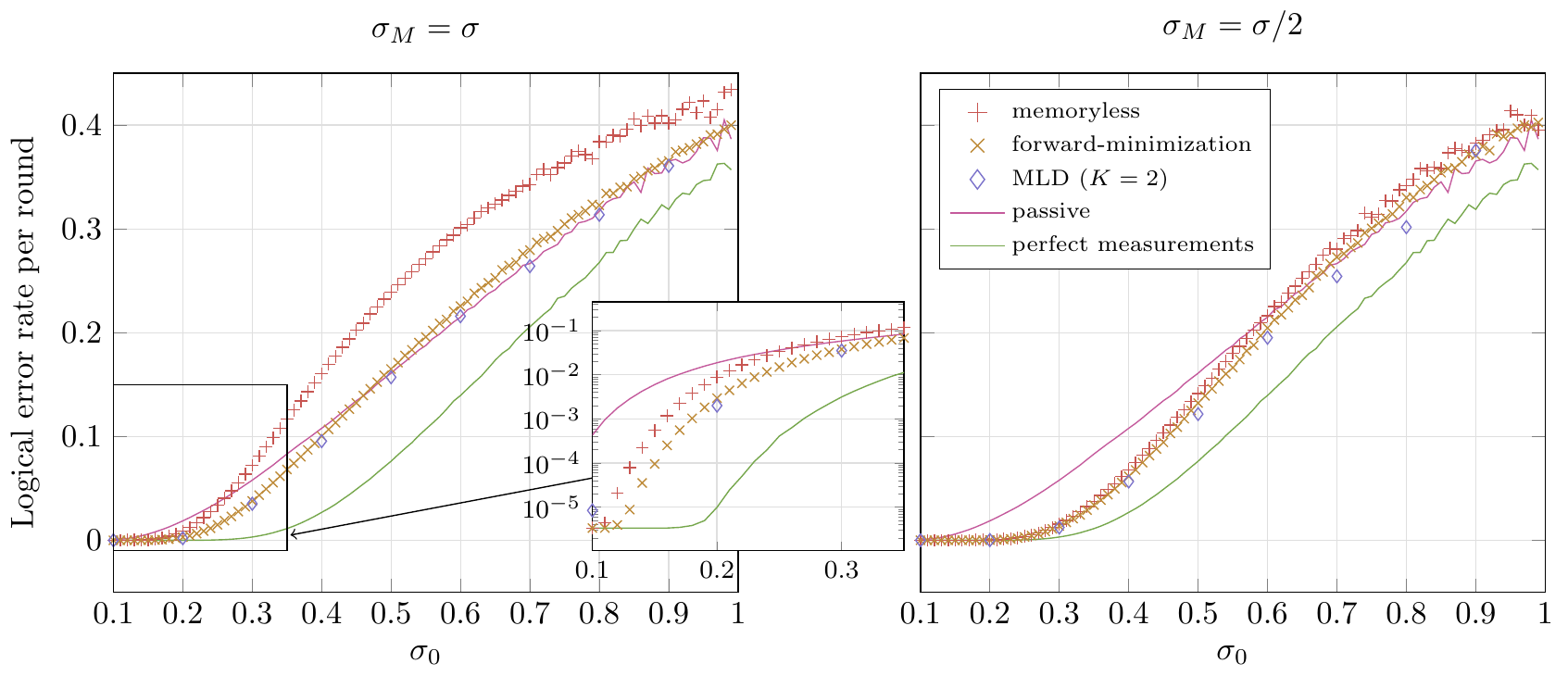}
        \caption{(Color online) Plots of the observed effective logical error rate per round for different decoding techniques.
        	In the left plot we have taken the same standard deviation for measurement and data errors.
        	On the right we have taken the measurement standard deviation equal to half that of the data errors.
        	The horizontal axis shows the bare standard deviation $\sigma_0$.}
			\label{fig:single_gkp_plots}
\end{figure} 

We have numerically simulated these different decoders: maximum likelihood with a cut-off, forward-minimization, and memoryless decoding.
We have also compared them to the scenario where the measurements are perfect, as well as the completely \emph{passive}
decoder where one lets shift errors happen without performing error correction measurements.
For each scenario we considered up to 11 rounds of measurements (M=7 for the maximum likelihood decoder), sampled errors, applied the decoder and gathered statistics of success or failure of the procedure for different bare standard deviation for the errors $\sigma_0\in[0.1, 1]$.

In every scenario we observe an exponential decay toward 1/2 of the probability of logical error as seen in Fig.~\ref{fig:expdecay}.
Thus, as expected, there is an eventual loss of logical information for any values of $\sigma$ and $\sigmaM$.
The decay can be fitted in order to extract an effective logical error rate per round which we then plotted in Fig.~\ref{fig:single_gkp_plots}.

One striking observation is that above a certain bare standard deviation, the measurement outcomes are simply not reliable enough, so that one cannot do substantially better than throwing away the measurements and passively letting errors accumulate.
Roughly speaking, this occurs for $\sigma_0\agt0.5$ when $\sigmaM=\sigma$ and for $\sigma_0\agt0.7$ when $\sigmaM=\sigma/2$.
Another observation for $\sigmaM=\sigma$ is that the memoryless technique actually quickly does more harm than the passive approach which forgoes error correction altogether.
Finally, we observe that in the range of parameters studied, the forward-minimization technique performs almost as well as the maximum-likelihood decoding, while having the advantage of being much simpler computationally.

\section{Concatenation: Toric-GKP Code}
\label{sec:concat}

\subsection{Setup}
\label{sec:setuptoricgkp}

\begin{figure}[htbp]
  \centering
  \includegraphics[width=0.8\textwidth]{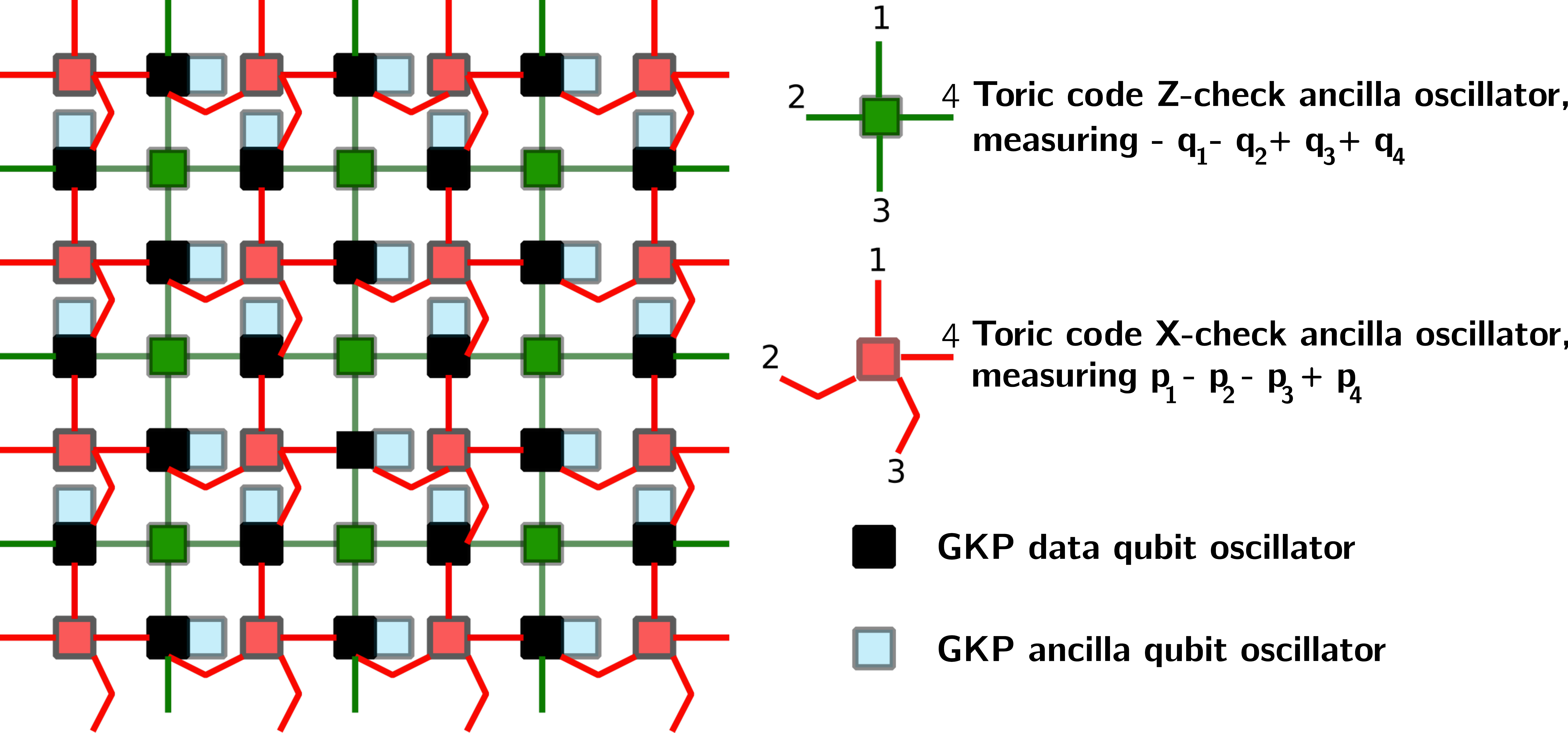}
  \caption{(Color online) The two-dimensional lay-out of oscillators, e.g. high-Q cavities, for the toric-GKP code.
  	Shown is a fragment of a surface or toric code lattice. The different $\pm$ signs are defined by the orientations as explained in the main text.}
  \label{fig:setup}
\end{figure}

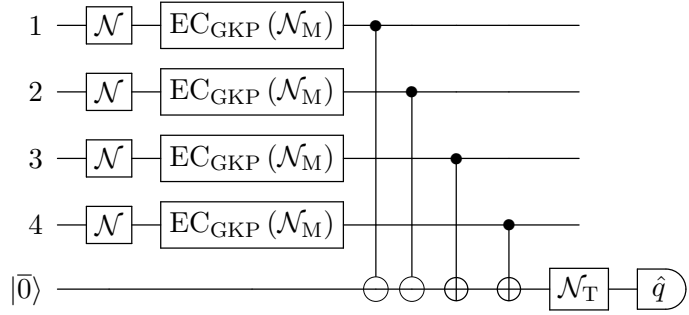
\begin{figure}[htbp]
  \centering
  \scalebox{1}{\Qcircuit @C=1em @R=.7em {
      \lstick{1} & \gate{\mathcal{N}} &  \gate{\rm EC_{GKP}\left(\mathcal{N}_{\rm M}\right )} & \ctrl{4} & \qw & \qw & \qw & \qw \\
      \lstick{2}& \gate{\mathcal{N}} & \gate{\rm EC_{GKP}\left(\mathcal{N}_{\rm M}\right )} & \qw & \ctrl{3} & \qw & \qw & \qw \\
      \lstick{3}  & \gate{\mathcal{N}} & \gate{\rm EC_{GKP}\left(\mathcal{N}_{\rm M}\right )} & \qw & \qw & \ctrl{2} & \qw & \qw \\
      \lstick{4}  & \gate{\mathcal{N}} & \gate{\rm EC_{GKP}\left(\mathcal{N}_{\rm M}\right )} & \qw & \qw & \qw & \ctrl{1} & \qw \\
      \lstick{\ket{\overline{0}}} & \qw & \qw & \scalebox{.9}{\filledcircle} \qw & \scalebox{.9}{\filledcircle} \qw  & \targ & \targ & \gate{\mathcal{N}_{\rm T}} &  \measureD{\hat{q}} 
	}}
\caption{A single round of error correction for a $Z$-check for the toric-GKP code in Fig.~\ref{fig:setup}, on oscillators numbered 1 to 4.
	The GKP error correction unit is given in Fig.~\ref{fig:EC_GKP}, $\ket{\overline{0}}$ is a $+1$ eigenstate of $Z$ and $S_p$.
	The inverse CNOT which induces the transformation $q_{\rm target} \rightarrow q_{\rm target}-q_{\rm control}$ (while $p_{\rm control} \rightarrow p_{\rm control}+p_{\rm target}$) is denoted using a $\ominus$ at the target qubit instead of a $\oplus$.
	A parallel execution of the CNOTs for the X-checks is possible in the toric code.
	}
	\label{fig:setup-concat1}
\end{figure}

We consider the following set-up shown in Figure~\ref{fig:setup}.
We have a 2D lattice of oscillators such that each oscillator encodes a single GKP qubit.
In order to error correct these GKP qubits by the repeated application of the circuits in Fig.~\ref{fig:EC_GKP}, a GKP ancilla qubit oscillator is placed next to each data oscillator, allowing for the execution of these circuits.
After each step of GKP error correction, we measure the checks of a surface or toric code: a single error correction cycle for one of the toric-code checks is shown in Fig.~\ref{fig:setup-concat1}.
Note firstly that we omit GKP error correction after each gate in the circuit in Fig.~\ref{fig:setup-concat1}: the reason is that we assume that these components are noiseless in this set-up so nothing would be gained by adding this.
Secondly, the check operators of the toric code are those of the continuous-variable toric code \cite{BMT:review} which are commuting operators on the whole oscillator space, see Appendix~\ref{sec:CV-toric}.
The reason for using these checks is that for the displacements $X$ and $Z$ of a GKP qubit, it only holds that $X=X^{-1}$ and $Z=Z^{-1}$ on the code space.
Expressed as displacement operators on two oscillators $1$ and $2$, it holds that $[X_1 X_2, Z_1 Z_2^{-1}]=0$.
The upshot is that one has to use some inverse CNOTs in the circuit in Fig.~\ref{fig:setup-concat1}.

In the following, we will denote vertices of the square lattice with letters $i,j,k,l$, and the directed edges with the corresponding vertex pairs, e.g., $e=(ij)$.
Quantities defined on the edges will be considered as vector quantities, e.g. $\hat{p}_{ij}=-\hat{p}_{ji}$ for the momentum operator.
The \emph{preferred orientation} is given by the direction of the coordinate axis.
For such a lattice vector field, say some field $\bs f$, defined on edges, with components $f_{ij}=-f_{ji}$, we will denote the sum of the vectors from a vertex $j$ as
\begin{equation}
(\nabla\cdot f)_j\equiv \sum_{k\sim j}f_{jk}, \label{eq:lattice-divergence}
\end{equation}
where the summation is over all vertices $k$ that are neighboring with $j$.
Note that this choice, recovers the $\pm$ signs for a $X$-check (red) in Fig.~\ref{fig:setup}, when the quantities $f_{jk}$ are written in their preferred orientation.
The circulation of a vector around a (square) plaquette $p\equiv (ijkl)$ is denoted as
\begin{equation}
(\nabla\times f)_p\equiv f_{ij}+f_{jk}+f_{kl}+f_{li}. \label{eq:lattice-curl}
\end{equation}
The preferred orientation for a plaquette, $p = (ijkl)$, is the one for which the closed path $i\rightarrow j\rightarrow k\rightarrow l\rightarrow i$ turns counter-clockwise.
With this choice Eq.~\eqref{eq:lattice-curl} recovers $\pm$ signs for a $Z$-check (green) in Fig.~\ref{fig:setup}, when the quantities $f_{jk}$ are written in their preferred orientation.
With these notations, the vertex $\hat{A}_j$ ($X$-type) and the plaquette $\hat{B}_p$ ($Z$-type) operators of the toric code in Fig.~\ref{fig:setup} can be denoted as,
\begin{equation}
  \label{eq:cv-toric-code-generators}
  \hat{A}_j\equiv e^{i\sqrt\pi(\nabla \cdot\hat{p})_{j}}\text{\ and\ } \hat{B}_p\equiv e^{i \sqrt\pi(\nabla\times\hat{q})_{p}}. 
\end{equation}

\subsection{Noiseless Measurements $\&$ Numerical Results}
\label{sec:noiseless}

We first examine the operation of the toric-GKP code in the channel setting.
This simplified error model is based on the assumption that there are no measurement errors, i.e. $\mathcal{N}_{\rm M} = \id$ and $\mathcal{N}_{\rm T} = \id$ in Fig.~\ref{fig:setup-concat1}, or equivalently, $\sigmaM=0$ and $\sigmaT=0$.
In other words, the assumption is that both the GKP syndrome and the toric code syndrome are measured perfectly at every round.

Generally, when two codes are concatenated, it is possible to pass error information of the lower-level code (in this case, the GKP code) to the decoder for the top-level code (here the toric code).
The information that can be passed on is an estimation of the error rate on the underlying GKP qubits based on the outcome of the GKP error correction measurement.
Intuitively, if the GKP measurement gives a $q \in [-\pi,\pi]$ which lies at the boundaries of the interval, say, beyond $-\pi/2$ or $\pi/2$, we are less sure that we have corrected this shift correctly \footnote{In fact, when one has approximate GKP states with finite photon number for ancilla and data oscillator, there is slightly more information in $q$ than in $q \cmod 2\pi$ but we neglect this extra information here.}.
In other words, the logical error rate depends on the measured value of the GKP syndrome, and this conditional error rate can be used in the standard minimum-weight matching decoder for the toric code.
If the conditional single-qubit error rate fluctuates throughout the lattice, then one can expect that using this information will substantially benefit the toric code decoder.

\begin{figure}[htbp]
	\centering
	\includegraphics[width=0.95\textwidth]{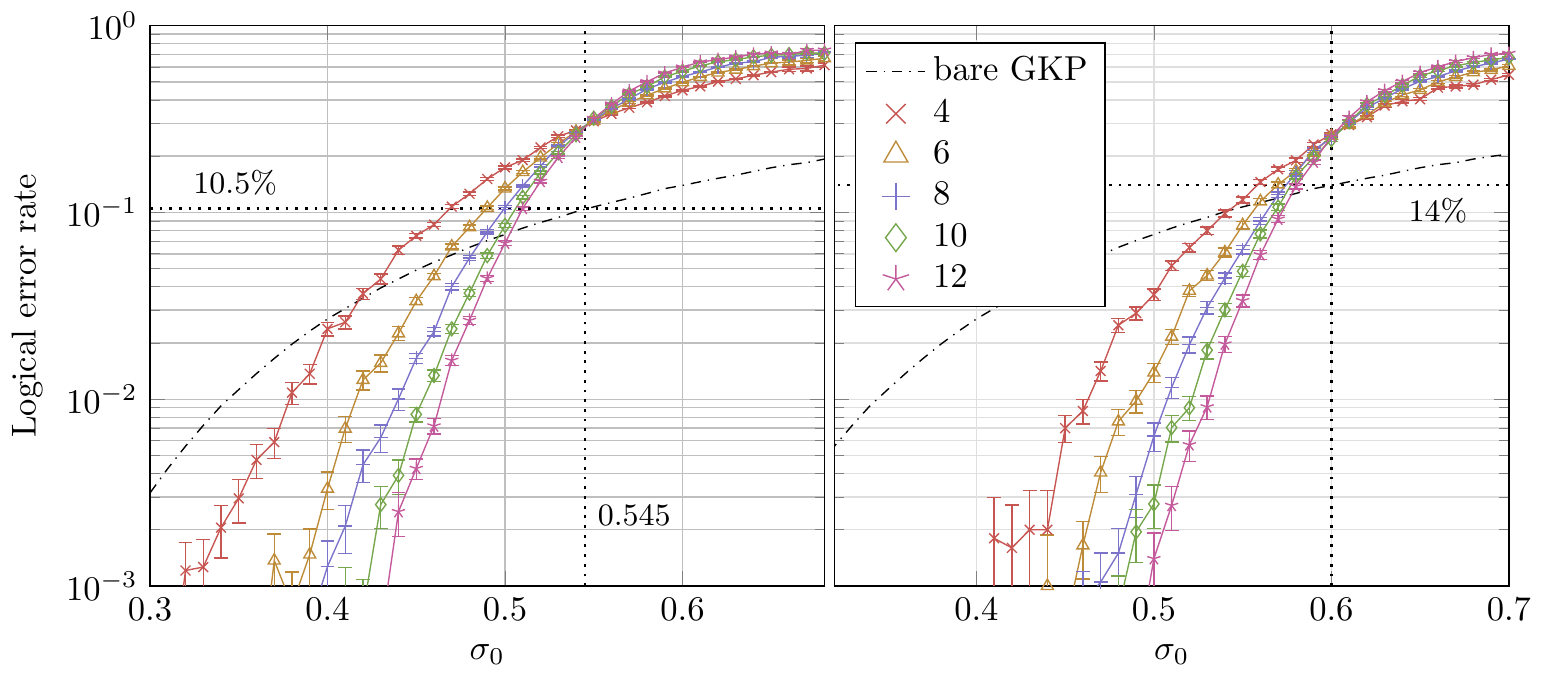}
	\caption{(Color online) Threshold comparison between decoding with or without GKP error information.
		On the left, the simulation only takes the average error rate into account and one obtains a threshold between $\sigma_0 \approx 0.54$ and $\sigma_0 \approx 0.55$ corresponding to $P(X) \approx 10\%$ and $P(X) \approx 10.7\%$,
		respectively.
		On the right, the simulation takes the GKP error information into account.
		In this case, the crossing point is around $\sigma_0^c \approx 0.6$ corresponding to $P(X) \approx 14\%$.
		The data are labeled by the distance of the toric code.
		``Bare GKP'' is the logical error rate for a single GKP qubit whose errors are processed perfectly without measurement errors (green line in Fig.~\ref{fig:single_gkp_plots}).}
	\label{fig:perfectEC}
\end{figure}

We numerically demonstrate that this is the case for the toric-GKP code, reproducing some of the results in Ref.~\cite{KTOF:concat}.
The threshold of the toric code without measurement error is about $11\%$ \cite{dennis+:top}.
If we are not using any GKP error information in the toric code decoding, then the threshold for $\sigma_0$ is set by the value for which $\mathbb{P}(X)=11\%$ with $\mathbb{P}(X)$ shown as the green line in Fig.~\ref{fig:single_gkp_plots}.
We can run a standard minimum-weight matching toric code decoder where qubit $X$-errors are generated by sampling Gaussian noise with standard deviation $\sigma$ followed by perfect GKP error correction on every GKP qubit.
The left plot in Fig.~\ref{fig:perfectEC} presents our numerical data in this scenario, showing a crossing point at $\sigma_0^c \approx 0.54$.

In order to use the GKP error information, we use Eq.~\ref{eq:succ} for the probability of an $X$ error conditioned on the outcome $q \in [-\pi,\pi)$.
Including normalization, this probability reads \footnote{Note that this conditional probability distribution is not identical to the likelihood function used in \cite{KTOF:concat}.
This might be why \cite{KTOF:concat} does not observe a single cross-over point in their threshold plots.}
\begin{equation}
\mathbb{P}(1|q)=\frac{\sum_{k \in \mathbb{Z}} \mathbb{P}_{\sigma}(-2\pi+q+4 \pi k)}{\sum_{k \in \mathbb{Z}} \mathbb{P}_{\sigma}(q+2\pi k)}.\label{eq:res} 
\end{equation}
To use these expressions we replace $k\in \mathbb{Z}$ by the corresponding sum with a cut-off $K$, restricting the summation to the interval $-K\le k\le K$.
This is warranted since the Gaussian weight for large $k$ is small.
In the numerics we used the cut-off $K=3$.

We numerically simulate the following process.
For each toric code qubit, $(ij)$, we first generate a shift error $\epsilon_{ij}$ according to the Gaussian distribution which leads to a GKP syndrome value $q_{ij} \in [-\pi,\pi)$.
Given $q_{ij}$, we infer a correction which may give rise to an $X$ error on qubit $(ij)$.
We evaluate the $Z$-checks of the toric code given this collection of errors and perform a minimum-weight matching algoritm to pair up the toric code defects.
Logical failure is determined when the toric decoder makes a logical $X$ error on any of the two logical qubits of the toric code.
To use the information about the logical error rates $\mathbb{P}(1| q_{ij})$, for each qubit $(ij)$, we define a weight:
\begin{equation}
  w_{ij}=\log\left[\frac{1 -\mathbb{P}(1|q_{ij})}{\mathbb{P}(1|q_{ij})}\right].
\label{eq:planar-weights}
\end{equation}
Then, we define a new weighted graph $G=(V,E)$, whose vertices, $p \in V$, are plaquette defects from the toric code graph and whose edges constitute the complete graph.
Given an edge, $(p,p^\prime)\in E$, its weight $\omega_{p,p^\prime}$ is the minimum weight of a path on the dual of the toric code graph connecting the defect plaquettes $p$ and $p^\prime$.
Here, the path weight, $\omega_{p,p^\prime}$, is the sum of the weights, $w_{ij}$, of all edges crossed by the path.
Minimum-weight-matching (Blossom) algorithm is then run on this $\omega$-weighted graph $G$, leading to a matching of defects and thus an inferred $X$ error.

Specifically, we used Dijkstra's algorithm for finding a minimum-weight path in a weighted graph as provided by the Python library \texttt{Graph-tools} \cite{graph_tools}, and the minimum-weight matching algorithm from the C++ library \texttt{BlossomV} \cite{blossom}.
The process of sampling from shift errors is repeated many times; the logical error rate plotted in Fig.~\ref{fig:perfectEC} is given by the fraction of runs which result in logical failure over the total number of runs.

\section{Noisy Measurements: 3D Space-Time Decoding}
\label{sec:3D}
\subsection{Error model}
\label{sec:3D-error-model}

In this section we consider how to use both GKP and toric code error information when both error correction steps are noisy, using repeated syndrome measurements.
This is the full error model in Fig.~\ref{fig:setup-concat1}, which represents one complete QEC cycle.
We only consider $p$-type shift errors (inducing shifts in $q$), the initial state at $t=0$ is assumed to be perfect, and the last of $M$ rounds of measurements noise-free, both for the GKP and the toric code ancillas.
We will address the question of whether or not there is a \emph{decodable} phase in the space of parameters, such that by increasing the size of the code and the number of measurement cycles, the probability of a logical error can be made arbitrarily small.

\begin{figure}[htbp]
	\centering
	\includegraphics[width=0.5\textwidth]{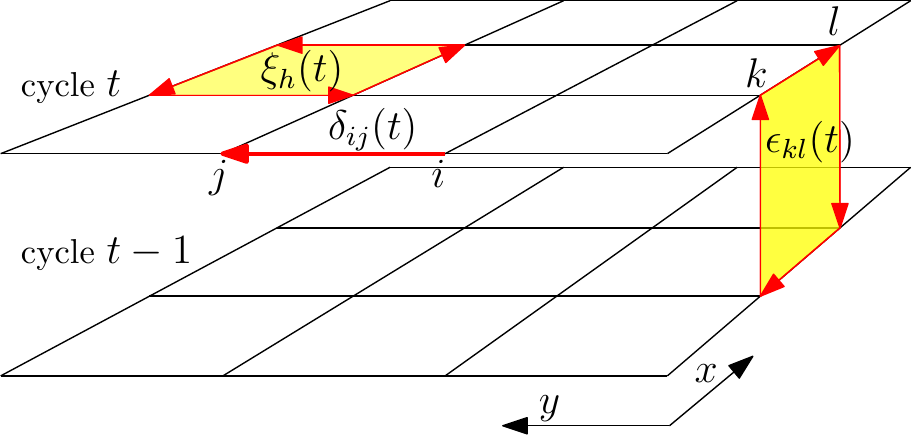}
	\caption{Notations for repeated toric-GKP errors.
		Data oscillators are located on the bonds of the square lattice.
		A GKP measurement error for the bond $b\equiv (ij,t)$ in the measurement cycle $t$ is denoted as $\delta_{ij}(t)\equiv \delta_b$, while the corresponding measurement error for the toric code generator on horizontal plaquette $p\equiv (h,t)$ is denoted as $\xi_h(t)\equiv\xi_p$.
		A data qubit error $\epsilon_{kl}(t)\equiv \epsilon_p$ is associated with the vertical plaquette $p$ directly below the bond $kl$ in the layer $t$.}
	\label{fig:cubic-lattice}
\end{figure}

To visualize errors of different origin, for the $M$-times repeated measurement of the toric code on an $L\times L$ square lattice, it is convenient to consider a three-dimensional cubic lattice, with periodic boundary conditions along $x$ and $y$ directions, separated into horizontal layers.
Each layer corresponds to a measurement round $t\in[M]$, see Fig.~\ref{fig:cubic-lattice}.
In each time layer $t$, we use the same notations and conventions of directed edges and vector quantities as in Section \ref{sec:setuptoricgkp}.
Hence we associate a GKP data qubit oscillator with each edge $(ij)$ of the square lattice.

We thus denote the shift occurring on a data oscillator just before the measurement at time $t$ as the shift $\epsilon_{ij}(t)$ (induced by channel $\mathcal{N}$).
Since the shifts accumulate, the net shift on the oscillator at bond $ij$ just before the measurement at time $t$ is [cf.\ Eq.~\ref{eq:qphie}]
\begin{equation}
\phi_{ij}(t)\equiv \sum_{t'=1}^t\epsilon_{ij}(t').\label{eq:phi-ij}
\end{equation}
Furthermore, we denote the GKP measurement error of this oscillator at time $t$ as $\delta_{ij}(t)$.
This is the shift error on the ancilla inside the corresponding $\rm EC_{GKP}$ unit (induced by channel $\mathcal{N}_{\rm M}$).
With these notations, we can write for the GKP syndrome $q_{ij}^{\rm GKP}(t)$ at time $t$, 
\begin{equation}
  \label{eq:GKP-syndrome-ij}
  q_{ij}^{\rm GKP}(t)=\delta_{ij}(t)+\phi_{ij}(t) \cmod 2\pi.
\end{equation}

In addition, we have the toric code syndrome.
Specifically, we consider the toric code plaquette operators, see Fig.~\ref{fig:setup} and Eq.~\ref{eq:cv-toric-code-generators}.
The result of the toric code syndrome measurement on the plaquette $ h\equiv (ijkl)$ at time $t$ is
\begin{equation}
  \label{eq:q-plaquette}
  q_h^{\rm tor}(t)=\xi_h(t)+(\nabla\times\phi)_{h}(t)
  \cmod 4\pi,
\end{equation}
where the bond vectors $\phi_{ij}(t)$ are the accumulated errors in Eq.~\ref{eq:phi-ij}, and $\xi_h(t)$ is the plaquette measurement error (induced by channel $\mathcal{N}_{\rm T}$).
Note that, unlike for the GKP measurements, the syndrome $\bs{q}^{\rm tor}$ is measured modulo $4\pi$, since the ancilla starts in the state $\ket{\overline{0}}$.

Since we assume that the measurement errors in the last layer, $t=M$, are absent, we have
$\xi_{h}(M)=\delta_{ij}(M)=0$.
Writing the product of the corresponding probability densities, we obtain an analog of Eq.~\ref{eq:GKP-path-integral} for the effective energy
\begin{eqnarray}
  H(\bs \phi;\bs q)
  &=&\sum_{t=1}^M\sum_{\langle i j\rangle }{\left (\phi_{ij}(t)-\phi_{ij}(t-1)\right )^2\over 2\sigma^2}\nonumber\\
  & & +\sum_{t=1}^{M-1}\sum_{\langle ij\rangle}V_{\sigmaM}\left(q_{ij}^{\rm GKP}(t)-\phi_{ij}(t)\right) + \sum_{t=1}^{M-1}\sum_{h}V_{\sigmaT/2}\left(\frac{{q_{h}^{\rm tor}(t)-(\nabla\times\phi)_{h}(t)}}{2}\right),
  \label{eq:total-ham-phi}
\end{eqnarray}
which depends on the accumulated field $\bs \phi$ with components $\phi_{ij}(t)$ and on the total measured syndrome $\bs q\equiv\{\bs q^{\rm GKP},\bs q^{\rm tor}\}$.
Here, $\sum_{\langle ij\rangle}$ indicates a summation over all bonds  of the square lattice, the summation over $h$ runs over all square-lattice faces (horizontal), and the structure of the last term accounts for the $4\pi$-periodicity of toric syndrome measurements, see Eq.~\ref{eq:q-plaquette}.  

The energy in Eq.~\ref{eq:total-ham-phi} defines the conditional probability $\mathbb{P}(\bs \phi\vert\bs q)\propto\exp(-H(\bs \phi;\bs q))$, up to a normalization factor.
The measurements in the last time-layer, $t=M$, constrain the values $\phi_{ij}(M)$ as follows.
From the GKP syndrome we have, as in Sec.~\ref{sec:singleGKP},
\begin{equation}
  \phi_{ij}(M)=q^{\rm GKP}_{ij}(M)-2\pi k_{ij},
  \label{eq:GKP-match}
\end{equation}
while the toric syndrome for each square-lattice face $h=(ijkl)$ gives
\begin{equation}
(\nabla\times \phi)_{h}(M)\equiv  \phi_{ij}(M)+\phi_{jk}(M)+\phi_{kl}(M)+\phi_{li}(M)=q_h^{\rm tor}(M)-4\pi k_h.\label{eq:tor-match}
\end{equation}
These equations can be solved to find a last-layer binary candidate error vector ${\bs b}\in \mathbb{F}_2^{2L^2}$, whose components $b_{ij}\equiv k_{ij}\bmod2$ give the parity of the integer shifts $k_{ij}$ in Eq.~\ref{eq:GKP-match}.
Just as for the usual toric code, Eqs.~\ref{eq:GKP-match}, \ref{eq:tor-match} determine ${\bs b}$ up to arbitrary cycles on the dual lattice, i.e. $X$-stabilizers (homologically trivial) and $X$ logical errors (homologically non-trivial) of the toric code.
Adding trivial cycles to $\bs b$ gives another equivalent last-layer candidate which should be summed as part of the same sector.
A trivial cycle can be written as the gradient of a binary field, $\bs{b}^{\rm triv}_{ij}= n_j-n_i$, with $n_i\in\mathbb{F}_2$.
To turn $\bs{b}$ into an inequivalent last-layer candidate error, one should add a homologically non-trivial cycle, $\bs{c}\in\mathbb{F}_2^{2L^2}$.

We can now write explicitly the partition function $Z_{{\bs c}}(\bs{b}\vert\bs q)$, equivalent to Eq.~\eqref{eq:part}, which determines the conditional probability, given the measurement outcomes $\bs{q}$ of the equivalence class of last-layer candidate error $[\bs{b} + \bs{c}]$,
\begin{equation}
  \label{eq:3D-Z}
Z_{\bs c}(\bs b\vert \bs{q})={N^\prime}^{-1}\int {\rm d}\bs\phi\, \e^{-H(\bs\phi;\bs q)}\prod_{\langle ij\rangle}\sum_{\{n_i\in\mathbb{F}_2\}}\sum_{m_{ij}\in\mathbb{Z}} \delta\left(\phi_{ij}(M)- q_{ij}^{\rm GKP}(M)+2\pi[b_{ij}+c_{ij}-n_j+ n_i+2 m_{ij}]\right).
\end{equation}
For ML decoding, given a $\bs b$ which satisfies Eqs.~\ref{eq:GKP-match} and \ref{eq:tor-match}, one needs to compare $Z_{\bs c}(\bs b \vert \bs q)$ for different ${\bs c}\in\mathbb{F}_2^{2L^2}$ which are inequivalent binary codewords of the toric code, i.e. the three homologically non-trivial domain walls on the square lattice or the trivial vector.
ML decoding then prescribes that we choose the error $\bs b + \bs c$ as the correction where $\bs c$ has the largest partition function $Z_{\bs c}(\bs b \vert \bs q)$.

\subsection{Equivalent formulation with $U(1)$ symmetry}
\label{sec:gaugemodel}

The partition function in Eq.~\ref{eq:3D-Z} with the Hamiltonian in Eq.~\ref{eq:total-ham-phi}, as a statistical-mechanical model, is not so convenient to analyze, since the components of the syndrome are not independent of each other.
So, we will consider an equivalent form of the partition function, that explicitly depends on the data errors $\epsilon_{ij}(t)\equiv \epsilon_p$, the measurement errors $\delta_{ij}(t)\equiv\delta_b$, and the toric code measurement errors $\xi_h(t)\equiv\xi_p$.
We group all these errors into one error record $\bs e=\{\bs \epsilon,\bs\delta,\bs \xi\}$.
Any error $\bs e'$ that is equivalent to $\bs e$ can be obtained by adding, so to say, stabilizer generators of the space-time code.
This can be expressed using a $2\pi$-periodic vector field $\bs A$ whose components are real-valued on horizontal bonds in layers $t\in[M-1]$, and $\{0,\pi\}$-valued on the vertical bonds connecting layers $t-1$ and $t$ for all $t\in[M]$.
For horizontal bonds $b$ in layers $t=0$ and $t=M$, $A_b=0$.
With these notations, we can express the partition function in Eq.~\ref{eq:part} as
\begin{eqnarray}
  H({\bs A};\bs e)
  &=&\sum_{b\|xy}V_{\sigmaM}\big(\delta_b-2A_b\big) + \sum_{p\|xy}V_{\sigmaT/2}\left({\xi_p\over 2}-(\nabla\times A)_p\right) + \sum_{p\perp xy}V_{\sigma/ 2}\left({\epsilon_p\over 2}+(\nabla\times A)_p\right),\qquad \label{eq:ham-epsilon-delta} \\  
  Z_{\bs 0}(\bs e)
  &=& {N^{\pp}}^{-1}\sum_{A_b\in\{0,\pi\}:b\perp xy}\; \prod_{b\| xy}\int_{-\pi}^\pi\!\! {\rm d}A_b \,\; e^{-H(\bs A;\bs e)}.
  \label{eq:Z-epsilon-delta} 
\end{eqnarray}
with some normalization $N^{\pp}$.
To derive these equations, we determine the shift errors that leave the syndrome record $\{\bs{q}^{\rm GKP}, \bs{q}^{\rm tor}\}$ unchanged without inducing a logical error.
These are called the gauge degrees of freedom and form the stabilizer group for the space-time code.
In our case there are five types of gauge degrees of freedom, four discrete and one continuous.
The discrete ones are genuine symmetries of the quantum states involved in the code or measurement circuits.
Namely, the input state of an ancilla in the GKP measurement circuit in Fig.~\ref{fig:repeated-QEC} is stabilized by an $X$ operator, whose action is equivalent to a $2\pi$-shift of the corresponding GKP measurement error $\delta_{ij}(t)$.
Similarly, application of an $X^2=S_p$ GKP stabilizer generator to a data qubit or a toric-code ancilla in Fig.~\ref{fig:setup-concat1} is equivalent to a $4\pi$-shift of the corresponding error, $\epsilon_{ij}(t)$ or $\xi_h(t)$, respectively.
We also have toric code vertex operators $\hat{A}_j$ [Eq.~\ref{eq:cv-toric-code-generators}] whose action corresponds to simultaneous $2\pi$-shifts on the four adjacent qubits, $\epsilon_{ij}(t)\to \epsilon_{ij}(t)+2\pi$.
This discrete gauge freedom will be captured by the two-valued field $A_b$ on the vertical bonds.

The only continuous degree of freedom is a space-time one: it corresponds to adding a continuous shift $a$ on a data oscillator at some time step and then canceling it at the next time step, while hiding the shift from the adjacent GKP and toric syndrome measurements by adding the shifts $\pm a$ as necessary on the corresponding ancillas. 

When applied to the Gaussian distribution, the discrete local shifts are responsible for forming the Villain potentials \ref{eq:Villain} in the effective Hamiltonian \ref{eq:ham-epsilon-delta}, where additional rescaling in the last two terms was necessary to account for the $4\pi$-periodicity.
The remaining two degrees of freedom are represented by the vertical (discrete) and horizontal (continuous) components of the doubled vector potential $2\bs A$.
The scale of the vector potential $\bs{A}$ was chosen to make easier contact with previous literature on related models.
With this choice, adding a $\pi$-shift to a component of $\bs{A}$ correspond to a GKP $X$-logical, which was a $2\pi$-shift in the previous sections.

Equations~\ref{eq:ham-epsilon-delta}, \ref{eq:Z-epsilon-delta} have to be supplemented with the appropriate boundary conditions to be used in decoding.
Given the toric-code codeword $\bs c$, in order to calculate $Z_{\bs c}(\bs e)$, corresponding to the sector $[\bs e+ \bs{c}]$, one can add $2\pi\bs c$ to the data error in the top layer, $t=M$.
This can be achieved by introducing a fixed non-zero vector potential in this layer, namely $A_{b}=\pi c_{ij}$ for all top-layer horizontal bonds $b=(ij,t=M)$, instead of zero for the trivial sector.

Equations~\ref{eq:ham-epsilon-delta}, \ref{eq:Z-epsilon-delta} look very different from the equivalent form that we first derived, i.e. Eqs.~\ref{eq:total-ham-phi} and \ref{eq:3D-Z}.
A map between these two formulations is given in the Appendix~\ref{app:derivation-Z0}.

\subsection{Anisotropic charge-two $U(1)$ gauge model with flux disorder}
\label{sec:phase-diag}

We would like to get some intuition about the constructed $U(1)$-symmetric model and its features that are relevant for decoding.
To this end, we are going to relax the constraint $A_b\in\{0,\pi\}$ for vertical bonds and consider the following anisotropic charge-two Villain $U(1)$ model in three dimensions, with quenched uncorrelated gauge and flux disorder,
\begin{eqnarray}
  \nonumber
  H&=&\sum_{b\|xy}V_{\sigmaM}(2A_b-\delta_b)+\sum_{b\perp xy} V_\eta(2A_b-\delta_b) \\
   & &\qquad\label{eq:gauge-model} 
       + \sum_{p\| xy} V_{\sigmaT/2}\Big((\nabla\times A)_p-\xi_p /2\Big) + \sum_{p\perp xy}V_{\sigma/2}\Big((\nabla\times A)_p+\epsilon_p/2\Big).
\end{eqnarray}
The model \ref{eq:ham-epsilon-delta}, \ref{eq:Z-epsilon-delta} is recovered with the help of the symmetry of the Villain potential, $V_\sigma(x)=V_\sigma(-x)$, by setting $\delta_b=0$ for vertical bonds, $b\perp xy$, and taking the limit $\eta\to +0$, in which case the field $A_b$ along the vertical bonds be only allowed to take the values $0$ or $\pi$.

In addition to making all components of the vector field $\bs A$ continuous, we will also relax the constraint on the parameters of the quenched disorder.
Specifically, instead of using the components of the fields $\{\bs \epsilon,\bs \delta,\bs \xi\}$ as normally-distributed with specific r.m.s.\ deviations $\sigma$, $\sigmaM$, and $\sigmaT$, respectively, we are going to treat the parameters of the disorder as independent from the parameters in the Hamiltonian \ref{eq:gauge-model}.
This is similar to the trick used originally for qubit-based surface codes \cite{dennis+:top}, where only the Nishimori line on the phase diagram of the disordered random-bond Ising model corresponds to ML decoding, while points away from that line correspond to a decoder given an incorrect input information, see Eq.~\ref{eq:decoding-cmp}.

We first examine the parent model \ref{eq:gauge-model} in the absence of background fields, by setting all $\delta_b=\epsilon_p=\xi_p=0$.
The case without anisotropy is relatively well studied, that is $\eta=\sigmaM$ and $\sigma=\sigmaT$.
The Wilson Hamiltonian of the compact charge-$q$ $U(1)$ lattice gauge model reads
\begin{equation}
\label{eq:3D-U1-model}
H=-\kappa\sum_{\langle uv\rangle }\cos(\theta_u-\theta_v+qA_{uv})-\lambda\sum_p\cos(\nabla\times A)_p, \quad \kappa={1\over \sigmaM^2}, \quad \lambda={4\over \sigma^2}, 
\end{equation}
where we restored local gauge symmetry $\theta_v\to \theta_v-q\chi_v$, $A_{uv}\to A_{uv}+\chi_u-\chi_v$ by adding a scalar \emph{matter field}, $U(1)$ phases $\theta_v$ on the vertices of the lattice.
Notice that the phases $\theta_v$ can be always suppressed by a gauge transformation with $\chi_v=q^{-1}\theta_v$.  For $U(1)$ symmetry, the charge $q$ in Eq.~\ref{eq:3D-U1-model} must be an integer; our original model with the Hamiltonian \ref{eq:gauge-model} corresponds to $q=2$.
In application to this model, the boundary conditions for the sectors with non-trivial codewords $\bs c\not\simeq\bs 0$ [see discussion below Eq.~\ref{eq:Z-epsilon-delta}] are equivalent to an externally applied uniform magnetic field $B=(\nabla\times A)_p$ (flux per plaquette), with the total flux of $\pi$ piercing the system along $x$, $y$, or both directions.
Quite generally, when the couplings $\kappa$ and $\lambda$ are sufficiently small, the net magnetic field remains uniform on average, with the total free energy cost \ref{eq:most-general-delta-F} proportional to the volume times $B^2$.
For a system with the volume $V=L^2M$ and $B=\pi/(LM)$, this gives the free energy cost $\Delta_BF\propto 1/M$, vanishing in the large-system limit.
The situation is different in the \emph{Meissner phase}, analogous in properties to that in type-II superconductors, where the magnetic field is expelled from the bulk, and is forced into \emph{vortices} (vortex lines) which can carry a flux quantized in the units of $2\pi/q$.
Such a vortex is a topological excitation, meaning that it cannot disappear without moving to the system boundary or annihilating with another vortex that carries the opposite flux, and it has a non-zero line tension (finite or logarithmically divergent with the system size), which gives a free energy cost proportional to the system size, $\Delta_BF\propto L$.

\begin{figure}[htbp]
  \centering
  \includegraphics[width=2.5in]{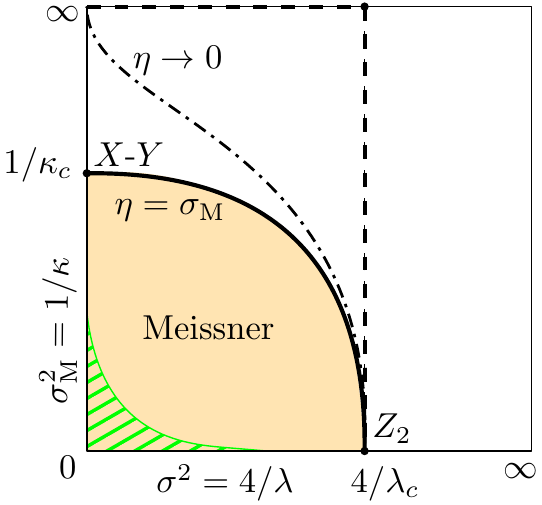}
  \caption{(Color online) Schematic phase diagram of the 3D anisotropic $q=2$ Villain gauge model \ref{eq:gauge-model} with $\sigmaT=\sigma$.
  	Solid line bounds the Meissner phase (light orange shading) in the clean isotropic limit, $\eta=\sigmaM$, which corresponds to the Villain version of the model \ref{eq:3D-U1-model}.
  	With $\eta\to0$, decodability condition for the toric-GKP code with $\sigma=\sigmaT$ can not be satisfied in the region to the right from the dashed line, $\sigma^2>4/\lambda_c(\mathbb{Z}_2)$, and along the upper boundary, $\sigmaM\to\infty$, and it is satisfied along the left boundary, $\sigma\to0$.
  	We therefore expect the boundary of the Meissner phase in the clean $\eta\to 0$ limit as shown with dash-dotted line; this region includes the entire Meissner phase of the isotropic model.
  	The green-hatched region represents the expected location of the ML decodable phase for the toric-GKP code; the sign of the curvature matches the bound in Ref.~\cite{Dumer-Kovalev-Pryadko-bnd-2015}.}
  \label{fig:qed3}
\end{figure}

The 3D lattice model \ref{eq:3D-U1-model} (along with related non-Abelian models) has been first discussed by Fradkin and Shenker \cite{Fradkin-Shenker-1979} as a toy model for quark confinement.
Subsequently, both the model \ref{eq:3D-U1-model} and its Villain version have been studied analytically and numerically in a number of papers, e.g., 
Refs.~\cite{Bhanot-Freedman-1981,Borgs-Nill-1987,
  Chernodub-Ilgenfritz-Schiller-2001,
  Chernodub-Ilgenfritz-Schiller-2002,
  Chernodub-Feldmann-Ilgenfritz-Schiller-2004A,
  Smiseth-etal-Subdo-2003,Borisenko-Fiore-Gravina-Papa-2010}.
The conclusion is that the model in 3D has only two phases, see Fig.~\ref{fig:qed3}.
The weak-coupling phase is characterized by the area law in the Wilson loop correlator and the absence of the Meissner effect.
In comparison, the strong-coupling phase, which requires both $\lambda$ and $\kappa$ sufficiently large, $\lambda>\lambda_c(q)>0$, $\kappa>\kappa_c>0$, is characterized by the presence of both the perimeter law in the Wilson loop and the Meissner effect.
Here, $\lambda_c(q)$ corresponds to the limit of $\kappa\to\infty$, which forces the gauge field $A_{uv}$ to take
values in $2\pi/q$ times an element of $\mathbb{Z}_q$.
In the case $q=2$, the corresponding critical point \cite{Balian-Drouffe-Itzykson-1975} $\lambda_c(2)\approx 0.7613$ is that of the three-dimensional $\mathbb{Z}_2$ lattice gauge theory \cite{Wegner-1971}.
Similarly, in the limit $\lambda\to\infty$ the fluxes are all frozen to zero, the charge $q$ is irrelevant and the remaining degrees of freedom are the on-site phases $\theta_v$.
This model is known as the $X$-$Y$ model, and its critical point in 3D is at $\kappa_c\approx 0.453$ (or $\kappa_c^{\rm (Villain)}\approx 0.4542$ for the corresponding Villain model \cite{Gottlob-Hasenbuch-Meyer-1993}).
Such a phase diagram shape with no reentrance as a function of either variable is consistent with the monotonicity of the correlation functions and free energy increments which follow from generalized GKS inequalities, see Appendix~\ref{app:phase-diag}.

We should notice that the perimeter law in the Wilson loop and the Meissner effect do not necessarily come together; examples are given by compact gauge models similar to Eq.~\ref{eq:3D-U1-model} in $D\ge4$ which also have small-$\kappa$ large-$\lambda$ phases characterized by the perimeter law but no Meissner effect \cite{Polyakov-1975,Fradkin-Shenker-1979}.
Of course, it is the Meissner phase that is associated with the formation of magnetic vortices with a non-zero line tension.

What do these results tell us about the anisotropic model \ref{eq:gauge-model} of interest, in particular, about the singular limit $\eta\to+0$?
To answer these questions, we notice that, in the absence of disorder, both the correlation functions and the response to external magnetic field (existence of the Meissner effect) are monotonically non-decreasing with respect to any coupling.  This follows from general correlation inequalities which are briefly discussed in Appendix D.  Moreover, these inequalities also predict an upper bound, Eq.~\eqref{eq:most-general-decoding-upper-bound}, on the ML decoding probability in terms of a similar quantity defined in the absence of disorder.  It follows that finite vortex line tension in the clean (Meissner effect) limit is a necessary condition for perfect decoding. 

Since decreasing $\eta$ corresponds to increasing some of the couplings, the entire strong-coupling (Meissner) phase of the 3D lattice gauge model with $q=2$ should be inside of the corresponding phase of the model \ref{eq:Z-epsilon-delta} with $\sigmaT\le \sigma$ and the values of $\sigmaM$, $ \sigma$ given by the map in Eq.~\ref{eq:3D-U1-model}.
Second, this phase cannot exist for $\sigma^2>4/\lambda_c(2)$, the limit $\kappa\to\infty$ which corresponds to taking \emph{both} $\eta$ and $\sigmaM$ to zero.

Furthermore, if we started with the model \ref{eq:total-ham-phi} in the limit of unusable GKP syndrome, $\sigmaM\to\infty$, the first term in Eq.~\ref{eq:gauge-model} would be absent.
In this case the continuous gauge symmetry $A_{uv}\to A_{uv}+\chi_u-\chi_v$ is not broken, which is sufficient to recover a continuous field $A_b$ along the vertical bonds $b\perp xy$; with $\sigma=\sigmaT$, the resulting model is the Villain version of Eq.~\ref{eq:3D-U1-model} with $\kappa=0$.
According to Polyakov's argument \cite{Polyakov-1975}, only one phase is expected in this limit; we expect no Meissner effect, and no decoding threshold.

On the other hand, the large-$\lambda$ (small-$\sigma$) limit of the model \ref{eq:3D-U1-model} corresponds to all fluxes $(\nabla\times A)_p$ frozen in the minimum-energy configuration.
The remaining degrees of freedom are the phases in the first term, which gives an $X$-$Y$ model.
However, if we look at the model \ref{eq:gauge-model} with $\sigma=\sigmaT\to0$, in the singular anisotropic limit $\eta\to0$, the phases $\theta_v\equiv \theta_{i,t}$ at the same square lattice position $i$ are forced to fluctuate together, which gives arbitrarily large effective $X$-$Y$ coupling $\kappa_\mathrm{eff}=M/\sigmaM^2$ as $M\to\infty$.
Assuming this argument also holds for $\sigma=\sigmaT$ small but finite, we expect the phase line as shown in Fig.~\ref{fig:qed3} with a dot-dashed line, with the region below it in the Meissner phase.

A different version of this argument can be obtained by examining the Hamiltonian in the form \ref{eq:total-ham-phi}, with $\bs q=\bs0$.
With $\sigma$ small, the fields $\phi_{ij}(t)$ in the neighboring $t$-layers are forced to move together, which is equivalent to increasing the couplings for the remaining terms.
The resulting model is a two-dimensional version of the gauge model~\ref{eq:3D-U1-model}, with in-plane vector potential $\bs A$.
Just like its 3D counterpart, this model is in a disordered phase except when the fluxes are suppressed in the limit $\sigmaT\to0$, which gives a 2D $X$-$Y$ model.
With $\sigma$ sufficiently small, the effective in-plane coupling $\kappa_{\rm eff}^{({\rm 2D})}$ can be made arbitrarily large, driving the model below the BKT transition.
 
We have discussed the expected phase boundary of the model \ref{eq:ham-epsilon-delta}, \ref{eq:Z-epsilon-delta} in the absence of disorder, $\bs e=\bs 0$.
We expect a Meissner phase to survive with $\bs e\neq\bs 0$.
The basic effect of a weak disorder on a topological excitation like a vortex line is to force its random displacements.
The displacements can be accounted in a simple linear approximation up to certain distance scale called the Larkin length \cite{Larkin-1970}.
Even though the displacements become non-linear beyond this scale, signifying the onset of glassiness, with weak enough disorder, topological excitations are not expected to be generated; this is called Bragg glass phase of an elastic solid \cite{Giamarchi-LeDoussal-1997}.
The absence of topological excitations would indicate a divergent free energy cost for a $\pi$ vortex excitation in the Meissner phase of the model \ref{eq:ham-epsilon-delta}, \ref{eq:Z-epsilon-delta}, with sufficiently weak disorder.

We thus expect a decodable phase for a toric-GKP code to exist in a finite region at sufficiently small $\sigma$, $\sigmaM$, and $\sigmaT$.
In the limit $\sigmaM\to0$, this phase should go continuously into a decodable phase of the regular (qubit) toric code \cite{dennis+:top}.
These expectations are confirmed by our numerical results with two (suboptimal) decoders presented in the next section.

\subsection{Decoder and numerical results}
\label{sec:decoder}

Maximum likelihood decoding can be done by comparing conditional probabilities in different sectors, see Eq.~\ref{eq:part}.
Just as in the case of a single GKP qubit, Gaussian integrations in the relevant partition functions, $Z_{\bs c}(\bs{b\vert}\bs q)$, in Eq.~\ref{eq:3D-Z}, or its equivalent form $Z_{\bs c}(\bs e)$, in Eq.~\ref{eq:Z-epsilon-delta}, can be carried through exactly.
This would leave expressions similar to Eq.~\ref{eq:mldexpr}, with an additional summation over $2L^2$ binary spins.
In principle, such expressions can be evaluated using Monte Carlo sampling techniques.
In practice, the complexity of such a calculation is expected to be high, because the corresponding coupling matrix is not sparse, just as the matrix $A$ in Eq.~\ref{eq:mldexpr} is not sparse, see Appendix~\ref{sec:ml-decoding}.
For this reason, we have not attempted ML decoding for toric-GKP codes.

We have constructed several decoders which approximate ME decoding.
The idea is to find a configuration of the field $\bs A$ minimizing the Wilson version of the Hamiltonian \ref{eq:ham-epsilon-delta}, i.e. with Villain potentials replaced by cosines, by decomposing it into a continuous part (to be guessed or found using a local minimization algorithm), and a binary field which represents frustration, to be found using minimum weight matching.
This decomposition relies on the analysis of the Hamiltonian \ref{eq:ham-epsilon-delta} in the limit of perfect GKP measurements, $\sigmaM\to0$.

\subsubsection{ME decoding in the limit of perfect GKP measurements}
\label{sec:perfectGKP}

Let us consider what happens with our model \ref{eq:ham-epsilon-delta}, \ref{eq:Z-epsilon-delta} in the limit $\sigmaM\to0$.
First, in this limit all GKP measurement errors $\delta_b$ vanish with probability one.
Second, the first term in Eq.~\ref{eq:ham-epsilon-delta} forces $A_b\in\{0,\pi\}$ for all horizontal bonds, the same as we already had for vertical bonds.
This forces all plaquette fluxes to take integer values times $\pi$.
We show here that in this limit we recover a version of the random-plaquette gauge model (RPGM) associated with decoding the usual qubit toric code in the presence of (toric) syndrome measurement errors \cite{WHP:threshold}.

Since the limit $\sigmaM\to0$ makes the vector potential $A_b\in\{0,\pi\}$, and in turn the plaquette flux $B_p\equiv (\nabla\times A)_p$, discrete, then one can interpret it as a spin degree of freedom, using the fact that the Villain potential is $2\pi$-periodic as well as even.
Indeed, considering for example a vertical plaquette, $p\perp xy$, in Eq.~\ref{eq:ham-epsilon-delta}, one can write,
\begin{align}
  V_{\sigma/2}\left({\epsilon_p\over2}+B_p\right)&=-\tau_p(\bs{e}) \e^{iB_p} + \mathrm{const},\nonumber\\
  p\perp xy,\quad\tau_p(\bs{e})&\equiv {1\over2}\left[V_{\sigma/2}\left({\epsilon_p\over2}-\pi\right) -V_{\sigma/2}\left({\epsilon_p\over2}\right)\right]\approx \frac{4}{\sigma^2}\cos\left (\frac{\epsilon_p}{2}\right ).
\label{eq:plaquette-weight-ver}
\end{align}
The additive constant has no effect and can be ignored.
Similarly for horizontal plaquettes, $p\parallel xy$, where one obtains the weights,
\begin{equation}
p\parallel xy,\quad\tau_p(\bs{e})\equiv {1\over2}\left[V_{\sigmaT/2}\left({\xi_p\over2}-\pi\right) -V_{\sigmaT/2}\left({\xi_p\over2}\right)\right]\approx \frac{4}{\sigmaT^2}\cos\left (\frac{\xi_p}{2}\right ).
\label{eq:plaquette-weight-hor}
\end{equation}
Then, if one defines from $\bs{A}$ some Ising spins, $\bs{s}$, using for each bond, $s_b\equiv \e^{iA_b}\in\{-1,1\}$, one obtains, in place of Eq.~\ref{eq:ham-epsilon-delta}, a RPGM very similar to that in Ref.~\cite{WHP:threshold},
\begin{equation}
  H({\bs s};\bs e) = -\sum_{p}\tau_p(\bs e) u_p,\quad u_p\equiv \prod_{b\in p} s_b. 
  \label{eq:RPGM-one}    
\end{equation}
Unlike in the usual RBGM obtained for the qubit toric code \cite{WHP:threshold} where plaquette weights can take only two values, $\tau_p = \pm J$, here quenched randomness leads to a continuous distribution of the weights.
This model is similar to the 2D random-bond Ising model constructed in Sec.~\ref{sec:noiseless} for decoding a toric-GKP code in the channel setting, also in the limit $\sigmaM=0$.
In fact, the weights concerning data errors in Eq.~\ref{eq:plaquette-weight-ver} (without the cosine approximation), are equivalent to those given by Eqs.~\ref{eq:res}, \ref{eq:planar-weights}.
Similar to its counterpart with sign disorder, the phase-diagram of the RPGM with the Hamiltonian \ref{eq:RPGM-one} will show a transition from an ordered to disordered phase as the temperature $\beta^{-1}$ or the strength of the quenched disorder is increased.
If we increase the disorder $\sigma$ along a line with any fixed ratio $r=\sigmaT/\sigma$, a version of the standard argument \cite{Nishimori-book} shows that no ordered phase can exist beyond the critical value of $\sigma$ reached along the ``Nishimori line,'' $\beta=1$ [cf. Eq.~\ref{eq:decoding-cmp}].
We expect that it is this critical value of $\sigma$ that is associated with the memory phase transition.

An important quantity that governs the structure of the minimum of the RPGM Hamiltonian \ref{eq:RPGM-one} is \emph{frustration}.
We call a cube frustrated when it has an odd number of its boundary plaquettes $p$ with $\tau_p(\bs e)<0$.
Since every spin, $s_b$, affects the sign of two plaquettes in a cube, for a frustrated cube, no spin configuration on the edges can simultaneously satisfy all the plaquette terms.
For the usual toric codes, frustrated cubes can be readily identified by stabilizer defects without referring to a candidate error.
For toric-GKP codes one has to be more careful.
In the limit $\sigmaM\to0$, the frustration cannot be read directly from the value of the toric code syndromes but all candidate errors exhibit the same frustration.
So picking any candidate error, $\bs{e}$, finds it.
When $\sigmaM\neq0$, this is no longer the case, the frustration can change between different candidate errors.

We examine the problem of finding the optimum spin configuration, $\bs s$, which minimizes the RPGM Hamiltonian \ref{eq:RPGM-one}, given the weights $\tau_p(\bs e)$.
For a given $\bs s$, we call a plaquette term, $\tau_p(\bs{e}) u_p$, \emph{satisfied} when $\tau_p(\bs{e}) u_p > 0$ and \emph{unsatisfied} when $\tau_p(\bs{e}) u_p < 0$.
Necessarily, a frustrated cube is incident to an odd number of unsatisfied plaquettes, in this sense frustrated cubes are the source of unsatisfied plaquettes.
Let $S$ be any set of plaquettes such that the frustrated cubes are incident to an odd number of plaquettes in $S$ and the unfrustrated cubes are incident to an even number of plaquettes in $S$.
One has
\[\min_{{\bs s}}\, H({\bs s};\bs e)=-\sum_p |\tau_p(\bs{e})|+2 \min_S
\sum_{p \in S} |\tau_p(\bs{e})|.\]
Hence minimum-weight matching on a 3D lattice with vertices representing the frustrated cubes determines the optimal set of unsatisfied plaquettes $S_{\rm min}$.
Given a candidate error $\bs e$ let $S_{\rm cand}$ be the set of plaquettes with $\tau_p(\bs e) < 0$.
The candidate error $\bs e$ is now modified using the minimum-weight matching by adding $2\pi$ for all plaquettes in $S_{\rm min}$ as well as adding $2\pi$ on all plaquettes in $S_{\rm cand}$ (remember that $4\pi$-shifts are elements of the stabilizer group).
This corresponds to an addition of a stabilizer or logical operator to the candidate error $\bs e$, leaving the syndrome unchanged.
This modified candidate error is the proposed correction for this decoder.

\begin{figure}[htbp]
	\centering
	\includegraphics[width=.6\textwidth]
	{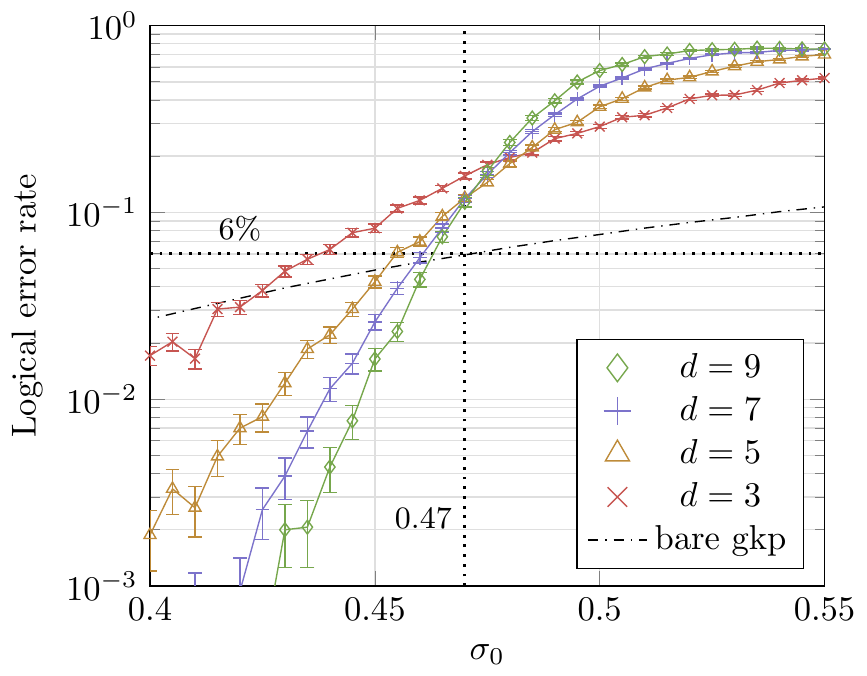}
	\caption{(Color online) Numerical results for perfect GKP measurements, with $\sigma = \sigmaT$ and $\sigmaM=0$.
		The logical error rate as a function of the bare standard deviation $\sigma_0$, see Eq.~\eqref{eq:scaled-variances}, is shown for toric code distances $d=L$ as indicated in the caption.
		The vertical dotted line indicates the position of the crossing point in the data, a threshold at $\sigma_0 \approx .47$, which corresponds to a logical error probability $p=6\%$ for single-qubit GKP code with $\sigmaM=0$.
		The latter value is recovered as the vertical position on the dash-dotted line which reproduces the green line $\sigmaM=0$ from Fig.~\ref{fig:single_gkp_plots} (left).
		The observed threshold is well above $\sigma_0=0.41$ which can be superficially expected from the $p=2.9\%$ threshold for the 2D toric code under phenomenological error model.}\label{fig:3dnumericsnomeaserr}
\end{figure}

We implemented the described decoder to minimize the energy \eqref{eq:RPGM-one}, with Villain potentials replaced with cosines, see Eqs.~\eqref{eq:plaquette-weight-ver},\eqref{eq:plaquette-weight-hor}.
A very simple estimate of the expected performance of the toric-GKP code in this setting is the following.
It is known that the threshold for the toric code is about $2.9\%$ under phenomenological noise with independent $X$ and $Z$ errors \cite{WHP:threshold}.
If we assume that all errors are due to the logical error on the underlying GKP qubits, see Fig.~\ref{fig:single_gkp_plots}, then one can ask what $\sigma_0$ leads to a probability for an $X$ error (or equivalently $Z$) equal to $2.9\%$.
This of course depends on the measurement error $\sigmaM$ as well as the decoding method for the GKP qubit.
In case of no measurement error ($\sigmaM=0$), the error probability can be found by averaging $\mathbb{P}_{M=1}(1 |q_1)$, see Eq.~\ref{eq:succ}, over the Gaussian error distribution, it is plotted as the green line in Fig.~\ref{fig:single_gkp_plots}.
An error probability of $2.9\%$ corresponds to $\sigma_0 \approx 0.41$.   
Our numerical results are shown in Fig.~\ref{fig:3dnumericsnomeaserr}.
One can see a crossing point around $\sigma_0 = .47$ which can be converted to around $6\%$ error rate per round for a GKP qubit with $\sigmaM=0$, see Fig.~\ref{fig:single_gkp_plots}.  
We can conclude that, similarly to the results in Sec.~\ref{sec:noiseless}, the continuous-valued weights $\tau_p(\bs{e})$, constitute valuable information to the decoder about the likelihood of errors and  permit to surpass decoders which do not have access to such information.

\subsubsection{Dealing with multiple competing minima}

A difficulty in minimizing the Hamiltonian \ref{eq:ham-epsilon-delta} is the existence of a large number of competing local minima.
This was already the case for a single GKP qubit, which we considered in Sec.~\ref{sec:singleGKP}.
We saw in the previous section that in the case of perfect GKP measurements, the solution can be found efficiently because the problem is equivalent to a minimum weight matching on a graph.
Our approach to minimizing Eq.~\ref{eq:ham-epsilon-delta} in the general case will be to decompose the vector potential $A_b\in[-\pi,\pi)$ for horizontal bonds $b=(ij,t)$ into a discrete field $A_b^{(0)}\in\{0,\pi\}$, and an auxiliary continuous field $a_b\in \left [-\frac{\pi}{2},\frac{\pi}{2}\right )$,
\begin{equation}
  \label{eq:aux-decomposition}
  A_b=A_b^{(0)}+a_b+\delta_b/2.  
\end{equation}
For vertical bonds there is no need of such a substitution since the corresponding field is already discrete, see Eq.~\ref{eq:Z-epsilon-delta} so we set $a_b=0$ for all $b\perp xy$.
The discrete part of the field, $\bs{A}^{(0)}$, can be used to define Ising spins, $\bs{s}$, similarily as before, $s_b \equiv \e^{iA_b^{(0)}}$.
The definition of the weigths, $\tau_p(\bs{e})$, has to be extended since now they depend also on $\delta_b\neq0$ and $a_b\neq0$, or more precisely the residual fluxes $b_p \equiv (\nabla\times a)_p$.
Adding these dependencies, the new weights, $\tau_p(\bs{a};\bs{e})$, given only in their cosine approximation, read
\begin{eqnarray}
\label{eq:RPGM-two-weight-horizontal}
\tau_p(\bs{a};\bs e)
&=& 
{4\over\sigmaT^2}\cos\left({\xi_p\over 2}-{1\over 2}(\nabla\times\delta)_p-b_p\right), \quad \qquad p\|xy,\\
\tau_p(\bs{a};\bs e)
&=& 
{4\over\sigma^2}\cos\left({\epsilon_p\over 2}-{1\over 2}\left (\delta_{ij}(t)-\delta_{ij}(t-1)\right )+b_p\right), \quad p\equiv (ij,t)\perp xy.
\label{eq:RPGM-two-weight-vertical}
\end{eqnarray}
Rewriting the Hamiltonian \ref{eq:ham-epsilon-delta}, using the substitution \ref{eq:aux-decomposition} and using the cosine approximation of the Villain potentials gives,
\begin{eqnarray}
  \label{eq:ham-a-s}
  H(\bs a,\bs s; \bs e)=-{1\over\sigmaM^2}\sum_{b\|xy}\cos(2a_b)-\sum_{p}\tau_p(\bs{a};\bs e)u_p.
\end{eqnarray}

Then, the minimization over the gauge field $\bs A$, or equivalently over $\{\bs{a},\bs{s}\}$, factors out into minimizing the RPGM similar to Eq.~\ref{eq:RPGM-one}, and a minimization over the continuous field $\bs{a}$.

In addition to the candidate error $\bs e$, the RPGM weights now also depend on $\bs{a}$ via the the flux field $\bs{b}$.
Moreover, even with the restriction on the auxiliary vector potentials, $|a_b|\le \pi/2$, the corresponding fluxes are not so restricted; in particular, both for horizontal and vertical plaquettes one may have $|b_p|=\pi$, sufficient to flip the sign of the RPGM weight, $\tau_p(\bs{a};\bs e)$, and flip the frustration of the two adjacent cubes.

Nevertheless, even though frustration depends on configuration of both the spins and the residual fluxes $b_p$, increasing the number of variables by the substitution \ref{eq:aux-decomposition} does simplify the minimization problem.
First, unlike in the isotropic model~\ref{eq:3D-U1-model}, the fields $a_b$ are uniquely defined by the fluxes $b_p$.
Indeed, since the gauge fields $a_b$ are only non-zero on the horizontal bonds, and are zero at the bottom layer, $t=0$, the gauge is fixed.
Furthermore, the GKP terms tend to suppress order-reducing fluctuations by favoring small $|a_b|$.
Thus, with $\sigmaM$ small compared to $\sigma$ and $\sigmaT$, we can hope to find a reasonably good solution just by setting $a_b=0$.

\subsubsection{Actual decoder algorithms and their performance} 
\label{sec:decod}

In order to design a syndrome-based decoder with the starting point $a_b=0$, we first need to come up with a candidate error $\bs e\equiv \bs e(\bs q)$.
Since we are not doing the full minimization of the corresponding energy, the method to find $\bs{e}$ will necessarily affect the performance of the resulting decoder.\\

\textsc{Algorithm 1:}
\begin{enumerate}
	\item For each plaquette $h$, starting from $t=M-1$ down to $t=1$, set the toric code measurement error $\xi_{h}(t)=q_h^{\rm tor}(t)-q_h^{\rm tor}(t+1)$, to suppress the increments of the toric syndrome.
	This leaves non-zero toric code syndromes only in the first layer, $q_h^{\rm tor}(t=1)$.
	\item Set data errors in the first layer, $\epsilon_{ij}(t=1)$, to move non-zero toric code syndromes to the left (with $ij\| y$), then down along the leftmost column.
	Due to the boundary conditions, the sum of all toric code syndromes is $0$, meaning that after this procedure, all toric code syndromes are removed including the one in the left-bottom plaquette.
	\item For each square lattice bond $(ij)$, starting with $t=1$ up to $t=M-1$, use Eq.~\ref{eq:GKP-syndrome-ij} to set GKP errors $\delta_{ij}(t)$ to suppress the GKP syndromes $q_{ij}^{\rm GKP}(t)$ (without changing $\phi_{ij}(t)$).
	This leaves non-zero GKP syndromes only in the top layer, $t=M$, with toric syndromes all zero.
	\item Make a gauge transformation on the top layer data errors, $\epsilon_{ij}(t=M)\to\epsilon_{ij}(t=M)+\chi_j-\chi_i$ to move non-zero GKP syndromes to the left (to $x=0$) and then down (to $y=0$). The last step works because equations \ref{eq:GKP-match}, \ref{eq:tor-match} are satisfied for the updated syndrome; with $q_h^{\rm tor}(t=M)=0$ these guarantee that the GKP syndrome is a gradient.
\end{enumerate}

Having determined the candidate error $\bs e$, we calculate the RPGM weights \ref{eq:RPGM-two-weight-horizontal}, \ref{eq:RPGM-two-weight-vertical} with $b_p=0$, and continue with minimum-weight matching decoding described in Sec.~\ref{sec:perfectGKP}.

\begin{figure}[htbp]
  \centering
  \includegraphics[width=0.95\textwidth]{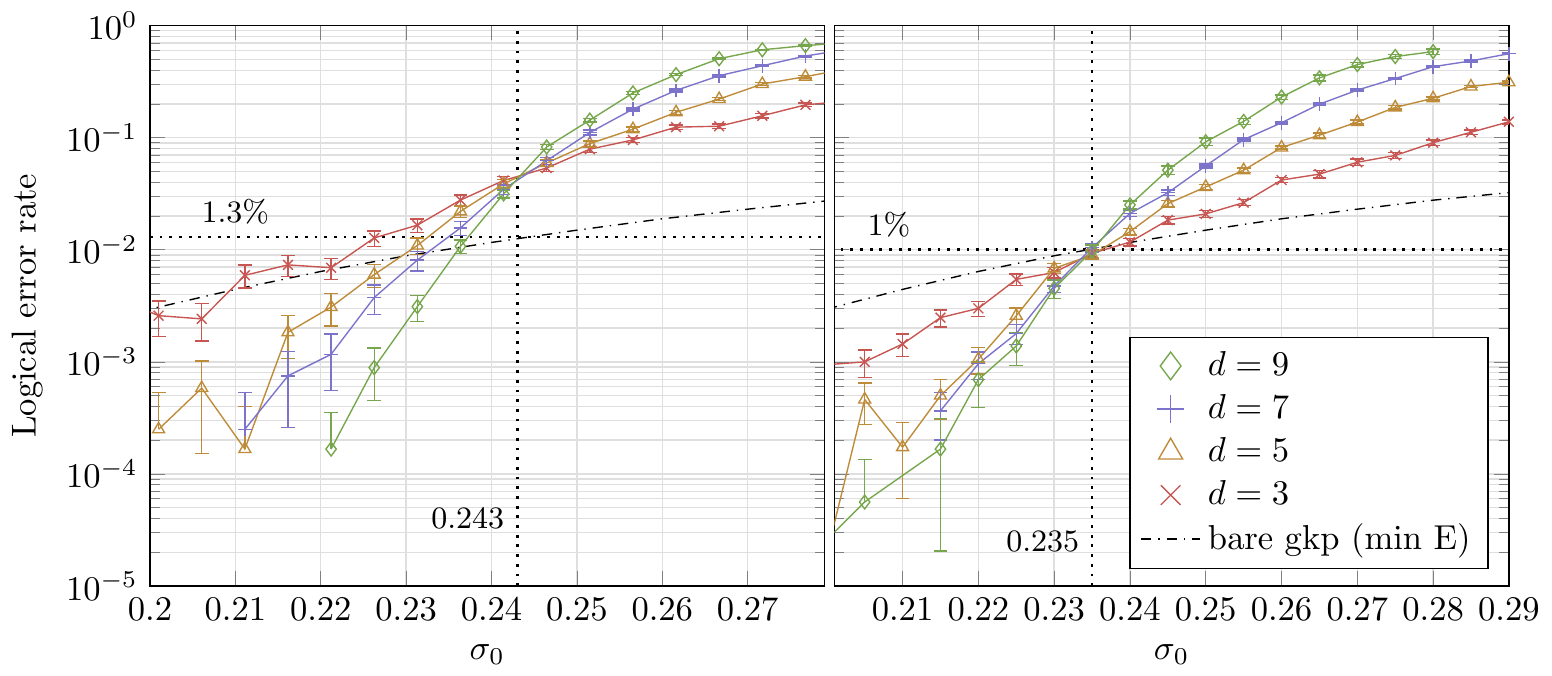}
  \caption{(Color online) For both plots the dash-dotted line shows the logical error rate per round for a single GKP qubit corrected with the forward-minimization decoder, with $\sigma = \sigmaT = \sigmaM$, as a function of the bare standard deviation $\sigma_0=\sigma/2\sqrt\pi$.
  	(Left) Numerical results for \textsc{Algorithm 1} without preprocessing of the GKP syndrome information (see text).
  	We observe a crossing point at $\sigma_0\approx0.243$ which can be translated to $p=1.3\%$.
  	(Right) Numerical results for the \textsc{Algorithm 2} with preprocessing of GKP syndrome information (see text).
  	We observe that this decoder improves the logical error rates.
  	On the other hand, the improvement being greater for smaller distances, the crossing point moves left to $\sigma_0\approx0.235$ which corresponds to an error rate $p=1\%$.
    }\label{fig:3dnumerics}
\end{figure}

The numerical results obtained with the described decoder at $\sigma=\sigmaM=\sigmaT$ are shown in Fig.~\ref{fig:3dnumerics} (left).
Despite the fact that this decoder does not make a particularly good use of the GKP syndrome information, $\bs{q}^{\rm GKP}$, and does not even try to find a good candidate error, the logical error rate rapidly goes down with increasing code distance and decreasing $\sigma_0$ below the crossing point at $\sigma_0\approx0.243$.
With the forward-minimization decoder on a single GKP qubit with $\sigma=\sigmaM$, this would correspond to a logical error rate of $p\approx 1.3\%$.
If the errors only come from having imperfect GKP states, then this also can be translated to having states with at least 4 photons.

It seems possible that, by making a better use of the GKP syndrome, one should be able to improve this decoder while preserving its computational efficiency.
To this end, we tried a \emph{preprocessing} algorithm.
The basic idea is, given the syndrome $\bs q^{\rm GKP}$, to find an initial approximation, $\bs e_0$, for the data errors which would bring back the GKP qubits closer to their code space.
Given $\bs e_0$, Algorithm 1 can be used to find an error $\bs e_1$ matching the updated syndrome, after which the RPGM weights can be computed using the full candidate error $\bs e_0+\bs e_1$.
The hope is then that the candidate error found is one for which the first term of the Hamiltonian in Eq~\ref{eq:ham-a-s} does not need to be minimized anymore.
In particular, we tried using our single-oscillator forward-minimization decoder from Sec.~\ref{sec:singleGKP} as the preprocessing step.
Using it directly produced a degradation of performance, seemingly resulting from the fact that the minimization of the RPGM Hamiltonian also tries to optimize the GKP measurement errors $\delta_b$.
Our solution was to drop the measurement errors from the decomposition of the field in \ref{eq:aux-decomposition}, which results in RPGM weights identical to those in Sec.~\ref{sec:perfectGKP}.
Our second decoder can then be summarized as follows.\\

\textsc{Algorithm 2:}
\begin{enumerate}
	\item For each bond $(ij)$, use the forward-minimization decoder from $t=1$ up to $M-1$ to calculate the accumulated data error $\phi_{ij}(t)$, calculate the corresponding $\phi_{ij}(t=M)$, and then go back from $t=M$ to $t=1$ with a version of the same algorithm, but using previously found values for a more accurate minimization.
	This gives the data errors $\bs \epsilon_{ij}(t)=\phi_{ij}(t)-\phi_{ij}(t-1)$, which we use to define the error vector $\bs e_0=\{\bs 0,\bs \epsilon,\bs0\}$.
	\item Calculate the residual syndrome, and run the entire \textsc{Algorithm 1}, to calculate the corresponding error $\bs e_1$.
	\item Calculate RPGM weights $\tau_p(\bs e)$ using Eqs.~\ref{eq:RPGM-two-weight-horizontal}, \ref{eq:RPGM-two-weight-vertical} with the combined error $\bs e=\bs e_0+\bs e_1$ and $\bs\delta$ set to zero.
	\item Use random-weight matching to minimize the RPGM Hamiltonian, and update the error $\bs e$, with the result being the output of the decoder.
\end{enumerate}

The results are shown in Fig.~\ref{fig:3dnumerics}, (right).
One can observe that for each distance, there is an improvement in the encoded logical error rate, compared to the results from Algorithm 1 on the left of Fig.~\ref{fig:3dnumerics}.
One can see, for each curve, a higher \emph{pseudo-threshold}, i.e. the point below which the logical error rate becomes smaller than the physical one, determined using the logical error rate for a single GKP qubit given by the forward-minimization decoder.
However, this improvement is greater for smaller distances, so the overall crossing point is shifted to the left, indicating a smaller threshold.
Specifically, the crossing point is at $\sigma_0\approx0.235$ which corresponds to $p=1\%$ for the single GKP qubit with the forward-minimization decoder.

We should note that even though both simple decoders we tried result in finite thresholds, with substantial reduction of the logical error rates with increasing distance below the crossing points, one could expect better performance.
Indeed, the toric code with a phenomenological error model shows a threshold at $p=2.9\%$.
The forward-minimization decoder with $\sigma=\sigmaM$ in Sec.~\ref{sec:singleGKP} reaches this error rate at $\sigma_0\approx 0.28$.  

Of course, a direct comparison is technically incorrect: forward-minimization or any other single-oscillator GKP decoder would return highly correlated errors and one cannot expect that the toric code would achieve the same performance.
Nevertheless, we expect that adding a minimization step for the continuous part of the potential $\bs a$ in Eq.~\ref{eq:aux-decomposition} would significantly improve the performance.

\section{No-Go Result for Linear Oscillator Codes}
\label{sec:nogo}

We turn to the class of codes defined by continuous subgroups of displacement operators. One can think of linear combinations of position and momentum operators as nullifiers for the code space, i.e. any code state is annihilated by these nullifiers, see Appendix \ref{sec:twoclasses}. 

It is known that one cannot distill entanglement from Gaussian states by means of purely Gaussian local operations and classical communication \cite{GC:distill}, \cite{ESP:distill}. In addition, the authors of Ref.~\cite{NFC:nogo} defined a quantity, ``entanglement degradation", for any single-mode Gaussian channel, such as the Gaussian displacement channel, and showed that it cannot decrease under Gaussian encoding and decoding.
In the setting here we consider any input state which is perfectly encoded into a linear oscillator code. This encoding map, ${\cal E}$, is a Gaussian operation as it is a linear transformation of the $\hat{p}$ and $\hat{q}$ variables.
After this encoding, the modes go through the Gaussian displacement channel $\mathcal{N}$, see Eq.~\eqref{eq:GDC}.
After this, linear combinations of $\hat{p}$ and $\hat{q}$ are measured to give rise to a syndrome $\bs{q}$.
Again, this operation is Gaussian.
Our results thus follow the model considered in \cite{NFC:nogo}.
However, our results do not require the definition of a new quantity but rather give a description of the logical noise model of Eq.~\eqref{eq:succ-decoding}.
Namely, we show in the following Theorem, that it can only lead to an effective squeezing of the original Gaussian displacement channel.
Hence, whatever protection is gained in one quadrature, is lost in the other quadrature.
This is a property of all linear oscillator codes, CSS or non-CSS, i.e. mixing $\hat{p}$ and $\hat{q}$ quadratures or not, with respect to the Gaussian displacement channel.
Since the result is a detailed expression of the logical Gaussian displacement channel, it does not immediately follow from earlier no-go results.

\begin{thm}[No-Go]
  Let $\mathcal{C}$ be a linear oscillator code on $n$ physical oscillators defined by a set of $n-k$ independent nullifiers, thus encoding $k$ logical oscillators.
  Let this code undergo independent Gaussian shifts in $\hat{p}$ and $\hat{q}$ of variance $\sigma_0^2$ on each of its physical oscillators, followed by a perfect (maximum-likelihood) decoding step.
  Then the remaining logical displacement noise model, Eq.~\eqref{eq:succ-decoding}, for logical shift errors $\bs{\epsilon} \in \mathbb{R}^{2k}$ is
  \begin{equation*}
  \mathbb{P}^{(\rm ML)}(\bs{\epsilon})=(2\pi \sigma_0^2)^{-2k} \exp\left(-\frac{1}{2\sigma_0^2} \bs{\epsilon}\, \Sigma^{-1} \bs{\epsilon}^T\right),\label{eq:nogo-res}
  \end{equation*}
  and the eigenvalues of the covariance matrix, $\Sigma^{-1}$, are paired by conjugated logical operators, $(\lambda_j^p,\lambda_j^q)_{j\in[k]}$, such that
  \begin{equation*}
  \forall j\in[k],\;\lambda_j^p\lambda_j^q = 1.
  \end{equation*} \label{thm:nogo}
  In particular one has $\det\Sigma^{-1} = 1$.
\end{thm}

The proof of this Theorem, which is rather lengthy, can be found in Appendix~\ref{sec:proof}.
The remaining logical displacement noise is obtained by working out Eq.~\ref{eq:succ-decoding}, using general properties of linear oscillator codes.
The theorem says that for each logical mode, in the basis given by the eigenvectors of $\Sigma$, the only effect of the encoding is to squeeze the displacement noise model between the conjugated operators of the mode.
The amount of squeezing can depend on code size but it is impossible to reduce the noise in both quadratures.

As a concrete example, we consider the linear oscillator code version of the 2D toric code in Appendix~\ref{sec:CV-toric} and show that the squeezing depends simply on the ratio $L_x/L_y$ for a $L_x \times L_y$ toric lattice.

\section{Discussion}

In this paper we have made the first strides in tackling the problem of error correction and decoding for the toric-GKP code.
Interestingly, the decoding problem maps onto a new class of physical continuous-variable models with quenched-disorder, going beyond the random plaquette gauge model corresponding to toric code decoding \cite{WHP:threshold}.
We have presented an efficient minimum-energy decoding method for a single GKP oscillator and an efficient decoder for the toric-GKP code.
We have also presented a combination of these two decoders for the toric-GKP code which improves the achieved logical error rates but in greater proportion for the small distances, hence achieving better pseudo-thresholds but a slightly worse threshold.
It would be interesting to design a better decoder to improve this threshold.

An interesting open problem is to study the phase diagram of the toric-GKP code numerically.
A particularly interesting question, even in the absence of disorder, is whether the Meissner phase is extended all the way to $\sigmaM\to\infty$ as we conjectured, see Fig.~\ref{fig:qed3}, or terminates at a point along the vertical axis.
This would indicate the singularity of the limit $\sigma\to0$.
An example of such a numerical study for 3D color codes is \cite{kubica+:part}.

Future studies could look at the question of decoding coherent errors and/or correcting both $p$ and $q$-shifts simultaneously.
Another question is whether it is possible to handle more realistic noise models, e.g. consider a model of repeated photon loss \cite{albert+:bos-codes} for a single GKP oscillator.
All such different error models will have a particular path integral representation and the idea of choosing an energy-minimizing path can be examined.

A variant of the toric-GKP code is the toric-rotor code.
This is the concatenation of a rotor space with integer $\hat{n}$ and $2\pi$-periodic $\hat{\varphi}$ with a rotor toric code whose nullifiers are linear combinations of $\hat{n}$s and $\hat{\varphi}$s of four rotors \cite{APD:overview}.
When one uses a two-dimensional rotor subspace such as a cat code \cite{ofek+:cat} or just a transmon qubit, one could still express the proper toric code checks in the entire rotor space and examine when a memory phase is present.
Note that the difference in such analysis versus the usual toric code analysis is that in this model the effect of leakage errors is automatically included.

A possible realization of a surface code variant of the toric-GKP code is an array of superconducting 2D or 3D resonators.
For a code such as Surface-17, this would require $2 \times 17=34$ (coplanar) microwave resonators.
One can compare this to the transmon + resonators lay-out for the regular surface code \cite{divincenzo:arch}, \cite{versluis+:layout} in which each CNOT gate is mediated via a coupling bus, hence one uses $8 \times 4=32$ such bus resonators for Surface-17.

\begin{acknowledgments}
BMT, YW and CV acknowledge support by the European Research Council (EQEC, ERC Consolidator Grant No: 682726) as well as a QuantERA grant for the QCDA consortium.
LPP acknowledges support from NSF Division of Physics via Grants No.\ 1416578 and 1820939.
\end{acknowledgments}

\bibliography{Toric-GKP-code-arxiv-v2}

\appendix 

\section{General multi-mode GKP codes}
\label{sec:twoclasses}

Here we provide a mathematical summary of the formalism of continuous-variable stabilizer codes, as introduced by Gottesman, Kitaev and Preskill (GKP) \cite{GKP}. We show that these codes contain two main classes, proper GKP codes, encoding discrete information and linear oscillator codes, encoding continuous-variable information. Hybrids between the two classes are also possible but do not seem particularly useful.\\

Recall that we write a displacement operator on a $n$-oscillator Hilbert space as follows
\begin{equation}
\hat{U}(\bs e)\equiv \prod_{k=1}^n \e^{i u_k\hat{p}_k+i
	v_k\hat{q}_k}, \quad \bs e\equiv (\bs u,\bs v)\in\mathbb{R}^{2n}.
\end{equation}
We also denote by $\hat{O}(\bs{e})$ the operator in the exponent of $\hat{U}(\bs{e})$,
\begin{equation}
\hat{O}(\bs{e}) \equiv \bs{u}\cdot\hat{\bs{p}} + \bs{v}\cdot\hat{\bs{q}} = \sum_{k=1}^nu_k\hat{p}_k + \sum_{k=1}^nv_k\hat{q}_k,\qquad\hat{U}(\bs e) = \exp\left (i\hat{O}(\bs{e})\right ).\label{eq:Oop}
\end{equation}
These operators, $\hat{U}(\bs{e})$, follow the product rule
\begin{equation}
\hat{U}(\bs e)   \hat{U}(\bs{e}') 
=\hat{U}(\bs e+\bs e')\e^{i\omega(\bs e,\bs e')},
\end{equation}
with the standard symplectic form $\omega(\bs e,\bs e') = \bs{u}\cdot\bs{v}'-\bs{v}\cdot \bs{u}'$.

Take $\mathcal{S}\subset\mathcal{H}_n$ an Abelian subgroup of the displacement operators with no element proportional to the identity  with a non-trivial phase, stabilizing some code space, $\mathcal{Q}$.
One can simply characterise the structure of such a subgroup.
Each element $\hat{S}\in\mathcal{S}$ can be uniquely written as
\begin{equation}
\label{eq:addgroup}
\hat{S} = \e^{i\theta}\hat{U}(\bs e),
\end{equation}
for some real vector $\bs{e}\in\mathbb{R}^{2n}$, and some phase $\theta\in[0,2\pi)$.
Each vector $\bs{e}$ can only appear with a unique phase $\theta$ in $\mathcal{S}$, otherwise an element proportional to the identity with a non-trivial phase would also be in $\mathcal{S}$.
Hence we have an isomorphism between $\mathcal{S}$ and the additive subgroup $G\subset\mathbb{R}^{2n}$, given by all the vectors, $\bs{e}$, obtained from the decomposition in Eq.~\ref{eq:addgroup}.
We can then use the following theorem to characterize the structure of $\mathcal{S}$.
\begin{thm}[\cite{book:Bourbaki}]
	Let $G$ be a closed additive subgroup of $\mathbb{R}^{2n}$, $G$ can be decomposed in the direct sum of a linear subspace, $G_0\subset\mathbb{R}^{2n}$, and a discrete lattice, $L$, generated by some vectors orthogonal to $G_0$, $\bs{u}_1,\dots,\bs{u}_m\in G_0^\perp$:
	\[G = G_0\oplus L,\quad L = \left \{\sum_{j=1}^mk_j\bs{u}_j\;\Big|\;k_1,\dots,k_m\in\mathbb{Z}\right \}.\]
	The linear subspace $G_0$ is the largest linear subspace contained in $G$.
\end{thm}
The condition for $G$ to be closed is without consequences in our case as an open set and its closure would stabilize the same space.
If $G$ is open then we can replace it by its closure and appropriately complete $\mathcal{S}$.
The limiting case when $G = G_0$ corresponds to the continuous case whereas $G = L$ corresponds to the discrete case.
Denote $\ell$ as the dimension of $G_0$, one can choose some basis vectors, $G_0 = \langle\bs{g}_1,\dots,\bs{g}_\ell \rangle$.
Consider one of the generators of $G_0$, w.l.o.g. $\bs{g}_1$, for this generator, given a scalar factor $\lambda\in\mathbb{R}$, there is some angle function $\theta_1(\lambda)$ such that
\[\e^{i\theta_1(\lambda)}\hat{U}\left (\lambda\bs{g}_1\right )\in\mathcal{S}.\]
It is easy to check that $\theta_1$ obeys Cauchy's functional equation and as such $\theta_1$ is automatically $\mathbb{Q}$-linear:
\[\forall(\lambda,\mu)\in\mathbb{R}^2,\;\theta_1(\lambda+\mu) = \theta_1(\lambda)+\theta_1(\mu)\;\Rightarrow\;\forall r \in\mathbb{Q},\;\theta_1(r) = r\theta_1(1) \equiv r\theta_1.\]
Hence, for any code state, $\ket{\overline{\Psi}}\in\mathcal{Q}$, and rational $r\in\mathbb{Q}$, we can write down
\begin{equation}
\e^{ir\theta_1}\hat{U}\left (r\bs{g}_1\right )\ket{\overline{\Psi}} = \exp\left (ir\left (\theta_1 + \hat{O}(\bs{g}_1)\right )\right )\ket{\overline{\Psi}} = \ket{\overline{\Psi}}.
\end{equation}
The previous equation means that code states are eigenstates of the operator $\hat{O}(\bs{g}_1)$ with eigenvalue $O(\bs{g}_1)$ which satisfies
\begin{align*}
\forall r \in\mathbb{Q},&\quad O(\bs{g}_1) + \theta_1 = 0 \mod{2\pi/r} \\
&\Leftrightarrow\; O(\bs{g}_1) =-\theta_1.
\end{align*}
Usually, $\mathcal{S}$ will be chosen such that $\theta_1=0$, and $\hat{O}(\bs{g}_1)$ will be called a \emph{nullifier} as it only takes eigenvalue $0$ on the code space.
Choosing some non-trivial $\theta_1$ just corresponds to shifting the whole code space by $\hat{U}(\theta_1\bs{d}_1)$, with $\bs{d}_1$ describing the conjugated pair to $\bs{g}_1$, i.e. such that $[\hat{O}(\bs{g}_1),\hat{O}(\bs{d}_1)] = i$.
Similarly, each generator $\bs{g}_j$ is a nullifier on the code space (up to some possible shift $\hat{U}(\theta_j\bs{d}_j)$).

At this point, if $G=G_0$ ($L=\emptyset$), then we have described a linear oscillator code.
It is defined by the $\ell$ nullifiers, $\hat{O}(\bs{g}_j)$, which each remove a single continuous degree of freedom from the system, leaving $k\equiv n-\ell$ logical oscillator modes.
The logical operators can be found by completing the $\bs{g}_j$ into a full symplectic basis.
For the stabilized code space to be non-trivial we therefore require that $\ell < n$.

Consider now that $L$ is non-trivial, then it will constrain the code space as described in Ref.~\cite[Sec.VI]{GKP} on the remaining $k=n-\ell$ modes available.
Take one of the generators of $L$, w.l.o.g. $\bs{u}_1$.
The difference with elements of $G_0$ is that it occurs in $\mathcal{S}$ only with integer multiples.
Similarly as previously, given a code state, $\ket{\overline{\Psi}}\in\mathcal{Q}$, and an integer $k\in\mathbb{Z}$ there will be some angle, $\vartheta_1$, such that,
\begin{equation}
\e^{ik\vartheta_1}\hat{U}\left (k\bs{u}_1\right )\ket{\overline{\Psi}} = \exp\left (ik\left (\vartheta_1 + \hat{O}(\bs{u}_1)\right )\right )\ket{\overline{\Psi}} = \ket{\overline{\Psi}}.
\end{equation}
This means that on the code states, the operator $\hat{O}(\bs{u}_1)$ can take now several values, given by
\begin{align*}
\forall k \in\mathbb{Z},&\quad O(\bs{u}_1) + \vartheta_1 = 0 \mod{2\pi/k} \\
&\Leftrightarrow\; O(\bs{u}_1) =-\vartheta_1 \mod{2\pi}.
\end{align*}
The effect is to discretize this mode.
As explained in Ref.~\cite{GKP}, one then needs $m = 2k$ lattice generators to fully discretize the remaining $k$ modes.

Summing up, the case where $G = G_0$ and $\ell < n$ ($m=0$) corresponds to linear oscillator codes defined by $\ell$ nullifiers, encoding $k = n-\ell$ oscillators.
On the other hand, the case $G = L$ and $m=2n$ ($\ell=0$) correspond to proper GKP codes described in \cite{GKP}.
Finally the hybrid case where $G = G_0\oplus L$, with $n = \ell + m/2 + k^\prime$ correspond to a case where $\ell$ nullifiers leave $n-\ell$ modes in the code space, among which $n-\ell-k^\prime$ are discretized into a qudit (depending on the characteristics of $L$) and $k^\prime$ remain as logical oscillators.
As we show in Section~\ref{sec:nogo}, under Gaussian noise, the logical oscillators defined by the nullifiers have essentially the same noise model as the physical oscillator modes, so there is little interest in going beyond proper GKP codes.
Note that linear oscillator codes can nevertheless correct erasure errors \cite{braunstein:CV, LS:CV}.

\section{Details About the Decoders for the GKP Qubit}
\subsection{Maximum Likelihood Decoder}
\label{sec:gauss-int}

Maximum likelihood decoding for a single GKP qubit requires the calculation of the partition functions $Z_c(\bs q)$, $c\in\{0,1\}$ given by Eq.~\ref{eq:Z0-one-gkp}.
Here we do the Gaussian integration.
Denote as $B$ the symmetric matrix with the components 
\begin{equation}
  B_{ij}\equiv {1\over 2}{\partial^2\over \partial \phi_i\partial\phi_j}R(\bs \phi,\bs0),\quad i,j\in[M-1], \label{eq:mat-B}
\end{equation}
associated with the first $M-1$ variables $\phi_t$ in the quadratic form 
\[R(\bs \phi,\bs q)={1\over {\sigmaM^2}}  \sum_{t=1}^{M-1}{\left (q_t - \phi_t\right )^2}+{1\over\sigma^2}\sum_{t=1}^{M}{\left (\phi_t - \phi_{t-1}\right)^2}\biggr|_{\phi_M=q_M}\]
in the exponent in the integrand of Eq.~\ref{eq:Z0-one-gkp}.
Collecting the remaining terms and completing the square we obtain an $M$-variable quadratic form $\bs q A\bs q^\T$ with the block matrix
\begin{equation}
  \label{eq:mat-A}
  A = \left (\begin{array}{c | c}
\tilde{A} & \bs{c}^\T \\\hline
\bs{c} & b
\end{array}\right ),
\end{equation}
expressed in terms of the $(M-1)\times (M-1)$ matrix $\tilde{A}$, row vector $\bs c$, and a scalar $b$:
\begin{align}
\tilde{A} &=
            \frac{1}{\sigmaM^2}\id-\frac{1}{\sigmaM^4}B^{-1}, 
            \quad  
  \bs{c}_i = -{1\over \sigma^2\sigmaM^2}\left [B^{-1}\right ]_{M-1,i},
            \quad 
  b = \frac{1}{\sigma^2}- \frac{1}{\sigmaM^4}\left[B^{-1}\right
      ]_{M-1,M-1}.
      \label{eq:MLDdetails}
\end{align}

\subsection{Dynamic Programming for the Minimum-Energy Decoder}
\label{sec:dynprog}

\begin{figure}[htbp]
	\centering
	\includegraphics[width=0.95\textwidth]{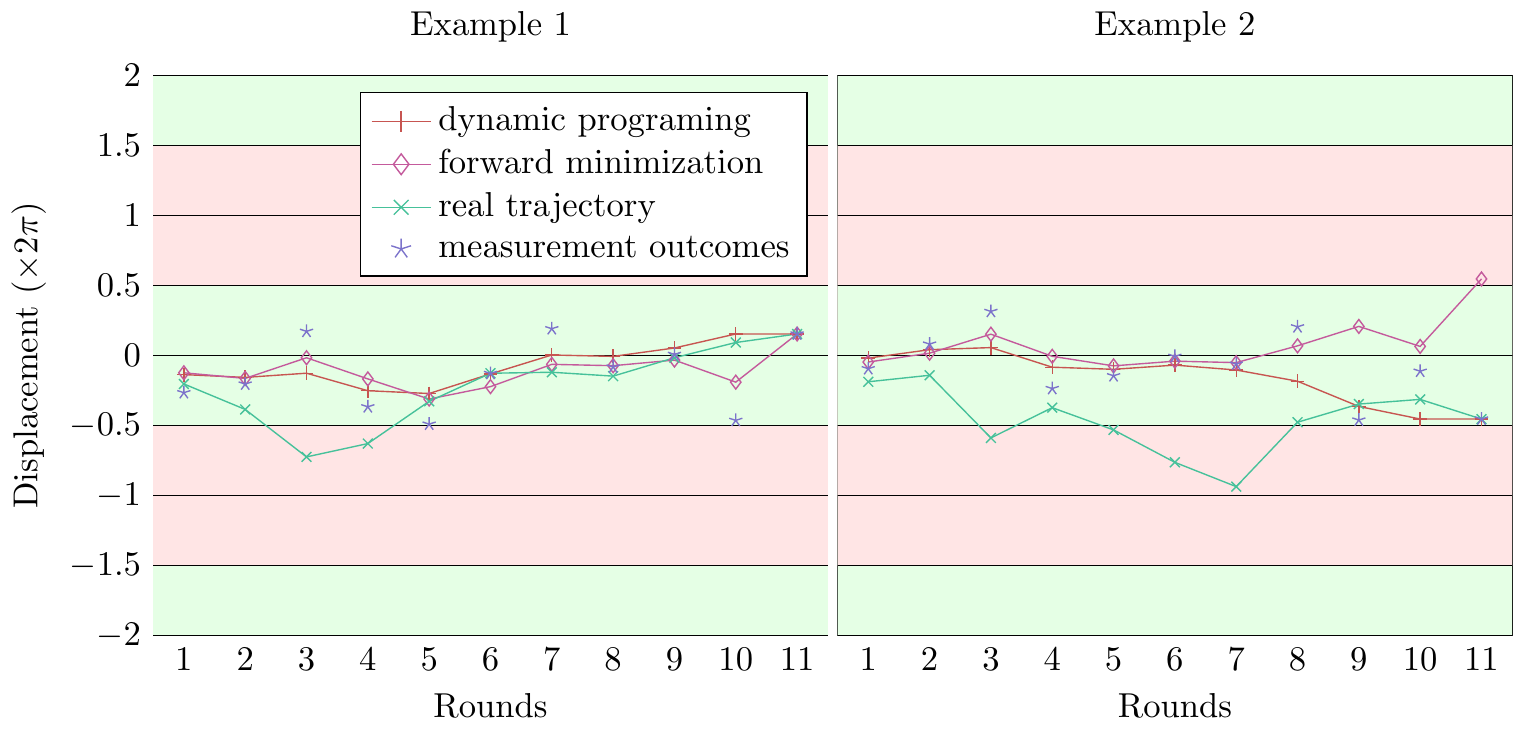}
	\caption{(Color online) Two examples of 11 rounds of error correction with data and measurement errors sampled from Gaussian distributions with $\sigma_0 = 0.4$.
		In the example on the left, forward-minimization and dynamic programming reach the same conclusion, whereas on the right forward-minimization reaches a wrong conclusion.
		One can compare this data with the sketched trajectories in Fig.~\ref{fig:winding-path-strategy}.}
	\label{fig:dynprogcompare}
\end{figure}
In this section we express the minimum-energy decoder for one GKP qubit as a dynamic programming problem.
This allows us to check how well the forward-minimization technique performs.
The goal is to minimize the energy from Eq.~\ref{eq:GKP-path-integral}, which we write using the Villain approximation in reverse,
\[H\left [\phi_1,\ldots,\phi_M\right ]=\sum_{t=1}^{M} \frac{\left(\phi_{t}-\phi_{t-1}\right)^2}{2\sigma^2}-\sum_{t=1}^{M-1}\frac{\cos(q_t-\phi_t)}{\sigmaM^2},\quad \phi_0\equiv 0.\]
We define the partial energy, $H_k\left[\phi_1,\ldots,\phi_k\right]$ as the contribution from the first $k$ terms from each sum,
\[
H_k\left [\phi_1,\ldots,\phi_k\right ]=\sum_{t=1}^{k} \frac{\left(\phi_{t}-\phi_{t-1}\right )^2}{2\sigma^2}-\frac{\cos(q_t-\phi_t)}{\sigmaM^2},
\] 
and a single-variable function of $\phi_k$ as 
\[M_k(\phi_k) = \min_{\phi_1, \ldots, \phi_{k-1}}\left \{H_k\left[\phi_1,\ldots,\phi_k\right ]\right \}.\]
Then $M_{k}$, $k<M$, can be defined recursively by 
\begin{align*}
M_1(\phi_1)&=\frac{\phi_1^2}{2\sigma^2}-\frac{\cos(q_{1}-\phi_{1})}{\sigmaM^2},\\
M_{k}(\phi_{k}) &= \min_{\phi}\left (M_{k-1}\left (\phi\right ) + \frac{\left(\phi_{k}-\phi\right )^2}{2\sigma^2}\right)-\frac{\cos(q_{k}-\phi_{k})}{\sigmaM^2}.
\end{align*}
If one discretizes the values of $\phi_k$ to a desired precision and restricts them to lie in a reasonable interval, the minimization with $M$ time steps amounts to computing $M$ lists of values of discrete functions $M_k(\phi_k)$.
Unlike the minimization technique based on solving Eq.~\ref{eq:GKP-FK-eqn}, there is no accuracy loss at larger $M$, and much less danger of missing the desired minimum with the present dynamic programming method.

We have compared our forward minimization technique with dynamic programming with a discretization of 200 points per period in a 4-periods window around the last measurement result.
The statistics of success or failure of both decoders agree pretty closely while the forward minimization is much faster.
In Fig.~\ref{fig:dynprogcompare} we show two of the obtained realizations for illustration purposes.

\section{Alternative derivation of the $U(1)$-symmetric model}
\label{app:derivation-Z0}

Here we derive Eqs.~\ref{eq:ham-epsilon-delta} and \ref{eq:Z-epsilon-delta} directly from Eqs.~\ref{eq:total-ham-phi} and \ref{eq:3D-Z}.
First, note that the integration over components of the field $\bs \phi$ in Eq.~\ref{eq:3D-Z} is done in an infinite interval, while the two last terms in the bulk Hamiltonian \ref{eq:total-ham-phi} are $2\pi$- and $4\pi$-periodic, respectively.
In addition, the background fluxes $(\nabla\times\phi)_{h}(t)$ in the last term of Eq.~\ref{eq:total-ham-phi} are explicitly symmetric with respect to the gauge transformation
\begin{equation}
  \phi_{ij}(t)\to \phi_{ij}(t)+\alpha_{j}(t)-\alpha_{i}(t),
  \label{eq:gauge-shift}
\end{equation}
where $\alpha_j(t)=\alpha_v$ is a real-valued scalar field associated with the vertices $v=(i,t)$ of the cubic lattice.
While these are \emph{not} the symmetries of the full Hamiltonian \ref{eq:total-ham-phi}, we can now render it in a more familiar $U(1)$-symmetric form with a simple change of variables.

For a fixed error $\bs e\equiv \{\bs\epsilon,\bs\delta,\bs\xi\}$ which corresponds to the syndrome $\bs q$, let us denote the corresponding accumulated error, see \ref{eq:phi-ij}, as $\bs\phi^{(0)}$.
We can make a change of integration variables in Eq.~\ref{eq:Z-epsilon-delta}, for all bonds $(ij)$ and layers $1\le t\le M$, following
\begin{equation}
  \phi_{ij}(t)=\phi^{(0)}_{ij}(t)+2A_{ij}(t)+4\pi m_{ij}(t)\pm 2\pi \sum_{t'=1}^t \big(s_{j}(t^\prime)-s_{i}(t^\prime)\big),\quad -\pi< A_{ij}(t)\le \pi,
  \label{eq:phi-decomposition}
\end{equation}
where the field $A_{ij}(t)$ is continuous, $m_{ij}(t)\in\mathbb{Z}$ is integer-valued, and an additional $2\pi$-shift is proportional to the lattice gradient of the binary field $s_{i}(t)\in\mathbb{F}_2$ accumulated over time.
Strictly speaking, the last term is unnecessary as it causes some double counting in the measure.
However, the corresponding factor is a constant that is finite on a finite lattice, so it does not cause any trouble.
On the other hand, these binary charges simplify the boundary conditions, since we can simply write $s_{i}(t)+\ldots+s_{i}(M)=n_i$, see Eq.~\ref{eq:3D-Z}, and thus trade the summation over the boundary field $n_i$ for the binary field $s_{i}(t)$ in the bulk.
To complete the derivation, also define $A_{ij}(t=0)=0$, and take $A_b=\pi s_{i}(t)$ for the vertical bond $b$ at the square-lattice vertex $i$, between layers $t-1$ and $t$ for all $t\in[M]$.
This gives for the vertical plaquette $p$ at the bond $(ij)$, between the same two layers,
\[
\phi_{ij}(t)-\phi_{ij}(t-1)\to \phi_{ij}^{(0)}(t)-\phi_{ij}^{(0)}(t-1)+2(\nabla\times A)_p\equiv \epsilon_p+2(\nabla\times A)_p,
\]
where the appropriate sign in Eq.~\ref{eq:phi-decomposition} needs to be chosen to recover the part of the flux that is missing in the first term of Eq.~\ref{eq:total-ham-phi}, and we absorbed the summation over the integer-valued $m_{ij}(t)$ into the definition of the Villain potential $V_{\sigma/2}\big(\epsilon_p/2+(\nabla\times A)_p\big)$, an even $2\pi$-periodic function of the argument, see Eq.~\ref{eq:Villain}.
In the remaining two terms we use Eqs.~\ref{eq:GKP-syndrome-ij} and \ref{eq:q-plaquette}, to recover Eq.~\ref{eq:ham-epsilon-delta} exactly \footnote{Strictly speaking we get $-(\nabla\times A)_p$ for a vertical $p$ along $y$. But we can push this sign to the disorder component using the parity of $V_{\sigma/2}$ and use the symmetry of our noise model to ignore this sign.}.
Comparing with Eqs.~\ref{eq:GKP-match}, \ref{eq:tor-match}, it is also easy to check that for this particular error $\bs e$, one can simply take the vector potential $A_b=0$ for the top-layer in-plane bonds, $b=(ij,M)$.
Similarly, for the three sectors where the error in the top layer is shifted by a non-trivial toric codeword ${\bs 0}\not\simeq {\bs c}\in\mathbb{F}_2^{2L^2}$, we can take $A_{ij}(M)=\pi c_{ij}$.

\section{Details about the phase-diagram derivation}
\label{app:phase-diag}
The model in Eq.~\ref{eq:gauge-model} is in a general class of compact $U(1)$ models whose partition functions can be written (in a conventional \emph{Wilson}, or non-Villain form) as
\begin{equation}
\label{eq:most-general-model}
Z(\bs \kappa,{\bs\varphi};P)= \int {\rm d}\bs \theta\, e^{-H},\quad H\equiv -\sum_b{\kappa_b\cos\Bigl(\varphi_b+\sum\nolimits_i \theta_i P_{ib }\Bigr)},
\end{equation}
where $P$ is an $r\times n$ coupling matrix with integer components $P_{ib}$ that determines the structure of the model, $\theta_i$, $i\in[r]$ are the $U(1)$ variables, $\theta_i\in[0,2\pi)$, $\kappa_b\ge0$, $b\in[r]$, are the coupling constants, and the additional phases $\varphi_b$ can represent quenched disorder and/or a uniform background field.
The only requirement on the integration measure for any component $\theta_j$ in Eq.~\ref{eq:most-general-model} is $2\pi$-periodicity.
For example, one can have the usual integration over the period, or a summation over discrete phases $\theta_j=2\pi m_j/q_j$, $m_j\in\mathbb{Z}_q$, with some integer $q_j\ge2$ that may differ for different ``spins'' $j$.
According to the Fourier theorem, the most general correlation function of the variables $\theta_j$ can be written as the average
\begin{equation}
\label{eq:most-general-corr-f}
{\cal C}_{\bs m}\equiv C_{\bs m}(\bs \kappa,\bs \varphi;P)=\left\langle e^{i\bs m\cdot\bs\theta}\right\rangle, 
\end{equation}
where the vector $\bs m$ has integer components.
The physics of the model \ref{eq:most-general-model} is characterized by the dependence of the free energy $F=-\ln Z$ and the correlation functions $C_{\bs m}$ on the parameters, the coupling constants $\kappa_b$ and the phases $\varphi_b$.

This dependence is restricted by several constraints.
Two of them, the first and the second generalized Griffiths-Kelly-Sherman (GKS) inequalities, concern the correlation functions in the absence of background phases, $\bs \varphi=\bs 0$ (in this case the averages \ref{eq:most-general-corr-f} are real-valued),
\begin{eqnarray}
\label{eq:GKS-one}
\langle f\rangle&\ge&0,\\
\label{eq:GKS-two}
\left\langle fg\right\rangle-\langle f \rangle \langle g\rangle&\ge&0,    
\end{eqnarray}
where $f$ and $g$ can be any nonnegative combination of products of $\cos (\bs m\cdot \bs \theta)$, with various integer vectors $\bs m\in\mathbb{Z}^n$.
In the case of the model~\ref{eq:most-general-model} in Wilson form these are called Ginibre inequalities \cite{Ginibre-1970}.
It is also easy to check that the average $C_{\bs m}(\bs \kappa,\bs \varphi;P)$ can only be non-zero if $\bs m$ is a linear combination of the rows of the matrix $P$.
Furthermore, the l.h.s.\ of the second inequality \ref{eq:GKS-two}, with $g=\cos(\bs m'\cdot\bs \theta)$, equals the derivative of $\langle f\rangle$ with respect to the coupling constant corresponding to the term $\cos (\bs m'\cdot \bs \theta)$.
This implies a monotonic non-decreasing dependence of any correlation function $\langle f\rangle$, including $C_{\bs m}(\bs\kappa,\bs 0;P)$, on any coupling constant $\kappa_b\ge0$.
The Villain version of the same model has similar properties, since the potential $V_\sigma(\varphi)$ can be approximated to an arbitrary precision with a chain of phases with pairwise Wilson couplings \cite{Frohlich-Spenser-1981}.
In particular, generalized GKS inequalities apply for the averages $C_{\bs m}(\bs\kappa,\bs0;P)$ in the Villain form of the model, which are also monotonically non-decreasing as a function of any coupling.

The second type of constraints concerns the free energy $F(\bs \kappa,\bs \varphi;P)\equiv -\ln Z(\bs \kappa,\bs \varphi;P)$, or, more precisely, the free energy cost associated with the background phases $\bs \varphi$,
\begin{equation}
\Delta_{\bs \varphi}F(\bs \kappa;P)\equiv F(\bs \kappa,\bs\varphi;P)-F(\bs \kappa,\bs 0;P).
\label{eq:most-general-delta-F}
\end{equation}
Again, the free energy cost is non-negative, $\Delta_{\bs \varphi}F(\bs \kappa;P)\ge0$, and it is a non-decreasing
function of the coupling constants $\kappa_b$.
This can be obtained from the generalized GKS inequalities for the dual form of the model, where the free energy cost $\Delta_{\bs \varphi}F$ is mapped into the logarithm of a correlation function parameterized by the phases $\bs \varphi$, up to a sign, with the coupling constants inverted.

An important (albeit nearly self-evident) consequence of the monotonicity of the free energy cost \ref{eq:most-general-delta-F} is a generalization of the inequalities (17) and (18) from Ref.~\cite{Kovalev-Prabhakar-Dumer-Pryadko-2018} to all GKP codes.
These inequalities imply an upper bound for the success probability of ML decoding, see Eq.~\ref{eq:succ-prob-init}, in terms of the partition functions in the absence of quenched disorder.
Namely, in the case of CSS-like codes, the bound reads
\begin{equation}
\label{eq:most-general-decoding-upper-bound}
\mathbb{P}_{\rm succ}^{(\rm ML)}\le {Z_{\bs 0}(\bs 0)\over \sum_{\bs c}Z_{\bs c}(\bs 0)},
\end{equation}
where the summation is over all $2^k$ inequivalent binary codewords that correspond to the entire group of $X$-type CSS logical operators.
When applied to the case of repeated syndrome measurement for the toric-GKP code, the numerator in Eq.~\ref{eq:most-general-decoding-upper-bound} is the partition function \ref{eq:Z-epsilon-delta} with zero argument, while the denominator is the sum of the same partition function $Z_{\bs 0}(\bs 0)$ with those for the three remaining non-trivial sectors, $Z_{\bs c}(\bs 0)$.
For the existence of an ML-decodable region it is necessary that these contributions vanish in the large-system limit, namely, when both the distance of the toric code $L$ and the number of layers $M$ diverge.

Typically examined are the Wilson loops, they are correlation functions in the form \ref{eq:most-general-corr-f},
\begin{equation}
{\cal W}_\Omega\equiv \left\langle e^{i\sum_{p\in \Omega}(\nabla\times A)_p}\right\rangle \equiv \left\langle e^{i\sum_{b\in \partial\Omega} A_b} \right\rangle, 
\label{eq:Wilson-loop}
\end{equation}
where $\Omega$ is some oriented surface, a set of plaquettes, with the boundary $\partial\Omega$, and the gauge-invariant two-point correlation function
\begin{equation}
{\cal C}_{\Pi}\equiv \left\langle e^{i(\theta_u-\theta_v)-iq\sum_{b\in\Pi}A_{b}}\right\rangle,
\label{eq:corr-function}
\end{equation}
where $\Pi\equiv \Pi(u,v)$ is a directed path (sequence of bonds) on the cubic lattice connecting vertices $u$ and $v$.
Quite generally, in a high-temperature phase, when coupling constants $\kappa$ and $\lambda$ in Eq.~\ref{eq:3D-U1-model} are sufficiently small, the correlation functions \ref{eq:Wilson-loop} and \ref{eq:corr-function} are characterized by the \emph{area law}: ${\cal W}_\Omega$ has an upper bound that decays exponentially with the area of $\Omega$, the minimum number of plaquettes that are needed to form the boundary $\partial\Omega$, while ${\cal C}_\Pi$ has an upper bound that decays exponentially with the distance between the points $u$ and $v$.
A low-temperature, or strong-coupling asymptotic form is qualitatively different and is characterized by the \emph{perimeter law}, where ${\cal W}_\Omega$ scales exponentially with the length $|\partial \Omega|$ of the perimeter, while ${\cal C}_\Pi$ becomes a constant or falls as a power law of the distance; the latter is the case below the Berezinskii-Kosterlitz-Thouless (BKT) transition in two dimensions.

\section{Proof of Theorem \ref{thm:nogo} }
\label{sec:proof}

For convenience, we rename the $\hat{p}_j$ and $\hat{q}_j$ operators of the oscillators as $\hat{r}_k$, i.e.
\[\forall k \in[2n],\quad\hat{r}_k = \left \{\begin{matrix}
\hat{p}_k\quad\text{ when }j\leq n\\
\hat{q}_{k-n}\quad\text{ when }j>n.
\end{matrix} \right .
\]
Let $\bs{g}_j$ be the real vector corresponding to the $j$th nullifier $\hat{O}(\bs{g}_j) = \bs{g}_j\cdot\hat{\bs{r}}$ of the code, see Appendix~\ref{sec:twoclasses}.
One can extend the set of nullifiers to a full canonical linear transformation given by a real $2n\times 2n$ matrix $A$ defining new variables $\hat{R}_k$, as follows
\[\bs{\hat{R}} = A\bs{\hat{r}}^\T,\qquad\text{i.e. }\hat{R}_k = \sum_{j=1}^{2n}A_{kj}\hat{r}_j.\]
In order to preserve the commutation relation the condition on the matrix $A$ is that it preserves the symplectic form $S$. This can be expressed in two ways:
\begin{equation}A S A^\T = S\quad\text{ or }\quad A^\T S A = S\quad\text{ with }\quad S = \begin{pmatrix}
0 & \id_{n\times n}\\
-\id_{n\times n} & 0
\end{pmatrix}.\label{eq:Asympl}\end{equation}
The matrix $A$ can be decomposed in blocks
\[A = \bordermatrix{
	& \leftarrow 2n \rightarrow \cr
	n-k\updownarrow & G\cr 
	\;\;\;k\quad\updownarrow & P\cr 
	n-k\updownarrow & D\cr
	\;\;\;k\quad\updownarrow& Q}
= 
\bordermatrix{
	& \leftarrow n \rightarrow & \leftarrow n \rightarrow \cr
	& G_p & G_q \cr 
	& P_p & P_q\cr 
	& D_p & D_q\cr
	& Q_p & Q_q},
\]
where the rows of $G$ are the nullifiers. The rows of $D$ represent the corresponding conjugated variables which will be called \emph{pure errors} (they are sometimes referred to as de-stabilizers).
These pure errors give a convenient basis for expressing an error which is compatible with a given syndrome which will be used below.
The rows of $P$ resp. $Q$ represent the logical $\hat{p}$ (resp. $\hat{q}$) operators of the code as linear combinations of the original $\hat{p}_i$ and $\hat{q}_j$.
The subscript $p$ (respectively $q$) indicates the $\hat{p}$ part (respectively $\hat{q}$ part) of the operators.
Inside the code space the operators $\hat{O}\left (\bs{g}_j\right )$ only take the value 0.
Let's assume that a displacement happens along a pure error direction $\bs{d}_j$, say $\hat{U}\left (-\lambda\bs{d}_j\right )$.  
Then, measuring $\hat{O}(\bs{g}_j)$ (equivalently $\hat{U}\left (\eta\bs{g}_j\right )$ for all $\eta\in\mathbb{R}$) would give outcome $\lambda \in \mathbb{R}$ called the syndrome, since
\[\hat{U}(\eta\bs{g}_j)\left [\hat{U}(-\lambda\bs{d}_j)\ket{\Psi}\right ] = \e^{i\lambda\eta}\hat{U}(-\lambda\bs{d}_j)\hat{U}(\eta\bs{g}_j)\ket{\Psi} = \e^{i\lambda\eta}\left [\hat{U}(-\lambda\bs{d}_j)\ket{\Psi}\right ].\]
Note that the logical operators $\hat{O}(\bs{p}_k)$ or $\hat{O}(\bs{q}_\ell)$, or nullifiers, act on the code space without affecting the measurement of $\hat{O}(\bs{g}_j)$ since they commute with it.

We can express the constraints on the matrix $A$ in Eq.~\eqref{eq:Asympl} in terms of the matrices $G,D,P$ and $Q$. The blocks which should be equal to zero are shown in blue and the blocks proportional to the identity are shown in green. The first condition gives
\begin{equation}
ASA^\T = \begin{pmatrix}
\highlight{cyan}{GSG^\T} & \highlight{cyan}{GSP^\T} & \highlight{green}{GSD^\T} & \highlight{cyan}{GSQ^\T}\\[.4em]
\highlight{cyan}{PSG^\T} & \highlight{cyan}{PSP^\T} & \highlight{cyan}{PSD^\T} & \highlight{green}{PSQ^\T}\\[.4em]
\highlight{green}{DSG^\T} & \highlight{cyan}{DSP^\T} & \highlight{cyan}{DSD^\T} & \highlight{cyan}{DSQ^\T}\\[.4em]
\highlight{cyan}{QSG^\T} & \highlight{green}{QSP^\T}& \highlight{cyan}{QSD^\T} & \highlight{cyan}{QSQ^\T}
\end{pmatrix}
= \begin{pmatrix}
0 & 0 & \id & 0\\
0 & 0 & 0 & \id\\
-\id & 0 & 0 & 0\\
0 & -\id & 0 & 0
\end{pmatrix} = S,\label{eq:Asympldetails}
\end{equation}
while the second implies
\begin{equation}
 \id = A^\T SAS^T =\begin{pmatrix}
\highlight{green}{D_q^\T G_p + Q_q^\T P_p - G_q^\T D_p - P_q^\T Q_p} & \highlight{cyan}{D_q^\T G_q + Q_q^\T P_q - G_q^\T D_q - P_q^\T Q_q}\\[.6em]
\highlight{cyan}{-D_p^\T G_p - Q_p^\T P_p + G_p^\T D_p + P_p^\T Q_p} & \highlight{green}{-D_p^\T G_q - Q_p^\T P_q + G_p^\T D_q + P_p^\T Q_q}.
\end{pmatrix}\label{eq:Asympldetailsinverse}
\end{equation}

\subsection{Logical Error Model under Displacement Errors}

We consider the Gaussian displacement noise model on every oscillator as in Eq.~\eqref{eq:GDC}.  Given a realization of the displacement error as a vector of real amplitudes, $\bs{e}^\prime\in\mathbb{R}^{2n}$, one can compute the vector of syndromes $\bs{q}\in\mathbb{R}^{n-k}$, and given $\bs{q}$, a candidate error with the same syndrome, $\bs{e}\in\mathbb{R}^{2n}$,
\[\bs{q} = \bs{e}^\prime SG^\T,\qquad\bs{e} = -\bs{q}D.\]

In addition, when any two errors $\bs{e}$ and $\bs{e}^{\prime}$ have the same syndrome $\bs{q}$, then they can only differ by a stabilizer and a logical operator and one has 
\[\exists\bs{u}_c,\bs{v}_c\in\mathbb{R}^k,\exists\bs{a}\in\mathbb{R}^{n-k},\quad\bs{e}^\prime = \bs{e} + \bs{u}_c P + \bs{v}_c Q+\bs{a}G=\bs{e} + \bs{c}C+\bs{a}G.\]
where we have used the notation
\[C := \begin{pmatrix}
P\\
Q
\end{pmatrix},\qquad \bs{c} = \begin{pmatrix}
\bs{u}_c & \bs{v}_c
\end{pmatrix}.
\]

We can compute the associated partition function, $Z_\bs{c}(\bs{e})$ from Eq.~\eqref{eq:part}, for the error $\bs{e}$ and the sector equivalent to $\bs{c}$,
\begin{equation}
Z_{\bs{c}}(\bs{e}) = \left (\frac{1}{\sqrt{2\pi\sigma_0^2}}\right )^{n}\int\prod_{j=0}^{n-k-1}\!{\rm d}a_j\exp\left (-\frac{\left (\bs{c}C+\bs{a}G+\bs{e}\right )\left (\bs{c}C+\bs{a}G+\bs{e}\right )^\T}{2\sigma_0^2}\right ). \label{eq:explZce}
\end{equation}
This integral can be evaluated since it is Gaussian, resulting, after some manipulations, in                                                                            
\begin{equation}
Z_{\bs{c}}(\bs{e}) =\mathcal{C}(\bs{e})\exp\left (-\frac{1}{2\sigma_0^2}\Big (\bs{c} - \bs{\mu}(\bs{e})\Big )\Sigma^{-1} \Big (\bs{c} - \bs{\mu}(\bs{e})\Big )^\T\right ),\label{eq:end}
\end{equation}
where $\mathcal{C}(\bs{e})$ only depends on $\bs{e}$. The covariance matrix $\Sigma$ and the off-set vector, $\bs{\mu}(\bs{e})$, are defined as
\begin{equation}
\Sigma^{-1} = C^\prime {C^{\prime}}^\T,\quad\text{and}\quad\bs{\mu}(\bs{e}) = - \Sigma C^\prime \bs{e}^\T\label{eq:covariancematrix}
\end{equation}
where
\begin{equation}
C^\prime=C \Pi_{G^\perp},\quad\text{and}\quad \Pi_{G^\perp} =\id - G^T (G G^T)^{-1} G.\label{eq:Cprime}
\end{equation}
Remark that $\Pi_{G^\perp}$ is the projector onto $\ker(G)$ along ${\rm im}(G^\T)$ \footnote{This is not an orthogonal projection as $\ker(G)$ and ${\rm im}(G^\T)$ are not orthogonal to one another.}.
Indeed it is easy to check that: $\Pi_{G^\perp}^2=\Pi_{G^\perp}$, $\ker(G)\subset{\rm im}(\Pi_{G^\perp})$, ${\rm im}(G^\T)\subset\ker(\Pi_{G^\perp})$, and the dimensions coincide.

One can see that Eq.~\eqref{eq:end} describes a multivariate Gaussian distribution over the logical variable $\bs{c}$, hence its maximum is readily given by the mean value, $\bs{\mu}(\bs e)$, see Eq.~\eqref{eq:covariancematrix}.
This means that given the error $\bs{e}$, one can directly express its most likely error class: it is $[\bs{e} + \bs{\mu}(\bs e)]$.
Using this, one can directly compute the probability density of a remaining logical error after ML decoding as given by Eq.~\eqref{eq:succ-decoding}:
\begin{equation}
\boxed{\mathbb{P}^{(\rm ML)}(\bs c) = \int \mathrm{d}\bs e\,\mathbb{P}(\bs e) \,\frac{Z_{\bs{\mu}(\bs{e}) + \bs c}(\bs e)}{
	\int{\rm d}{\bs b}\,Z_{\bs b}(\bs e)} = \frac{1}{N}\exp\left (-\frac{1}{2\sigma_0^2}\bs{c}\Sigma^{-1}\bs{c}^\T\right ),}\label{eq:logical_displace}
\end{equation}
where $N$ is a normalization constant.
All possible dependence on the choice and size of the code is contained in the covariance matrix $\Sigma$, leading to some rescaled displacement noise model.
Recall that the logical variable vector $\bs{c}\in\mathbb{R}^{2k}$ represents all $k$ pairs of conjugated logical operators.
We will prove in Section~\ref{sec:det} below that for any linear oscillator code, the eigenvalues of $\Sigma^{-1}$ can be paired between corresponding conjugated logical operators, denoted $(\lambda_j^p,\lambda_j^q)_{j\in[n]}$, and are such that
\begin{equation}
\forall j \in [n],\;\lambda_j^p\lambda_j^q = 1.\label{eq:squeezedlambdas}
\end{equation}
This means that the noise remaining after ML decoding on the logical variables is identical to the original physical noise except for a possible squeezing between each logical operator and its conjugated pair.

\subsection{Eigenvalues of the Covariance Matrix}
\label{sec:det}

First we remark that $C^{\prime}$, appearing in $\Sigma^{-1}=C^\prime{C^\prime}^\T$ in Eq.~\eqref{eq:covariancematrix}, correspond to a valid basis for the logical operators, ---we call this the spread-out logical basis---.
This basis is obtained by adding linear combination of stabilizer generators to $C$.
Such addition can be summarized, using Eq.~\eqref{eq:Cprime}, by the matrix equation,
\[C^\prime = C - CG^\T\left (GG^\T\right )^{-1}G = C + \Lambda G,\]
where $\Lambda$ is a $2k\times(n-k)$ matrix defining the linear combination of stabilizer generators added to the logical operators. One can verify that one can replace $C$ by $C^\prime$ and still satisfy the constraints of Eq.~\eqref{eq:Asympl}, if one appropriately redefines the pure errors to be \[D^\prime = D + \Lambda^\T S_{2k}^\T C,\]
where $S_{2k}$ is the symplectic form of size $2k\times 2k$.
One can also check that this choice of basis is the only choice which enforces the following constraint
\begin{align}
C^\prime G^\T = \left (C + \Lambda G\right )G^\T = 0.
\label{eq:spread-cond}
\end{align}
Indeed, solving for $\Lambda$, using the fact that $G$ is full rank and therefore $GG^\T$ is invertible, gives
\begin{equation}
C^\prime = C - CG^T\left (GG^\T\right )^{-1}G.\label{eq:spreatoutlog}
\end{equation}

Now we use this spread-out logical basis to prove Eq.~\eqref{eq:squeezedlambdas}.
We thus consider that we already have chosen the spread-out logical basis, so $C^\prime=C$, and want to get information about the eigenvalues of $\Sigma^{-1} = CC^\T$.\\

We use the diagonal block of Eq.~\eqref{eq:Asympldetailsinverse} as well as the block only about logical operators in Eq.~\eqref{eq:Asympldetails}:
\begin{empheq}[left=\empheqlbrace]{align}
-D_p^\T G_q - Q_p^\T P_q + G_p^\T D_q + P_p^\T Q_q &= \id\label{eq:orth1}\\[.6em]
D_q^\T G_p + Q_q^\T P_p - G_q^\T D_p - P_q^\T Q_p &= \id\label{eq:orth2}\\[.6em]
PSQ^\T = P_pQ_q^\T - P_qQ_p^\T  &= \id,\label{eq:logicalsympl}
\end{empheq}
We multiply Eq.~\eqref{eq:orth1} on the left by $P_p$ and on the right by $Q_q^T$. We also multiply Eq.~\eqref{eq:orth2} on the left by $P_q$ and on the right by $Q_p^T$. Then we take the difference, i.e. we consider
\[P_p\ref{eq:orth1}Q_q^\T - P_q\ref{eq:orth2}Q_p^\T.\]
By Eq.~\eqref{eq:logicalsympl}, the right hand-side is still the identity, so we have
\[
\begin{array}{c c c c c}
\id = && P_pG_p^\T D_qQ_q^\T & + & P_qG_q^\T D_pQ_p^\T\\[.6em]
& - & P_qD_q^\T G_pQ_p^\T & - & P_pD_p^\T G_qQ_q^\T\\[.6em]
& + & P_pP_p^\T Q_qQ_q^\T & + & P_qP_q^\T Q_pQ_p^\T\\[.6em]
& - & P_pQ_p^\T P_qQ_q^\T & - & P_qQ_q^\T P_pQ_p^\T
\end{array}.
\]
Since the logical operators are in their spread-out basis we have the corresponding equations
\begin{empheq}[left=\empheqlbrace]{align*}
P_pG_p^\T + P_qG_q^\T &= 0\\[.6em]
G_pQ_p^\T + G_qQ_q^\T &= 0,
\end{empheq}
and can write
\begin{align*}
\id &= P_q\left (D_q^\T G_q - G_q^\T D_q\right )Q_q^\T\\
&+ P_p\left (D_p^\T G_p - G_p^\T D_p\right )Q_p^\T\\
& +  P_p\left (P_p^\T Q_q - Q_p^\T P_q\right )Q_q^\T\\
& + P_q\left (P_q^\T Q_p - Q_q^\T P_p\right )Q_p^\T.
\end{align*}
One can now recognize that the off-diagonal terms in Eq.~\eqref{eq:Asympldetailsinverse} can be used to make a equation only about logical variables, using
\begin{empheq}[left=\empheqlbrace]{align*}
D_q^\T G_q - G_q^\T D_q &= P_q^\T Q_q - Q_q^\T P_q\\
D_p^\T G_p - G_p^\T D_p &= P_p^\T Q_p - Q_p^\T P_p.
\end{empheq}
Hence we have the following matrix identity
\begin{equation}
\id = PP^\T QQ^\T - PQ^\T PQ^\T\label{eq:fulldeteq}
\end{equation}

In the next section, \ref{sec:spread-ortho}, we show that there always exist a logical basis which is both spread-out as well as orthogonal.
Hence for this new basis, Eq.~\ref{eq:spread-cond} is satisfied, as well as
\begin{equation}
\tilde{P}\tilde{P}^\T ={\rm Diag}\left (\lambda_j^p\right ),\quad \tilde{Q}\tilde{Q}^\T ={\rm Diag}\left (\lambda_j^q\right ),\quad\text{and}\quad \tilde{P}\tilde{Q}^\T = 0\label{eq:orthobasis}.
\end{equation}
Therefore deriving Eq.~\eqref{eq:fulldeteq} in this basis yields
\begin{equation}
\id = \tilde{P}\tilde{P}^\T \tilde{Q}\tilde{Q}^\T= {\rm Diag}\left (\lambda_j^p\lambda_j^q\right ),\label{eq:finaleq}
\end{equation}
and therefore
\begin{equation}
\forall j\in[k],\;\lambda_j^p\lambda_j^q = 1.
\end{equation}
Lastly recall that starting with this choice of basis, one has
\begin{equation}
\Sigma^{-1} =\tilde{C}\tilde{C}^\T = \begin{pmatrix}
\tilde{P}\tilde{P}^\T & \tilde{P}\tilde{Q}^\T\\[.6em]
\tilde{Q}\tilde{P}^\T & \tilde{Q}\tilde{Q}^\T
\end{pmatrix} = \begin{pmatrix}
{\rm Diag}\left (\lambda_j^p\right ) & 0\\[.6em]
0 & {\rm Diag}\left (\lambda_j^q\right )
\end{pmatrix}.\label{eq:mat-CCT}
\end{equation}\qed

\subsection{Existence of a Spread-out $\&$ Orthogonal Logical Operator Basis}
\label{sec:spread-ortho}

We start from a given stabilizer matrix $G$ and denote the rows by $\left \{\bs{g}_1,\ldots,\bs{g}_{n-k}\right \}$.
We showed above that one can always find a logical basis, $P$ and $Q$ with rows $\{\bs{p}_1, \ldots, \bs{p}_k\}$ and $\{\bs{q}_1, \ldots, \bs{q}_k\}$, which is orthogonal to the stabilizer generators, i.e
\[\mathcal{L}_k = {\rm Span}\left \{\bs{p}_1, \ldots, \bs{p}_k,\bs{q}_1, \ldots, \bs{q}_k\right \} \perp \mathcal{G} = {\rm Span}\left \{\bs{g}_1, \ldots, \bs{g}_{n-k}\right \}.\]
We want to construct a new symplectic and orthogonal basis, $\left \{\tilde{\bs{p}}_1, \ldots, \tilde{\bs{p}}_k,\tilde{\bs{q}}_1, \ldots, \tilde{\bs{q}}_k\right \}$, for the $2k$-dimensional space $\mathcal{L}_k$.
One can first find an orthogonal basis for $\mathcal{L}_k$ by the Gram-Schmidt process for the regular inner product, denote it $\{\bs{c}^k_1,\ldots,\bs{c}^k_{2k}\}$.
We will now construct a new basis pair by pair, decreasing at each step the dimension of the space $\mathcal{L}_k$ by $2$.
We can choose \[\tilde{\bs{p}}_1 = \bs{c}_1^k.\]
There has to exist a conjugated pair, $\tilde{\bs{q}}_1$, in the space spanned by $\left \{\bs{c}_2^k,\ldots,\bs{c}_{2k}^k\right \}$.
Indeed, all stabilizer generators, pure errors as well as $\bs{c}_1^k$ have a trivial symplectic product with $\tilde{\bs{p}}_1$, hence any conjugated $\tilde{\bs{q}}_1\in{\rm Span}\left \{\bs{c}_2^k,\ldots,\bs{c}_{2k}^k\right \}$.
So the following equation 
\[\bs{\lambda}O_kS\tilde{\bs{p}}_1^\T = -1,\]
where $O_k$ are the basis vectors $\left \{\bs{c}_2^k,\ldots,\bs{c}_{2k}^k\right \}$ stacked in rows, has a solution for $\bs{\lambda}$.
Then we can choose \[\tilde{\bs{q}}_1 = \bs{\lambda}O_k.\]
Note that the created pair is indeed conjugated for the symplectic inner product as well has orthogonal.
Moreover they are composed only of linear combinations of the original $\bs{p}_i$s and $\bs{q}_j$s, so they do have trivial symplectic product with stabilizer generators and pure errors and they are orthogonal to the stabilizer generators.
Now we define the $2(k-1)$-dimensional subspace $\mathcal{L}_{k-1}$, as 
\begin{align*}
\mathcal{L}_{k-1} &= {\rm Span}\left \{\bs{c}^k_2\left (\id - \Pi_{\tilde{\bs{q}}_1}\right ),\ldots,\bs{c}^k_{2k}\left (\id - \Pi_{\tilde{\bs{q}}_1}\right )\right \}\\
&={\rm Span}\left \{\bs{c}^{k-1}_1,\ldots,\bs{c}^{k-1}_{2(k-1)}\right \},
\end{align*}
where $\Pi_{\tilde{\bs{q}}_1}$ is the orthogonal projector onto $\tilde{\bs{q}}_1$, and the $\bs{c}^{k-1}_{j}$ form a new orthogonal basis for this space.
We can then repeat the procedure until we reach $\mathcal{L}_0 = \{0\}$.
To summarize we have constructed a new symplectic and orthogonal basis $\tilde{C} = \begin{pmatrix}
\tilde{P}\\\tilde{Q}
\end{pmatrix}$ obeying Eq.~\eqref{eq:orthobasis}.

\section{Continuous-Variable Toric-Code}
\label{sec:CV-toric}

\begin{figure}[htbp]
	\begin{minipage}[l]{0.49\textwidth}
		\centering
		\includegraphics[height=14em]{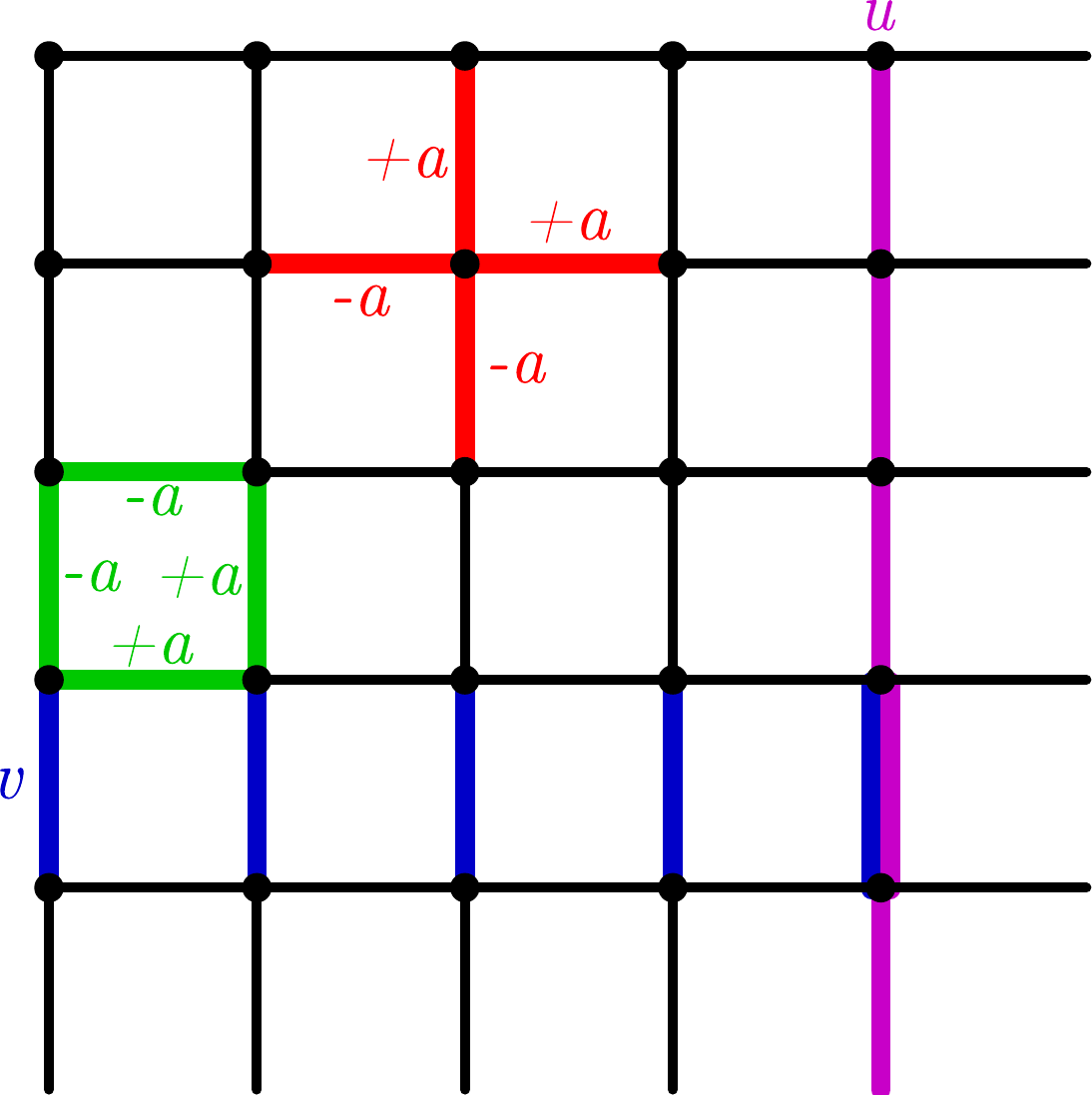}
	\end{minipage}
	\begin{minipage}[r]{0.49\textwidth}
		\centering
		\includegraphics[height=14em]{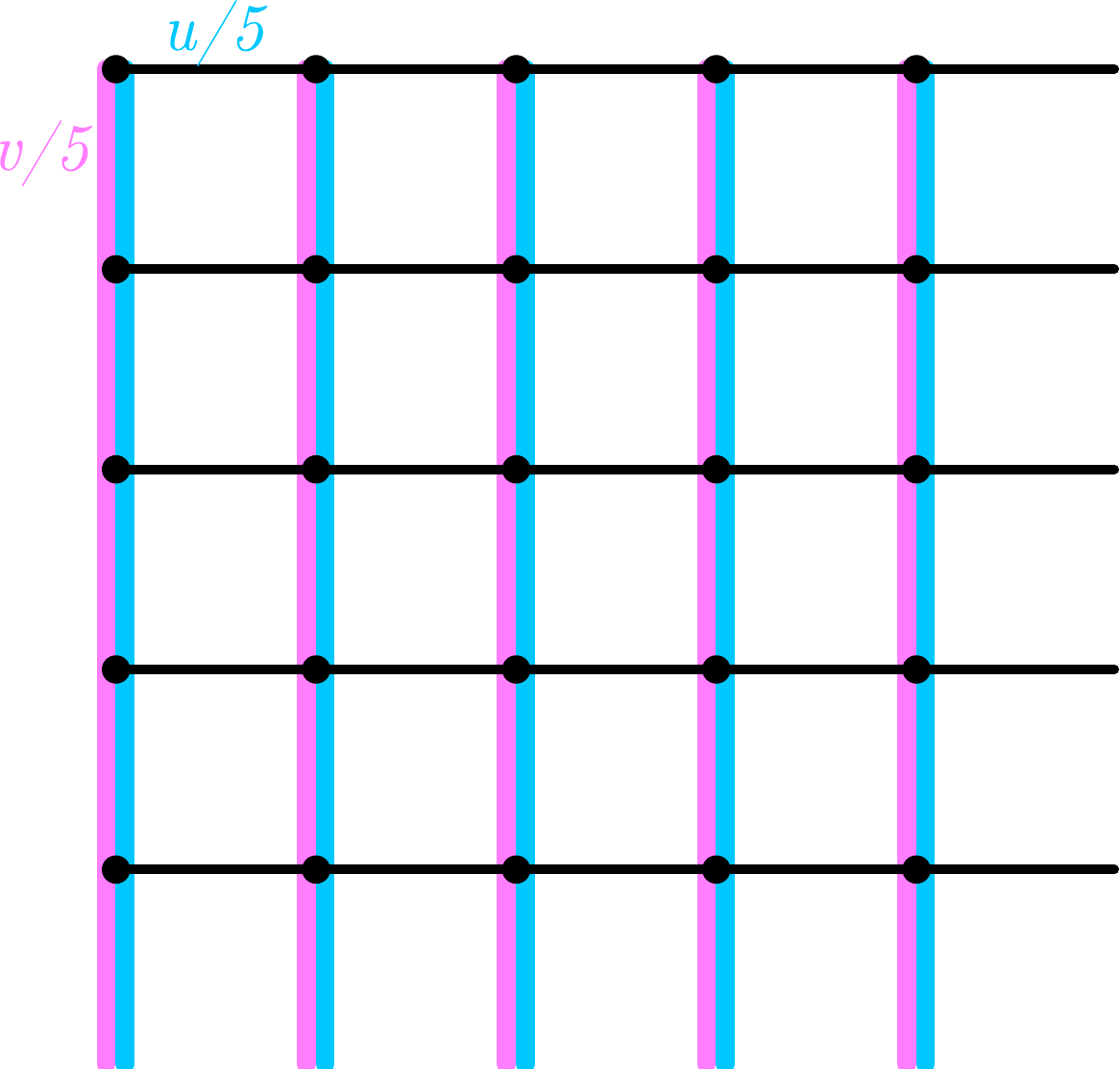}
	\end{minipage}
	\caption{In both figures periodic boundary conditions are assumed.
		(Left) Example of stabilizer checks and logical operators for the distance-5 continuous-variable toric-code. The stabilizer checks are also shown in Fig.~\ref{fig:setup-concat1}. The support of the logical $\bs{p_1}$ (in dark blue) and $\bs{q}_1$ (in dark purple) of one of the encoded oscillators is depicted. Shifts of strength $v$ on the support of $\bs{p}_1$ is a logical $\bs{p}$-shift of strength $v$ of the first oscillator.
		(Right) The support of the spread-out version of the logical operators $\bs{p}_1$ and $\bs{q}_1$. Both operators have the same support (in light blue and light purple) and one can verify that these logical operators are orthogonal to the stabilizer checks as well as commuting with them. A logical $\bs{p}$-shift of strength $v$ on the first oscillator is now realized by applying $v/5$ on the support of the spread-out $\bs{p}_1$.
		In general, if the torus had dimension $L_x \times L_y$, the spread-out $\bs{p}_1$ would have a shift rescaling of $\frac{1}{L_x}$ over the whole lattice while $\bs{q}_1$ would a shift rescaling of $\frac{1}{L_y}$. At the same time $\bs{p}_2$ (resp. $\bs{q}_2$) would have rescaling $\frac{1}{L_y}$ (resp. $\frac{1}{L_x}$).}
	\label{fig:cv_toric}
\end{figure}
In this section we examine the continuous-variable toric code \cite{zhang+:anyon_kitaev,BMT:review} as an example of a continuous-variable topological code. 
Since the toric code is a homological code, it is easy to convert it from a $\mathbb{Z}_2$-code to a $\mathbb{R}$-code using orientation to add appropriate minus signs to the stabilizer checks. 

The stabilizer checks are shown on the left in Fig.~\ref{fig:cv_toric}. Since the toric code is a CSS code we will always have the orthogonality condition $PQ^\T=0$. The vectors $\bs{p}_1$ and $\bs{p}_2$ are the two rows of $P$ while the vectors $\bs{q}_1$ and $\bs{q}_2$ are the two rows of $Q$. On the left, one sees the usual string-like $\bs{p}_1$ and $\bs{q}_1$. The spread-out basis for the code can be computed and is shown on the right of Fig.~\ref{fig:cv_toric}. The name {\em spread-out basis} comes from the fact that these logical operators have support over the full lattice: the continuous-variable stabilizer checks have been used to spread out or distribute their support over the lattice.

Computing $C C^T$, Eq.~\ref{eq:mat-CCT}, with $C$ being the spread-out logical operators in the general case of different dimensions $L_x$ and $L_y$ gives directly a diagonal matrix:
\begin{equation}
C C^T={\rm diag}\left(\lambda_1^p, \lambda_2^p,\lambda_1^q, \lambda_2^q\right) = {\rm diag}\left(\bs{p}_1^2, \bs{p}_2^2,\bs{q}_1^2, \bs{q}_2^2\right),
\end{equation}
with
\begin{alignat*}{2}
\lambda_1^p = \bs{p}_1^2 &= \sum_{e_x}\frac{1}{L_y^2} = \frac{L_xL_y}{L_y^2} = \frac{L_x}{L_y}, \qquad& \lambda_2^p = \bs{p}_2^2 &= \sum_{e_y}\frac{1}{L_x^2} = \frac{L_xL_y}{L_x^2} = \frac{L_y}{L_x},\\
\lambda_1^q = \bs{q}_1^2 &= \sum_{e_x}\frac{1}{L_x^2} = \frac{L_xL_y}{L_x^2} = \frac{L_y}{L_x},\qquad& \lambda_2^q = \bs{q}_2^2 &= \sum_{e_y}\frac{1}{L_y^2} = \frac{L_xL_y}{L_y^2} = \frac{L_x}{L_y}.
\end{alignat*}
One sees that it is possible to choose the squeezing amount by
choosing the ratio of the dimensions $L_x/L_y$. In particular, with
Eq.~\ref{eq:logical_displace}, one sees that the first encoded oscillator
experiences a Gaussian displacement noise model with variance
$\sigma_p^2=\frac{L_y}{L_x}\sigma_0^2$ for $\hat{p}$ and variance
$\sigma_q^2=\frac{L_x}{L_y}\sigma_0^2 $ for $\hat{q}$. For the second
oscillator the quadrature squeezing goes in the other direction.

\end{document}